\documentclass{article}
\pdfoutput=1
\usepackage{jheppub}

\usepackage{amsmath}
\usepackage{ulem}
\usepackage{float}

\newcommand{\roughly}[1]{\mathrel{\raise.3ex\hbox{$#1$\kern-0.85em
\lower1ex\hbox{$\sim$}}}}

\newcommand{\lsim}{\roughly<}

\def\exd{{\hbox{d}}}

\def\bea{\begin{eqnarray}}
\def\eea{\end{eqnarray}}
\def\be{\begin{equation}}
\def\ee{\end{equation}}

\def\tr{{\rm tr}\,}
\def\Tr{{\rm Tr}\,}

\def\bfx{{\bf x}}

\def\ssD{{\scriptscriptstyle D}}

\def\ssI{{\scriptscriptstyle I}}

\def\ssM{{\scriptscriptstyle M}}

\def\ssT{{\scriptscriptstyle T}}

\def\cA{\mathcal{A}}

\def\cC{\mathcal{C}}

\def\cG{\mathcal{G}}
\def\cH{\mathcal{H}}

\def\cK{\mathcal{K}}
\def\cL{\mathcal{L}}

\def\cO{\mathcal{O}}
\def\cP{\mathcal{P}}
\def\cQ{\mathcal{Q}}

\def\cS{\mathcal{S}}

\def\mathi{i}

\def\llangle{\langle\hspace{-2mm}\langle}
\def\rrangle{\rangle\hspace{-2mm}\rangle}

\def\nn{\nonumber}

\def\({\left(}
\def\){\right)}

\def\pref#1{(\ref{#1})}

\usepackage{braket}
\newcommand{\bR}{ {\mathbb{R}} }
\newcommand{\bC}{ {\mathbb{C}} }
\newcommand{\bx}{ {\mathbf{x}} }

\newcommand{\vac}{ {\Omega} }
\newcommand{\Mink}{{\mathrm{M}}}
\newcommand{\WM}{ {\mathcal{W}_{\ssM}} }
\newcommand{\RM}{ {\mathcal{R}_{\ssM}} }
\newcommand{\CM}{ {\mathcal{C}_{\ssM}} }
\newcommand{\CMp}{ {\mathcal{C}_{\ssM}^{\prime}} }
\newcommand{\DMp}{ {\Delta_{\ssM}^{\prime}} }
\newcommand{\SM}{ {\mathcal{S}_{\ssM}} }
\newcommand{\DM}{ {\Delta_{\ssM}} }
\newcommand{\WO}{ {\mathcal{W}_{\Omega}} }
\newcommand{\RO}{ {\mathcal{R}_{\Omega}} }
\newcommand{\CO}{ {\mathcal{C}_{\Omega}} }
\newcommand{\SO}{ {\mathcal{S}_{\Omega}} }
\newcommand{\DO}{ {\Delta_{\Omega}} }
\newcommand{\COp}{ {\mathcal{C}_{\Omega}^{\prime}} }
\newcommand{\DOp}{ {\Delta_{\Omega}^{\prime}} }

\newcommand{\tD}{ {\xi_\ssD} }

\newcommand{\tT}{ {\xi_\ssT} }

\def\smath#1{\text{\scalebox{.85}{$#1$}}}
\def\sfrac#1#2{\smath{\frac{#1}{#2}}}
            
\usepackage{footmisc} 

\usepackage{graphicx}
\usepackage{dcolumn}
\usepackage{bm}
\usepackage{color}
\usepackage{ulem}

\title{Hot Accelerated Qubits: Decoherence, Thermalization, Secular Growth 
and Reliable Late-time Predictions}

\author{Greg Kaplanek}
\author{and C.P.~Burgess}
\affiliation[a]{Department of Physics \& Astronomy, McMaster University, Hamilton, Ontario, L8S 4M1, Canada}
\affiliation[b]{Perimeter Institute for Theoretical Physics, Waterloo, Ontario, N2L 2Y5, Canada }
\emailAdd{kaplaneg@mcmaster.ca}
\emailAdd{cburgess@perimeterinstitute.ca}

\date{}

\abstract {We compute how an accelerating qubit coupled to a scalar field -- {\it i.e.}~an Unruh-DeWitt detector --  evolves in flat space, with an emphasis on its late-time behaviour. When calculable, the qubit evolves towards a thermal state for a field prepared in the Minkowski vacuum, with the approach to this limit controlled by two different time-scales. For a free field we compute both of these as functions of the difference between qubit energy levels, the dimensionless qubit/field coupling constant, the scalar field mass and the qubit's proper acceleration. Both time-scales differ from the Candelas-Deutsch-Sciama transition rate traditionally computed for Unruh-DeWitt detectors, which we show describes the qubit's early-time evolution away from the vacuum rather than its late-time approach to equilibrium. For small enough couplings and sufficiently late times the evolution is Markovian and described by a Lindblad equation, which we derive in detail from first principles as a special instance of Open EFT methods designed to handle a breakdown of late-time perturbative predictions due to the presence of secular growth. We show how this growth is resummed in this example to give reliable information about late-time evolution including both qubit/field interactions and field self-interactions. By allowing very explicit treatment, the qubit/field system allows a systematic assessment of the approximations needed when exploring late-time evolution, in a way that lends itself to gravitational applications. It also allows a comparison of these approximations with those -- {\it e.g.}~the `rotating-wave' approximation -- widely made in the open-system literature (which is aimed more at atomic transitions and lasers). }

\begin{document}

\maketitle
\section{Introduction}
It is an old observation that physical processes occurring in spacetimes with horizons share many features of open systems. This resemblance is based on the fact that any parts of the system that cross the horizon become eternally beyond the reach of some observers (those outside the horizon) \cite{Hawking:1976ra,Gibbons:1977mu,Israel:1976ur,Kiefer:2003nw,Giulini:1996nw,StochInf,Starobinsky:1994bd,Tsamis:2005hd}.  Open systems are the natural description of this because they (by definition) are systems for which measurements are only performed on some subsystem (call it sector $A$) and so for which it is possible to marginalize over the rest (the `environment,' sector $B$) when making predictions \cite{DaviesOQS,Alicki,Kubo,Gardiner,Weiss,Breuer:2002pc,Rivas,Schaller}. 

In a gravitational context sector $B$ might consist of degrees of freedom on the far side of an observer's horizon, with sector $A$ representing the degrees of freedom on the near side.  This makes the effective description of systems outside a horizon more like the effective description of a particle moving through a medium ({\it e.g.}~photons moving through water, or neutrinos within the Sun) than a traditional Wilsonian effective field theory. The difference arises because although a Wilsonian description also divides a system into observed and unobserved sectors (low and high energies), this division is based on a conserved quantity (energy). The same is not true for a horizon (or a medium), where no selection rules prevent particles and information from being exchanged and entangled between the observed and unobserved sectors.

Several less well-appreciated side-effects come along with such an open-system perspective, including phenomena potentially of relevance to predictions in both cosmology and within black-hole spacetimes. The one of most interest in this paper is the phenomenon of {\it secular growth}, and the related inevitability of the breakdown of perturbation theory at very late times. Strictly speaking secular growth is the phenomenon where the coefficients, $c_n(t)$, of a perturbative evaluation of some observable, 
\be
   \cO(t) = \sum_{n} c_n(t) \, g^n \,,
\ee
in powers of some small coupling $|g| \ll 1$, are time-dependent and grow without bound at late times ({\it i.e.}~$|c_n(t)|$ remains unbounded  as $t \to \infty$) \cite{Nayfeh,Tanaka:1975fn,Chen:1994zza,Chen:1995ena,Bender:1996je,Berges:2004yj,Urakawa:2009my}. Secular growth such as this is disturbing because it represents a breakdown of the ability to predict late-time behaviour using perturbative methods. It is also generic to open systems, for which $g$ is typically a measure of the strength of the coupling between sectors $A$ and $B$. 

This kind of secular perturbative breakdown is actually generic in almost all of physics, and ultimately arises because the time-evolution operator is given by $U(t) = \exp[-i(H_0 + H_{\rm int})t]$. No matter how small an interaction Hamiltonian $H_{\rm int}$ might be, there is always a time after which perturbative evaluation of $U(t)$ breaks down. Even very small effects can accumulate to become significant over long enough periods of time. The scattering of wave-packets is an exception to this generic late-time observation, because in this case interactions turn off once the overlap of the scattering wave-packets goes to zero. As a result, late-time perturbative breakdown tends to be less familiar to particle physicists, for whom scattering is often the main calculational focus.

The good news is that there are well-developed tools for making reliable late-time predictions without having to exactly solve the full theory. These involve resummations of various types that turn perturbative calculations into reliable late-time inferences. These usually rely on a renormalization-group type of argument, in which a perturbative calculation computed in powers of $g$ is resummed to all orders in (say) $g^2 t$ while dropping contributions of order $g^n t$ with $n > 2$. Such a resummation is performed by deriving a differential evolution equation that ultimately has a broader domain of validity --- and so whose solutions can be trusted to later times --- than did the initial perturbative calculation. 

Perhaps the simplest example along these lines is the prediction of exponential laws for radioactive decay. In this case the number of atoms surviving un-decayed at a time $t$ is given by 
\be \label{ExpDecay}
   n(t) = n_0 \, \exp[- \Gamma(t-t_0)] \,, 
\ee
where $n_0$ is the number of atoms present at the initial time $t_0$. In this expression the decay rate, $\Gamma$, is usually computed in perturbation theory and the question arises why \pref{ExpDecay} is trusted rather than just the expression $n(t) = n_0 [1 - \Gamma (t-t_0)]$ that directly emerges from a leading-order perturbative calculation. Ultimately eq.~\pref{ExpDecay} is justified by the statistical independence of the decay for each atom, which very generally\footnote{Of course, it is possible to ask questions about decays for which \pref{ExpDecay} is not the right description, and to do so one must choose questions that invalidate the reasoning leading to \pref{ExpDecayDiffer}.} implies the differential relation
\be \label{ExpDecayDiffer}
   \frac{\exd n}{\exd t} = - \Gamma \, n \,,
\ee
for all $t$. This differential relation is sufficient to justify \pref{ExpDecay}, and perturbation theory is then simply used to derive the value of the coefficient $\Gamma$.

An argument similar in spirit to this -- though different in detail -- is also often available for computing the late-time limit of open systems \cite{OpenEFT1, OpenEFT2, OpenEFT3,OpenEFT4,OpenEFT5,OpenEFT6,OpenEFT7,OpenEFT8,Agon:2014uxa, Boyanovsky:2015tba, Boyanovsky:2015jen, Nelson:2016kjm, Hollowood:2017bil, Shandera:2017qkg, Agon:2017oia, Martin:2018zbe, Martin:2018lin}. We argue here that for many OpenEFT applications it is the Lindblad equation \cite{Lindblad:1975ef, Gorini:1976cm} that is the desired evolution equation for these purposes. The evolution equation obtained for qubits differs from \pref{ExpDecayDiffer} because of unitarity-based feedback on the decay rate of the initial state as the other state becomes significantly occupied. 

A central purpose of this paper is to develop and explore these arguments in a particularly simple example for which all steps can be made explicit and concrete. To this end we examine the late-time limit of a qubit --- {\it i.e.}~a two-level system whose energies are split by an amount $\omega$ --- coupled to a quantum scalar field, $\phi$, within flat spacetime. The field is prepared in its (Minkowski) vacuum state and the qubit is assumed to move along a uniformly accelerated trajectory, and the resulting evolution is followed as functions of the scalar mass $m$, the qubit energy spacing $\omega$, the acceleration parameter $a$ and the qubit/field coupling constant $g$.

These tools allow the following late-time inferences about the accelerating qubit coupled to a field:
\begin{itemize}
\item There is a robust asymptotic evolution to a late-time, static thermal state. (This is a general result for systems coupled to any environment that exhibits thermal properties, in the sense that correlation functions obey the Kubo-Martin-Schwinger (KMS) condition  \cite{Kubo:1957mj,Martin:1959jp} described in later sections.) 
\item In general the evolution of the qubit's reduced $2\times 2$ density matrix, $\boldsymbol{\varrho}(t)$, can be developed explicitly in powers of its coupling with the field (or environment). This evolution is described by a {\it Nakajima-Zwanzig} equation \cite{Nakajima, Zwanzig} for which $\partial_t \boldsymbol{\varrho}(t)$ depends on the details of an integral over $\boldsymbol{\varrho}(t')$ over its entire evolution history at earlier times.

\item At late-enough times the slow evolution of the qubit's reduced density matrix becomes Markovian inasmuch as $\partial_t \boldsymbol{\varrho}(t)$ at a given time eventually can be predicted given only $\boldsymbol{\varrho}(t)$ at the same time (no longer depending on its entire past history). We find the general constraints on the parameters of the problem which control this regime of Markovian evolution. 
\item Although the problem of secular growth prevents directly calculating the evolution of $\boldsymbol{\varrho}(t)$ at late times, the differential evolution relating $\partial_t \boldsymbol{\varrho}(t)$ to $\boldsymbol{\varrho}(t)$ during the Markovian regime proves to have a broader domain of validity than its perturbative derivation, and so lends itself to the same kind of arguments that allow the robust inference of a decay law like \pref{ExpDecay} from \pref{ExpDecayDiffer}. This allows the inference of late-time behaviour to all orders in $g^2 t$ as $t \to \infty$ and $g \to 0$.
\item Diagonal and off-diagonal components of $\boldsymbol{\varrho}(t)$ turn out to evolve independent of each other, and the Markovian regime that dominates at very late times consequently reveals two separate relaxation time-scales that govern the exponential approach to the late-time thermal state. We call these time-scales $\tD$ and $\tT$, and they respectively describe the evolution of the off-diagonal and diagonal parts of the qubit density matrix. Both $\tD$ and $\tT$ differ from the classic transition rate for Unruh-DeWitt detector excitation computed many years ago \cite{Unruh:1976db, DeWitt:1980hx, Sciama:1981hr}. 
\end{itemize}

Because the uniformly-accelerated-qubit/free-field system is particularly simple, calculations for it can be extremely explicit. This allows the above general remarks to be quantified in more detail in terms of the system parameters $m$, $a$, $g$ and $\omega$. In particular:
\begin{itemize}
\item The temperature of the late-time static thermal limit for the qubit is the standard Unruh result: $T = {a}/{(2\pi)}$. This temperature provides the natural correlation scale for the qubit's environment.
\item Markovian evolution emerges when two conditions are satisfied. First, attention must be focussed on late enough times; which for the accelerated qubit means proper times $\tau \gg 1/a$. Second, the proper time-scale $\xi = 1/\Gamma$ of the evolution must also be large, again compared with $1/a$. 
\item Late-time evolution generically becomes Markovian in perturbation theory (in powers of $|g | \ll 1$) because the predicted evolution rate vanishes at zero coupling, and at weak coupling is $\Gamma \sim g^2 a F(m/a, \omega/a)$ for a calculable dimensionless function of two arguments, $F(x,y)$. The condition $\Gamma \ll a$ is therefore automatic whenever $F$ is order unity. (As described explicitly below, small $g$ need not suffice in extreme parameter limits for which $F$ is not order unity.) 
\item In the Markovian limit the evolution equation for $\boldsymbol{\varrho}(\tau)$ can be written as a Lindblad equation, and we show why its solutions have a broader domain of validity at late times than does straight-up perturbation theory itself. In particular, although direct calculation of $\boldsymbol{\varrho}(\tau)$ in perturbation theory naively breaks down once $a \tau \simeq 1/g^2$ (due to secular-growth effects), solutions to the Lindblad equation are nevertheless trustworthy even when $a \tau \sim \cO(1/g^2)$. Integrating the Lindblad equation turns out to resum all orders in $g^2 a \tau$ while dropping terms of order $g^4 a \tau$. 
\item The Lindblad equation we find satisfies automatically the positivity conditions (explained below) required by unitarity over its entire evolution, provided one stays strictly within the domain of validity of its derivation. Apparent positivity violations only arise if one strays outside of this domain. Because of this there is no need to impose an extra coarse-graining, as is often done in the literature -- called there the rotating-wave approximation (RWA). The RWA is commonly used to remedy the appearance of positivity-violating terms in the evolution equations for $\boldsymbol{\varrho}(\tau)$. We find this unnecessary if one stays ruthlessly within the domain of one's approximations.
\item The predicted Markovian time-scales can be explicitly solved in terms of Bessel functions and robustly satisfy $\tD > \tT$, so the off-diagonal components of $\boldsymbol{\varrho}(\tau)$ relax to equilibrium more slowly than do the diagonal components.
\item Relaxation becomes exponentially inefficient in the limit $m \gg a$, for want of thermally occupied excited field states. In the large-mass limit the relaxation times take the asymptotic form 
\be
\tD  =  2 \tT  \simeq \frac{8\pi}{a g^2} \,\mathrm{sech} \left( \frac{\pi \omega}{a} \right) \; e^{{2m}/{a}} \,,
\ee
and so diverge in the limit $m\to \infty$, as expected as the scalar field decouples. 

\item For small masses, $m \ll a$, the relaxation times instead become
\be 
\tD  =  2 \tT  \simeq \frac{4\pi}{\omega g^2} \tanh\left( \frac{\pi \omega }{a} \right) \Bigl[ 1 + \cO(m^2/\omega^2) \Bigr]   \,,\label{introMinkxis}
\ee
where the explicit form for the subdominant $\cO(m^2/\omega^2)$ term is given in eq.~\pref{massivexiRind} below.  
\end{itemize}

All of the above statements apply when the accelerating qubit is coupled to a free field. In a final section we also briefly explore the effects of field self-interactions of the form ${\cal H}_{\rm int} = \lambda \phi^4/4!$ for a massless scalar. As discussed in \cite{Burgess:2018sou}, secular-growth effects also obstruct the validity of perturbing in $\lambda$ at late times, although it is also argued there that late-time behaviour can be controllably resummed by incorporating a small mass shift $\delta m^2 = {\lambda a^2}/{96\pi^2}$ into the zeroeth-order theory. 

Using this observation in the previous small-mass results then gives the leading power of $\lambda$ in the late-time relaxation times, which turn out to have the form
\be 
\tD = 2 \tT \simeq \frac{4\pi}{g^2 \omega} \tanh\left( \frac{\pi \omega }{a} \right) + \frac{\lambda a^2}{48 \pi g^2 \omega^3} \tanh\left( \frac{\pi \omega}{a} \right)  \left\{  1 - \frac{\cos\left[ \frac{\omega}{a} \log\left( \frac{\lambda}{384 \pi^2} \right) - \zeta \right]}{\sqrt{ ( \omega / a )^2 + 1 }} \right\} \,,
\ee
with $\zeta$ as given in \pref{zetadef}. Notice that although this correction is small when $\lambda$ is small, it is also not analytic at $\lambda = 0$ -- a consequence of the non-perturbative nature of the late-time resummation.

In the remainder of this paper these results are presented in the following way. In \S\ref{sec:QiS} the dynamics for the basic qubit/scalar-field system is set up and solved perturbatively in powers of the qubit-scalar coupling $g$. Several general properties of the scalar-field correlation functions of interest are displayed, including the `KMS' conditions \cite{Kubo:1957mj, Martin:1959jp} that encode detailed balance and so are sufficient for the late-time limit to be thermal. \S\ref{sec:LTL} follows this up with a summary of how the evolution of the reduced qubit density matrix can be described by a Nakajima-Zwanzig equation from which can be derived the Markovian-Lindblad limit. (This derivation is amplified somewhat in Appendix \ref{App:NakZwan}.) \S\ref{sec:AcceleratedMarkovian} applies these general techniques to the specific example of an accelerating qubit in Minkowski space, and then \S\ref{sec:LateTimeLimit} focuses on identifying reliable statements that can be made at late times. Our conclusions are briefly summarized in \S\ref{sec:Conc} and a several intermediate steps and results are given in a series of Appendices.

 \section{Qubits in space}
 \label{sec:QiS}

Our goal in this --- and a companion paper \cite{qubitdS} --- is to follow the evolution of the state of a qubit that moves along various world-lines in simple spacetimes while interacting with a quantum field. Of particular interest is the reliable calculation of its behaviour at very late times. Following earlier studies of Unruh-DeWitt detectors we here work perturbatively in the qubit/field coupling, $g$, although unlike early work \cite{Unruh:1976db, Sciama:1981hr, DeWitt:1980hx} our results are not implicitly restricted to the regime $1 \ll a \tau \ll \cO(1/g^2)$ (where $\tau$ is the qubit's proper time); in resummed form they are also valid for time-scales $a \tau \simeq \cO(1/g^2)$. 

\subsection{The setup}
\label{sec:setup}

We study a Unruh-DeWitt detector along the lines of that first introduced in \cite{Unruh:1976db,DeWitt:1980hx} and consider a 2-level qubit with free Hamiltonian
\begin{eqnarray}
\mathfrak{h}  =  \frac{\omega}{2} \, {\boldsymbol{{\sigma_3}}}  =  \left( \begin{matrix} {\omega}/{2} & 0 \\ 0 & - {\omega}/{2} \end{matrix} \right) \,,
\end{eqnarray}
which denotes the difference of the two qubit energies by $\omega >0$. We suppose the qubit moves along a trajectory $x^\mu = y^\mu(\tau)$ within a given spacetime geometry, along which $\tau$ is the proper time as measured with the spacetime metric $\exd s^2 = g_{\mu\nu} \, \exd x^\mu \, \exd x^\nu$, so that
\begin{eqnarray}
  g_{\mu\nu}[y(\tau)] \, \frac{\exd y^\mu}{\exd \tau} \, \frac{\exd y^\nu}{\exd \tau} = -1 \,.
\end{eqnarray}

The qubit is assumed to couple to a free real Klein-Gordon scalar field $\phi$ with mass $m$. Assuming the spacetime is static and admits a foliation with metric $\exd s^2 = - \exd t^2 + \gamma_{ij} \, \exd x^i \, \exd x^j$, the scalar's Klein-Gordon Hamiltonian can be written
\begin{eqnarray}
\cH & = & \int_{\Sigma_t} \exd^3\bx\ \sqrt{\gamma} \left[ \frac{1}{2} \dot{\phi}^2 + \frac{1}{2} \gamma^{ij}  \partial_i \phi\, \partial_j \phi + m^2 \phi^2 \right] \label{freefield}
\end{eqnarray}
where $\Sigma_t$ is a sheet of fixed $t=x^0$ and over-dots denote $\partial_t$.

The Hilbert space of states for the combined qubit/field system is the product of the Fock space for the field with the qubit's two-dimensional space of states. The free Hamiltonian (before adding a qubit-field coupling) acting on the full Hilbert space is then
\begin{eqnarray}
H_0 & = & \cH \otimes \boldsymbol{I} + \mathcal{I} \otimes \mathfrak{h}\, \frac{\exd\tau}{\exd t} \,, \label{H0definition}
\end{eqnarray}
where $\boldsymbol{I}$ and $\mathcal{I}$ are identity operators, and the factor ${\exd\tau}/{\exd t}$ is included so that $H_0$ generates translations in $t$ (whereas $\mathfrak{h}$ generates translations in the qubit's proper time $\tau$). 

Finally, the qubit/field coupling is described by the interaction Hamiltonian 
\begin{eqnarray}
  H_{\mathrm{int}} & = & g \, \phi\big[ y(\tau)\big] \otimes \mathfrak{m} \, \frac{\exd\tau}{\exd t} \label{Hint}
\end{eqnarray}
where the dimensionless coupling $0 < g \ll 1$ is small enough to justify a perturbative treatment and we follow a choice often made in the literature by picking the $2\times 2$ matrix here to be $\mathfrak{m} = \boldsymbol{\sigma_{1}}$, so that $H_{\rm int}$ drives transitions between the eigenstates of $\mathfrak{h}$. The complete hamiltonian\footnote{In \S\ref{sec:LTL} we also include a counter-term $H_{\rm ct} = \frac12 \, g^2 \omega_1\, ( \mathcal{I} \otimes \boldsymbol{\sigma_{3}} )({\exd \tau}/{\exd t}) \subset H_{\mathrm{int}}$ in the interaction Hamiltonian because at $\cO(g^2)$ the qubit/field interaction \pref{Hint} alters the qubit gap size $\omega \to \omega + g^2 \omega_1$. Inclusion of this counterterm ensures the parameter $\omega$ continues to represent the gap size to this order. \label{FootnoteCounterterm}} is then given by $H = H_0 + H_{\rm int}$.

The rest of this paper computes how the state of the qubit responds to its motion through the spacetime while interacting with the quantum field, with the field assumed to be prepared in its vacuum $\ket{\vac}$. The initial qubit state is taken to be uncorrelated with the field degrees of freedom, with
\begin{eqnarray}
\rho(0) \ = \ \ket{ \vac } \bra{ \vac } \otimes \boldsymbol{\varrho_0} \label{uncorrelated1}
\end{eqnarray}
where $\boldsymbol{\varrho_0}$ is the qubit's initial $2\times 2 $ hermitian density matrix, that satisfies $\tr \boldsymbol{\varrho_0} = 1$. To the extent that only qubit observables are measured the problem of time evolution is completely solved once the time-dependence of the reduced density matrix is known where
\begin{eqnarray} \label{RedRhoDef}
\boldsymbol{\varrho}(t) := \underset{\phi}{\mathrm{Tr}}\big[ \rho(t)\big]  
\end{eqnarray}
This takes a partial trace over the field theory subspace of the full density matrix $\rho(t)$ describing the quantum state of the entire system, and given \pref{uncorrelated1} has the initial condition
\be \label{varrhoIC}
  \boldsymbol{\varrho}(0) = \boldsymbol{\varrho_0} \,.
\ee

\subsection{Perturbative time evolution}

The strategy is to compute $\boldsymbol{\varrho}(t)$ directly from its definition after first computing $\rho(t)$ perturbatively in powers of the small coupling $g$. To this end we switch to the interaction picture, and suppose when doing so that the time coordinate $t$ and the qubit proper time $\tau$ are sychronized to ensure $\tau(t=0)=0$. In this case the time-evolution operator for the free system becomes
\begin{eqnarray}
U_0(t,0) \ = \ \mathcal{T}\exp\left( - i \int_0^t \exd s\ H_0 \right) \ = \ e^{- i \cH t} \otimes e^{- i \mathfrak{h} \tau(t)} \,,
\end{eqnarray}
and so the interaction-picture interaction Hamiltonian $V(t)$ is given by
\begin{eqnarray}
V(t) \ = \ U^{\dagger}_0(t,0) H_{\mathrm{int}} U_0(t,0) \ = \ g \, \phi^\ssI[y (\tau)] \otimes \mathfrak{m}^\ssI(\tau) \, \frac{\exd\tau}{\exd t} \,, \label{intintH}
\end{eqnarray}
where (as usual) the interaction-picture field is related to the Schr\"odinger-picture field by $\phi^\ssI (\bfx,t) := e^{+ i \cH t} \phi(\bfx) e^{- i \cH t}$ and the interaction-picture qubit interaction is given by 
\be
  \mathfrak{m}^\ssI(\tau) := e^{+ i \mathfrak{h} \tau} \mathfrak{m} \,e^{- i \mathfrak{h} \tau} \,. 
\ee

The interaction-picture density matrix is similarly given by $\rho^\ssI (t) := U^{\dagger}_0(t,0) \rho(t) \,U_0(t,0)$ and we use the notation
\begin{eqnarray}
\boldsymbol{\varrho}^\ssI (t) \ := \  e^{+ i \mathfrak{h} \tau(t) } \boldsymbol{\varrho}(t) \,e^{- i \mathfrak{h} \tau(t)} \label{INTpicturereducedstate}
\end{eqnarray}
for the interaction-picture reduced density matrix in the qubit sector. With these definitions the evolution of $\rho^{\ssI}(t)$ is then found by integrating the Liouville equation, which in the interaction picture states
\be \label{FullLiouville}
 \frac{\partial \rho^{\ssI}}{\partial t} = -i \Bigl[ V(t) \,, \rho^\ssI(t) \Bigr] \,.
\ee
Standard arguments give the perturbative solution to this equation, which to second order in $V$ is
\be \label{2ndOrderLiouville}
    \rho^\ssI(t) = \rho(0) -i \int_0^t \exd s_1\; \big[V(s_1), \rho(0) \big]  + (-i )^2 \int_0^t \exd s_1 \int_0^{s_1} \exd s_2 \; \big[ V(s_2), \big[ V(s_1),  \rho(0) \big] \big] + \cO(V^3) \,.
\ee
Taking the trace of this expression and using the definition \pref{RedRhoDef} gives the desired perturbative prediction for the time-dependence of the reduced density matrix. 

Specializing to the qubit-field hamiltonian considered here and using the uncorrelated initial condition \pref{uncorrelated1} gives --- after noting that $\langle \Omega | \phi^\ssI(\bfx,t) | \Omega \rangle = 0$ kills the first-order term --- the comparatively simple result
\begin{eqnarray} \label{PertVarRho}
\boldsymbol{\varrho}^\ssI (\tau) & = & \boldsymbol{\varrho_0} - g^2 \int_0^\tau \exd s_1 \int_0^{s_1} \exd s_2  \bigg\{ \WO(s_2-s_1) \ \big[ \mathfrak{m}^\ssI(s_2) , \mathfrak{m}^\ssI(s_1) \, \boldsymbol{\varrho_0} \big] \\
& & \qquad\qquad\qquad\qquad\qquad\qquad + \WO^{\ast}(s_2-s_1) \ \big[ \mathfrak{m}^\ssI(s_2) , \mathfrak{m}^\ssI(s_1) \, \boldsymbol{\varrho_0} \big]^{\dagger} \bigg\} + \cO(g^4) \,,\nn
\end{eqnarray}
which changes integration variable from $t$ to the qubit's proper time $\tau(t)$ and uses $\WO$ to denote the Wightman function evaluated along the qubit's trajectory
\begin{eqnarray} \label{WightmanDef}
   \WO(\tau_1 - \tau_ 2) & := & \braket{ \vac | \phi\big[  y (\tau_1) \big] \phi\big[ y (\tau_2) \big] |\vac } \,.
\end{eqnarray}
The static nature of the spacetime -- and the fact that $|\Omega \rangle$ is the ground state -- ensures the Wightman function depends only on $t_2 - t_1$. In what follows spacetime trajectories are used for which this ensures the proper times also only appear through the difference $\tau_2 - \tau_1$. 

Eq.~\pref{PertVarRho} can be made more explicit by choosing the specific interaction $\mathfrak{m} = \boldsymbol{\sigma_1}$ and supposing the qubit begins in its ground state, $\boldsymbol{\varrho_0} = \ket{ \downarrow} \bra{\downarrow} = \frac12(\boldsymbol{I}- \boldsymbol{\sigma_3})$. These choices imply
\begin{eqnarray}
\left[ \boldsymbol{\varrho}(0) \mathfrak{m}(s_1), \mathfrak{m}(s_2) \right] & = & - e ^{- i \omega (s_1 - s_2)} \boldsymbol{\sigma_3}
\end{eqnarray}
which (switching back to the Schr\"{o}dinger-picture) results in the simple expression
\begin{eqnarray}
\boldsymbol{\varrho}(\tau) & = & \ket{\downarrow} \bra{\downarrow} + g^2 \boldsymbol{\sigma_3} \int_0^\tau \exd s_1 \int_0^{\tau} \exd s_2 \ \WO(s_1 - s_2) \, e^{- i \omega (s_1 - s_2)} + \cO(g^4) \,. \label{perturbation1}
\end{eqnarray}
As many authors have observed \cite{Sciama:1981hr, Takagi:1986kn, DeWitt:1980hx, Birrell:1982ix, Hinton:1984ht}, the second term in this expression need not vanish and when it does not the qubit is in general excited for $\tau > 0$ by its interaction with the field even though both qubit and field begin in their respective ground states. 

\subsection{Wightman function}

Notice the definition \pref{WightmanDef} and the hermiticity of $\phi$ ensure that $\WO$ enjoys the symmetry
\begin{eqnarray}
\WO^{\ast}(\tau) & = & \WO(-\tau) \,. \label{conjugateW1}
\end{eqnarray}
In the cases examined in later sections $\WO$ also satisfies a skew periodicity in imaginary time 
\begin{eqnarray}
\WO(\tau -  i \beta) \ = \ \WO(-\tau) \ , \label{KMS1}
\end{eqnarray}
known as the Kubo-Martin-Schwinger (KMS) condition \cite{Kubo:1957mj,Martin:1959jp}. This property turns out to be sufficient to ensure that $\boldsymbol{\varrho}(\tau)$ asymptotes to a thermal state, with temperature $T = 1/\beta$, as we verify explicitly in later sections.

The Wightman function also has a universal singularity \cite{Hadamard} in the limit that its arguments become light-like separated, with $\langle \Omega | \phi(x) \phi(x') | \Omega \rangle$ diverging proportional to 
\begin{eqnarray}
\langle \Omega | \phi(x) \phi(x') | \Omega \rangle \sim \frac{\Delta^{1/2}}{4\pi^2}\; \frac{1}{ \hat{\sigma}(x,x') + 2 i \epsilon [ T(x) - T(y) ] + \epsilon^2} \,,
\end{eqnarray}
up to logarithmic terms, where $\hat{\sigma}$ is the square of the geodesic distance between $x$ and $x'$ while $\Delta$ is the Van Vleck-Morette determinant \cite{DeWitt:1960fc}, $T$ is a globally-defined future-increasing function of time \cite{Kay:1988mu,Radzikowski:1996pa}, and $\epsilon$ is an infinitesimal that defines how to handle singularities associated with integrating through $\hat{\sigma} = 0$. With our later choices for qubit trajectories this implies 
\be \label{Wsmalltau}
 \WO(s) \propto \frac{1}{(s - i \epsilon)^2} \,,
\ee
in the limit of small proper-time separation.

\subsection{Integration issues}

Considerable effort has been put into computing (\ref{perturbation1}) for various trajectories and spacetimes. There are two kinds of difficulties when evaluating the integrals, with potential divergences arising because the integrand is singular as $s_1 \to s_2$ and when taking the late-time limit $\tau \to \infty$. 

In what follows we find that the potential divergence at $s_1 \to s_2$ is less severe than it might have been, largely due to the $i \epsilon$ behaviour appearing in \pref{Wsmalltau}. There is a residual logarithmic divergence in this limit that we find gets cancelled when one renormalizes the bare parameter $\omega$ (as might be expected for a short-distance divergence). These divergences are much discussed in the literature \cite{Svaiter:1992xt,Higuchi:1993cya,sriramkumar,Louko:2006zv}, which sometimes approaches them differently than we do here.  

Our main focus is on problems associated with the long-time limit, $\tau \to \infty$. Part of the problem in this regime is well-understood, and is generic to time-translation invariant systems. Because transition {\it rates} are time-independent for such systems, transition {\it probabilities} grow linearly with time. Unbounded growth of \pref{perturbation1} at $\tau \to \infty$ should therefore be avoided if one simply computes the late-time transition rate \cite{Sciama:1981hr} by differentiating \pref{perturbation1}, leading to:
\begin{eqnarray}
\lim_{\tau \to \infty} \frac{\partial \boldsymbol{\varrho}(\tau) }{\partial \tau} \ = \ g^2 \boldsymbol{\sigma_3} \, \RO(\omega) \label{uprate1}
\end{eqnarray}
where $\RO$ denotes the Wightman function's Fourier transform
\begin{eqnarray}
\RO(\omega) & := & \int_{-\infty}^{\infty} \exd\tau \ \WO(\tau) \, e^{- i \omega \tau} \ . \label{ROrate}
\end{eqnarray}

This approach of computing the derivative of the transition probability indeed leads to the correct result in other physical situations, such as when computing decay rates for unstable particles. We argue below that for the accelerating-qubit/quantum-field system considered here the same approach is only partly successful, since it only properly captures evolution for times $1 \ll a \tau \ll \cO(1/g^2)$ and does not properly capture the later-time limit when $a \tau \simeq \cO(1/g^2)$. Our goal is to reliably infer evolution in this regime at much-later times.

The property (\ref{conjugateW1}) allows the above (and later) formulae to be written in other useful ways. It implies in particular that $\mathrm{Re}[\WO(\tau)]$ is an even function of $\tau$ while $\mathrm{Im}[\WO(\tau)]$ is an odd function of $\tau$. As a result the above Fourier transform can be decomposed as the sum
\begin{eqnarray}
\RO(\omega) & = & \CO(\omega) + \SO(\omega) \label{RCSsum}
\end{eqnarray}
where we define the useful integrals
\begin{eqnarray} 
\CO(\omega) & \equiv & \int_{-\infty}^{\infty} \exd\tau \ \mathrm{Re}[ \WO(\tau) ] \cos(\omega \tau) \label{Crate} \\
\hbox{and} \quad
\SO(\omega) & \equiv & \int_{-\infty}^{\infty} \exd\tau \ \mathrm{Im}[ \WO(\tau) ] \sin( \omega \tau) \,.\label{Srate}
\end{eqnarray} 

More can be said about these integrals when the Wightman function satisfies the thermal KMS relation (\ref{KMS1}), as does the Minkowski-vacuum Wightman function when evaluated along an accelerating world-line (as is well-known, and we see below explicitly). Whenever this is true the rate integral $\RO(\omega)$ obeys the detailed-balance relation \cite{Takagi:1986kn}
\begin{eqnarray}
 \RO(\omega) - e^{-\beta  \omega } \RO( - \omega) \ = \ 0 \ . \label{detailedbalance}
\end{eqnarray}
Since $\CO(\omega)$ and $\SO(\omega)$ are even and odd in $\omega$ respectively, the detailed-balance relation also implies a relation between $\CO(\omega)$ and $\SO(\omega)$:
\begin{eqnarray}
\frac{\SO(\omega)}{\CO(\omega)} & = & - \tanh\left( \frac{\beta \omega }{2} \right) , \label{ratioSCgeneral}
\end{eqnarray}
from which several other useful relations also follow: 
\begin{eqnarray}
\RO(\omega) & = & \frac{2}{e^{\beta\omega} + 1}\; \CO(\omega)  =  -\, \frac{2}{e^{\beta\omega} - 1}\; \SO(\omega) \,. \label{rateSC1}
\end{eqnarray}

\section{The Nakajima-Zwanzig equation and the Markovian limit}
\label{sec:LTL}

Although eq.~\pref{uprate1} is a standard result, something must be wrong with it. In particular, \pref{uprate1} does not describe an approach to a static late-time thermal state, as might be expected at late times when coupled to a thermal environment. 

This (and the following) section develop the tools needed to see why \pref{uprate1} goes wrong, and to see what must be done to reliably access the qubit's late-time behaviour. In particular we argue that (unlike for particle decays) rates like $g^2 \RO$ of \pref{uprate1} only accurately capture the transition rate for an intermediate range of times and not the evolution of $\boldsymbol{\varrho}(\tau)$ at very late times.

We now argue that a better perturbative approach to the late-time evolution of the reduced density matrix $\boldsymbol{\varrho}(\tau)$ is given by the Nakajima-Zwanzig equation \cite{Nakajima,Zwanzig,DaviesOQS,Alicki,Kubo,Gardiner,Weiss,Breuer:2002pc,Rivas,Schaller} (whose derivation is briefly summarized in Appendix \ref{App:NakZwan}).  This equation uses the full Liouville equation, \pref{FullLiouville}, to marginalize the rest of the system once and for all, and thereby derive an expression for $\partial_\tau \boldsymbol{\varrho}(\tau)$ that refers only to the interaction hamiltonian and to $\boldsymbol{\varrho}(\tau')$. The result is an integro-differential equation that is displayed explicitly for our qubit system in \S \ref{sec:NZ}. Although this equation in general remains difficult to solve, it is useful because it can be used to show -- as is done in \S \ref{sec:Markovian} -- how things simplify in the late-time limit when evolution becomes approximately Markovian. \S \ref{sec:AcceleratedMarkovian} then explicitly solves the resulting Markovian evolution for the concrete example of an accelerated qubit in Minkowski space.

\subsection{The Nakajima-Zwanzig equation}
\label{sec:NZ}

The derivation of the Nakajima-Zwanzig equation starts with the observation that the traced Liouville equation is hard to use directly because the right-hand side depends on the full density matrix $\rho(\tau)$ rather than just the reduced matrix $\boldsymbol{\varrho}(\tau)$. The Nakajima-Zwanzig equation fixes this by solving for the rest of $\rho$ in terms of $\boldsymbol{\varrho}$ so as to get an equation that involves only the reduced density matrix. This can be done quite generally, but at the expense of making the Liouville equation into an integro-differential equation in which the right-hand side involves an integral over the entire history of $\boldsymbol{\varrho}(\tau)$.

For the qubit/field system described above the result (see Appendix \ref{App:NakZwan}) at second order in $g$ is given in the interaction picture by
\begin{eqnarray} \label{INTpictureNZ} 
\frac{\partial \boldsymbol{\varrho}^\ssI (\tau)}{\partial \tau} &\simeq& g^2 \int_0^\tau \exd s\ \bigg( \WO(\tau - s) \big[ \mathfrak{m}^\ssI(s) \, \boldsymbol{\varrho}^\ssI (s), \mathfrak{m}^\ssI(\tau) \big]  \\
 && \qquad\qquad\qquad\qquad  + \WO^{\ast}(\tau - s) \big[ \mathfrak{m}^\ssI(\tau) , \boldsymbol{\varrho}^\ssI (s)\, \mathfrak{m}^\ssI(s)  \big] \bigg) - i \left[ \frac{g^2 \omega_1}{2} \boldsymbol{\sigma_3} , \boldsymbol{\varrho}^{\ssI}(\tau) \right] \, . \nn
\end{eqnarray}
It is the late-time implications of this equation that are explored for much of the rest of this paper.

The last term of \pref{INTpictureNZ} naively has no counterpart in \pref{PertVarRho}, and so deserves some explanation. It arises because of the inclusion in the interaction Hamiltonian of an $\mathcal{O}(g^2)$ counter-term to the qubit gap $\omega$ \pref{Hint}
\be 
  H_{\mathrm{int}} \to  g \, \phi\big[ y(\tau)\big] \otimes \boldsymbol{\sigma_1} \, \frac{\exd\tau}{\exd t} \ + \  \mathcal{I} \otimes \frac{g^2 \omega_1}{2} \boldsymbol{\sigma_3} \frac{\exd \tau}{\exd t} \,, \label{adjustedHint}
\ee
and introduction that carries with it an associated adjustment of the interaction picture.  As mentioned in footnote \ref{FootnoteCounterterm}, this counter-term arises because at second order in $g$ the qubit/field interaction shifts the energy difference between the two qubit levels, so that $E_\uparrow - E_\downarrow \simeq \omega + g^2 \DO$, with
\be
  \DO(\omega) := 2 \int_0^\infty \exd s\ \mathrm{Re}[\WO(s)] \sin(\omega s) \,. \label{DivergentShift}
\ee
The counterterm of eq.~\pref{adjustedHint} is obtained by redefining the parameter $\omega \to \omega_0 := \omega + g^2 \omega_1$, with $\omega_1 = - \DO$ chosen so that $E_\uparrow - E_\downarrow \simeq \omega_0 + g^2 \DO = \omega + g^2(\omega_1 + \DO ) = \omega$, which ensures the parameter $\omega$ continues to denote the physical qubit level-difference at this order. (This redefinition has the added bonus of cancelling the short-distance logarithmic ultraviolet divergence\footnote{In intermediate-stage manipulations to follow we imagine the divergence in $\DO$ to be regulated, making $\DO$ finite but logarithmically sensitive to the UV regularization scale. In the end our physical predictions do not depend on precisely how this regularization is carried out.} that $\DO$ would otherwise introduce into the evolution of $\boldsymbol{\varrho}^\ssI$ at second order in $g$.) It is important when doing this redefinition to recall that this also involves a slight redefinition of the interaction picture, since the free Hamiltonian, $H_0$, appearing in \pref{intintH} is defined by \pref{H0definition} with $\mathfrak{h}$ built using the physical qubit gap $\omega$, in addition to $H_{\mathrm{int}}$ being adjusted to \pref{adjustedHint}.

Returning to eq.~\pref{INTpictureNZ}, it is clear that this expression agrees with the time derivative of \pref{PertVarRho} if the replacement $\boldsymbol{\varrho}(t) \to \boldsymbol{\varrho_0}$ were made, as would be natural to do given that $\boldsymbol{\varrho}(t)$ and $\boldsymbol{\varrho_0}$ themselves only differ by higher orders in $g$. This shows how the Nakajima-Zwanzig equation reproduces the strict perturbative result at order $g^2$. It also shows how \pref{INTpictureNZ} can also carry information beyond leading order, because of the replacement of $\boldsymbol{\varrho_0}$ with a convolution over $\boldsymbol{\varrho}(t)$. This difference can be important, particularly at late times once even $\cO(g^2)$ changes have had time to modify $\boldsymbol{\varrho}(t)$ significantly from $\boldsymbol{\varrho}_0$. It is this difference that allows \pref{PertVarRho} and \pref{INTpictureNZ} to differ from one another at late times.

Switching the integration variable $s \to \tau - s$ in \pref{INTpictureNZ} and performing some matrix algebra yields the component equations of motion
\be
\frac{\partial {\varrho}^{\ssI}_{11}}{\partial \tau} =  g^2 \int_{-\tau}^{\tau} \exd s \; \WO(s) \, e^{- i \omega s} - 4 g^2 \int_0^{\tau} \exd s\ \mathrm{Re}[\WO(s) ] \, \cos(\omega s) \varrho^{\ssI}_{11}(\tau - s)  \,, \label{rho111} 
\ee
and
\bea
\frac{\partial \varrho^{\ssI}_{12}}{\partial \tau} & = & - i g^2 \omega_1 \; \varrho^{\ssI}_{12}(\tau) - 2 g^2 \int_0^\tau \exd s \ \mathrm{Re}[\WO(s)] e^{+ i \omega s} \varrho_{12}^{\ssI}(\tau - s) \ \label{rho121} \\
&\ & \quad \quad  \quad \quad  \quad \quad  \quad \quad \quad  \quad \quad  \quad \quad + \ 2 g^2  e^{+ 2 i \omega \tau} \int_0^\tau \exd s \ \mathrm{Re}[\WO(s)] e^{- i \omega s} \varrho_{12}^{\ssI\ast}(\tau - s)\,, \nn 
\eea
which use the identities\footnote{ These properties follow from the identities $\tr\boldsymbol{\varrho}(\tau)  = 1$ and $\boldsymbol{\varrho}^{\dagger}(\tau) = \boldsymbol{\varrho}(\tau)$, which are preserved for all $\tau>0$ by \pref{INTpictureNZ}. (Proving the hermiticity identity is easiest using the method of Laplace transforms.)} $\varrho_{22} = 1 - \varrho_{11}$ and $\varrho_{21} = \varrho_{12}^{\ast}$ to eliminate $\varrho_{21}$ and $\varrho_{22}$. Further simplification comes from using the properties $\tr\boldsymbol{\varrho} = 1$ and $\boldsymbol{\varrho}^{\dagger} = \boldsymbol{\varrho}$ 

Eqs.\pref{rho111} and \pref{rho121} are the main results of this section. In particular, they show that the components $\varrho^\ssI_{11}$ and $\varrho^\ssI_{22}$ evolve completely independently of the components $\varrho^\ssI_{12}$ and $\varrho^\ssI_{21}$. It is this independent evolution that implies the existence of two independent relaxation time-scales $\tD$ and $\tT$ in the approximations that follow, with $\tD$ describing the rate with which $\boldsymbol{\varrho}$ diagonalizes, while the other $\tT$ captures the time-scale with which the diagonal elements approach thermal values.

\subsection{The Markovian limit}
\label{sec:Markovian}

So far so good, but eqs.~\pref{rho111} and \pref{rho121} do not yet bring us closer to integrating the system to determine the evolution of $\boldsymbol{\varrho}(t)$. After all, the Nakajima-Zwanzig equation really contains much the same information as does the underlying Liouville equation; just better organized. Its main virtue is to manipulate the right-hand side of the Liouville equation in order to write it completely in terms of the reduced density matrix $\boldsymbol{\varrho}(t)$. This is accomplished at the expense of introducing convolutions over the evolution history, thereby introducing memory effects into the system (which show why both equations are in general difficult to solve). 

Things become simpler, however, if it happens that $\WO(\tau)$ falls off sharply for large $\tau$. In the example to follow it happens that $\WO(\tau)$ falls off exponentially fast on sufficiently large time-scales
\begin{eqnarray}
\WO(\tau) \ \sim \ e^{ - \tau / \tau_c} \ \ \ \ \ \ \ \mathrm{for \ }\tau \gg \tau_c \label{WFallOff}
\end{eqnarray}
for some time-scale $\tau_c$.  When this happens the evolution for $\boldsymbol{\varrho}(t)$ simplifies provided one only tries to predict behaviour that is slow in comparison with the scales over which $\mathrm{Re}[\WO(\tau)]$ varies.\footnote{The relative simplicity coming from a hierarchy of scales between the variations of $\boldsymbol{\varrho}$ and $\WO$ is the `effective' part of Open Effective Field Theories \cite{OpenEFT1, OpenEFT2, OpenEFT3,OpenEFT4,OpenEFT5,OpenEFT6,OpenEFT7,OpenEFT8}.} In this case the function $\boldsymbol{\varrho}^{\ssI}(\tau-s)$ within the integral can be Taylor expanded in powers of $s$ such that
\begin{eqnarray}
\boldsymbol{\varrho}^{\ssI}(\tau - s) \simeq \boldsymbol{\varrho}^{\ssI}(\tau) - s \; \frac{\partial\boldsymbol{\varrho}^{\ssI}(\tau)}{\partial \tau}  + \ldots \label{MarkovSeries}
\end{eqnarray}
with higher terms generically suppressed by a derivative expansion of the form $(\tau_c\, \partial_\tau)^n \boldsymbol{\varrho}^{\ssI}(\tau)$ \cite{Montroll,Barnett} once the integration over $s$ is performed. Because \pref{INTpictureNZ} ensures the derivative of the {\it interaction-picture} state $\partial_\tau{\boldsymbol{\varrho}}^{\ssI}(\tau)$ in \pref{MarkovSeries} is $\mathcal{O}(g^2)$, for small $g$ each power of $\tau_c\, \partial_\tau$ tends automatically to be small.

Because small $g$ automatically suppresses derivatives of $\boldsymbol{\varrho}^\ssI(\tau)$, the key ingredients required for this expansion to be useful are: ($i$)  the existence of a characteristic scale $\tau_c$ beyond which $\WO$ falls to zero, and ($ii$) the requirement that \pref{INTpictureNZ} be evaluated at sufficiently late times, $\tau \gg \tau_c$, that the falloff in $\WO$ is important when evaluating the integral over $s$. In this case the upper limit of the $s$ integration can also be placed at infinity rather than $\tau$, because the integral's support dominantly comes from $s  \lsim \tau_c \ll \tau$. With these approximations the evolution equation \pref{MarkovSeries} becomes 
 \begin{eqnarray} \label{Markov1st} 
 \frac{\partial \boldsymbol{\varrho}^\ssI (\tau)}{\partial \tau} & \simeq & g^2 \int_0^\infty \exd s\ \bigg( \WO(s) \big[ \mathfrak{m}^\ssI(\tau -s) \, \boldsymbol{\varrho}^\ssI (\tau), \mathfrak{m}^\ssI(\tau) \big] \\
&& \qquad\qquad\qquad\qquad + \WO^{\ast}(s) \big[ \mathfrak{m}^\ssI(\tau) , \boldsymbol{\varrho}^\ssI (\tau)\, \mathfrak{m}^\ssI(\tau - s)  \big] \bigg)   - \frac{ig^2 \omega_1}{2}  \left[ \boldsymbol{\sigma_3} , \boldsymbol{\varrho}^{\ssI}(\tau) \right] \,, \nn 
\end{eqnarray}
which is {\it Markovian}, in the sense that $\partial_\tau \boldsymbol{\varrho}^\ssI(\tau)$ depends only on the instantaneous value of $\boldsymbol{\varrho}^\ssI(\tau)$ at the same time, and not on its entire past history. 

In what follows we next explicitly solve the equations of motion for $\varrho^{\ssI}_{11}(\tau)$ and $\varrho^{\ssI}_{12}(\tau)$ in this Markovian regime to quantify the size of $\tau_c \partial_\tau \boldsymbol{\varrho}(\tau)$ and thereby provide more precise conditions for the validity of the Markovian limit of the Nakajima-Zwanzig equation. To this end it is worth focussing on the $(ij)$ components of \pref{Markov1st} (or, equivalently, specializing \pref{rho111} and \pref{rho121} to the Markovian regime).

\subsubsection*{The diagonal component}

We start by solving for the Markovian evolution of the diagonal components of $\boldsymbol{\varrho}^\ssI$. Writing $\varrho_{ij}^{\ssI}(\tau - s) \simeq \varrho_{ij}^{\ssI}(\tau)$ in \pref{rho111} and replacing $\tau \to \infty$ in the integration limits leads to the following equation
\begin{eqnarray}
\frac{\partial \varrho^{\ssI}_{11}}{\partial \tau} & \simeq & g^2 \RO - 2 g^2 \CO \varrho_{11}^{\ssI}(\tau)  \, ,\label{rho11three} 
\end{eqnarray}
where the definitions \pref{ROrate} and \pref{Crate} define the $\tau$-independent coefficients $\RO$ and $\CO$.  Comparing this to the perturbative expression \pref{uprate1} derived earlier shows agreement on the first term of \pref{rho11three}, while the perturbative expression misses the second term. It is the absence of this second term that causes the naive perturbative expression to grow indefinitely and so to fail at late times.

When $\WO$ satisfies the KMS condition \pref{KMS1} the solution to \pref{rho11three} is found to be
\begin{eqnarray}
\varrho_{11}^{\ssI}(\tau) \simeq  \frac{1}{e^{\beta \omega} + 1}  + \left[ \varrho_{11}(0) - \frac{1}{e^{\beta \omega} + 1} \right] e^{ - \tau / \tT } \ . \label{Solution11Markovian}
\end{eqnarray}
where the identity \pref{rateSC1} is used and where
\begin{eqnarray}
\tT : = \frac{1}{2g^2 \CO} \label{GeneralTT}
\end{eqnarray}
defines the solution's relaxation time-scale. Notice that $\CO$ is always positive in the examples that follow. Eq.~\pref{Solution11Markovian} describes exponential relaxation towards the static solution: $\varrho^\infty_{11} = 1/(e^{\beta \omega} + 1)$. Furthermore, the condition $\varrho_{11} + \varrho_{22} = 1$ implies $\varrho^\infty_{22} = e^{\beta \omega}/(e^{\beta \omega} + 1)$ and so $\varrho^\infty_{11}/\varrho^\infty_{22} = e^{-\beta \omega}$, showing that the static solution populates the qubit levels thermally. 

This solution allows more precise quantification of the regime of validity for the Markovian approximation. Keeping the first sub-dominant term of the expansion \pref{MarkovSeries} in \pref{rho111} gives
\begin{eqnarray}
\frac{\partial \varrho^{\ssI}_{11}}{\partial \tau} & \simeq & g^2 \RO - 4 g^2 \int_0^{\tau} \exd s\ \mathrm{Re}[\WO(s) ] \, \cos(\omega s) \bigg[ \varrho^{\ssI}_{11}(\tau) - s \frac{\partial \varrho^{\ssI}_{11}}{\partial \tau}  + \ldots \bigg] \,,
\end{eqnarray}
which (again taking $\tau \to \infty$ in the integration limit) leads to the more compact expression
\begin{eqnarray}
\frac{\partial \varrho^{\ssI}_{11}}{\partial \tau} & \simeq & g^2 \RO - 2 g^2 \bigg[ \CO \varrho^{\ssI}_{11}(\tau) - \frac{\exd \DO}{\exd \omega} \, \partial_\tau{\varrho}^{\ssI}_{11}(\tau) + \ldots  \bigg] \,,
\end{eqnarray}
which also uses the definition \pref{DivergentShift}. Using the solution \pref{Solution11Markovian} to evaluate $\partial_\tau{\varrho}_{11}^{\ssI}(\tau) \sim - {\varrho_{11}^{\ssI}(\tau)}/{\tT}$ shows that neglect of the last term requires $\tT$ to satisfy
\begin{eqnarray}
\frac{1}{\tT} \ll \left| \frac{\CO}{\exd\Delta_{\Omega}/\exd\omega} \right| \,. \label{11condition1}
\end{eqnarray}
Equivalently, using $1/\tT = 2 g^2 \CO$ in \pref{11condition1} yields
\begin{eqnarray}
2 g^2 \left| \frac{\exd\Delta_{\Omega}}{\exd \omega } \right| \ll 1 \,, \label{11condition2}
\end{eqnarray}
as the condition to be satisfied when using the Markovian limit.

\subsubsection*{The off-diagonal component}

A similar procedure gives a Markovian solution for $\varrho^{\ssI}_{12}(\tau)$, with an important complication: the Markovian approximation produces a differential equation which oscillates as well as damps. Since the oscillations are driven with frequency $\omega$ this makes dropping derivatives in the Taylor series \pref{MarkovSeries} less straightforward in the large-$\omega$ limit than is the case for the diagonal equation.

To see this in detail we again use $\varrho_{12}^{\ssI}(\tau - s) \simeq \varrho_{12}^{\ssI}(\tau)$ in the Nakajima-Zwanzig equation \pref{rho121} (and send $\tau \to \infty$ in the integration limits), leading to 
\be 
\frac{\partial \varrho^{\ssI}_{12}}{\partial \tau}  \simeq  - g^2 (\CO + i [ \DO + \omega_1 ] ) \varrho_{12}^{\ssI}(\tau) + e^{+ 2 i \omega \tau} g^2 (\CO - i \DO) \varrho_{12}^{\ssI\ast}(\tau) \ . \label{intpic12Markov}
\ee
The new complication in this equation is the potentially rapid time-dependence coming from the factor $e^{2 i \omega \tau}$. This can be removed from the differential equation by redefining the dependent variable, which in this case simply amounts to converting eq.~\pref{intpic12Markov} back to the Schr\"odinger picture. Recalling that the interaction-picture component is related to the Schr\"odinger-picture component by $\varrho_{12}(\tau) = e^{- i \omega \tau} \varrho_{12}^{\ssI}(\tau)$, eq. \pref{intpic12Markov} in the Schr\"odinger picture becomes 
\bea
\frac{\partial \varrho_{12}}{\partial \tau}  &\simeq&  - i \omega \varrho_{12}(\tau) - g^2 (\CO + i [ \DO + \omega_1 ] ) \varrho_{12}(\tau) + g^2 (\CO - i \DO) \varrho_{12}^{\ast}(\tau) \nn \\
 &\simeq&  - (i \omega + g^2 \CO ) \varrho_{12}(\tau) + g^2 (\CO - i \DO) \varrho_{12}^{\ast}(\tau) \,, \label{Schropic12Markov}
\eea
where the first line shows that it is the combination $\omega + g^2(\omega_1 + \DO)$ that enters this equation the way the qubit gap would appear. The second line uses the counter-term condition $\omega_1 = - \DO$ to ensure that this gap is given just by $\omega$. 

The solutions to \pref{Schropic12Markov} are straightforwardly found by writing it in matrix form:
\be 
\frac{\exd \mathbf{x}(\tau)}{\exd \tau}  =  \mathbb{S}\, \mathbf{x}(\tau) \quad \hbox{with solutions} \quad
\mathbf{x}(\tau) = e^{\mathbb{S}\tau}\, \mathbf{x}(0) \,,\label{SchroMatrix}
\ee
where 
\begin{eqnarray}
\mathbf{x}(\tau) := \left[ \begin{matrix} \varrho_{12}(\tau) \\ \varrho^{\ast}_{12}(\tau) \end{matrix} \right]  \quad \mathrm{and} \quad \mathbb{S} := \left[ \begin{matrix} - g^2 \CO - i \omega   & g^2 (\CO - i  \DO) \\
g^2 (\CO + i  \DO) & - g^2 \CO + i  \omega \end{matrix} \right] \,.  
\end{eqnarray}
When calculable ({\it ie.} in the non-degenerate {\it Case I} below), these solutions describe exponential relaxation towards a late-time static solution, with the static solution this time being $\varrho_{12}^\infty = 0$. The relaxation times in this case are governed by the eigenvalues of the matrix $\mathbb{S}$, with explicit solutions given by
\begin{eqnarray}
\varrho_{12}(\tau) & = & e^{-g^2 \CO \tau} \left\{ \varrho_{12}(0) \left[ \cos(\Sigma\tau) - i \frac{\omega}{\Sigma} \sin(\Sigma \tau) \right] + \varrho_{12}^{\ast}(0) \,\frac{g^2 \CO - i g^2 \DO}{\Sigma} \,\sin(\Sigma\tau) \right\} \label{exact12Schro}
\end{eqnarray}
where
\begin{eqnarray}
\Sigma & = & \omega \sqrt{ 1  - \frac{g^4(\mathcal{C}_{\Omega}^2 + \Delta_{\Omega}^2)}{\omega^2}  } \ .\label{Sigamvsomega}
\end{eqnarray}

When interpreting this equation care must be taken to remain within the domain of validity of all approximations. In particular, since the Nakajima-Zwanzig equations, \pref{rho111} and \pref{rho121}, were obtained after expanding to second-order in the coupling $g$, so we cannot reliably keep $\mathcal{O}(g^4)$ effects\footnote{There is nothing fundamental that stops one from working to higher order in $g$ with the Nakajima-Zwanzig equation. Tracking $\mathcal{O}(g^4)$ effects would involve expanding the kernel $\mathcal{K}$ (defined in Appendix \ref{App:NakZwan}) to fourth-order in $V$ and studying the master-equation that arises in this case.} in $\partial_\tau \boldsymbol{\varrho}$. The implications of this observation depend on what is assumed about the size of $\omega$, so we consider two cases separately.

\bigskip\noindent{\it Case I: $\omega \gg g^2 \sqrt{\mathcal{C}_{\Omega}^2 + \Delta_{\Omega}^2 }$}

\medskip\noindent
Consider first the parameter regime where $\omega \gg g^2 \sqrt{ \mathcal{C}_{\Omega}^2 + \Delta_{\Omega}^2 }$ in which case \pref{Sigamvsomega} shows the difference between $\Sigma$ and $\omega$ can be dropped. Note in particular that this automatically implies both $g^2 \CO / \omega$ and $g^2 \DO / \omega$ are both small, though possibly not negligibly small in $\partial_\tau \boldsymbol{\varrho}$. 

In this regime the Schr\"odinger-picture solution therefore becomes 
\be 
\varrho_{12}(\tau) \simeq e^{-g^2 \CO \tau} \bigg[ \varrho_{12}(0) e^{- i \omega \tau} + \varrho_{12}^{\ast}(0) \, \frac{g^2 \CO - i g^2 \DO}{\omega} \, \sin(\omega\tau) \bigg] \,, \label{approx12Schro}
\ee
which in the interaction picture yields
\be 
\varrho^{\ssI}_{12}(\tau)  \simeq  e^{ -g^2 \CO \tau} \left[ \varrho_{12}(0) + \  \varrho_{12}^{\ast}(0) \, \left( \frac{g^2 \DO}{2\omega} + \frac{ig^2 \CO}{2\omega} \right)  \left( 1  - e^{ 2 i \omega \tau }  \right) \right] \,. \label{approx12int}
\ee
This last expression describes very slow damping with relaxation time-scale
\begin{eqnarray}
\tD := \frac{1}{g^2 \CO} \,,
\end{eqnarray}
on which is superimposed much faster oscillations whose amplitude is small.

As before, this solution can be used to determine more precisely when the Markovian equation \pref{intpic12Markov} is valid. This means rederiving the conditions under which it is sufficient to drop the derivatives in the expansion \pref{MarkovSeries} of $\varrho_{12}^{\ssI}(\tau - s)$. Keeping the first subdominant term in the expansion \pref{MarkovSeries}, the Nakajima-Zwanzig equation for $\partial_\tau \varrho^{\ssI}_{12}(\tau)$ contains on its right-hand-side the terms
\begin{eqnarray}
 && - 2 g^2 \int_0^\tau \exd s \ \mathrm{Re}[\WO(s)] e^{+ i \omega s} \bigg[ \varrho_{12}^{\ssI}(\tau) - s \frac{\partial \varrho^{\ssI}_{12}}{\partial \tau}  + \ldots \bigg]  \\
&& \quad \quad  \quad \quad  \quad \quad  + \ 2 g^2  e^{+ 2 i \omega \tau} \int_0^\tau \exd s \ \mathrm{Re}[\WO(s)] e^{- i \omega s} \bigg[ \varrho_{12}^{\ssI\ast}(\tau) - s \frac{\partial \varrho^{\ssI\ast}_{12}}{\partial \tau}  + \ldots \bigg]\ , \notag 
\end{eqnarray}
which, once $\tau \to \infty$ is taken in the integration limits, becomes 
\begin{eqnarray}
 && -  g^2 \bigg[ \CO \varrho_{12}^{\ssI}(\tau) + \left( \frac{\exd \DO}{\exd \omega} - i \frac{\exd \CO}{\exd \omega} \right) \frac{\partial \varrho^{\ssI}_{12}}{\partial \tau} + \ldots \bigg] \\
&& \quad \quad  \quad \quad  \quad \quad + g^2 e^{+ 2 i \omega \tau} \bigg[ ( \CO - i \DO) \varrho_{12}^{\ssI}(\tau) + \left( \frac{\exd \DO}{\exd \omega} + i \frac{\exd \CO}{\exd \omega} \right) \frac{\partial \varrho^{\ssI}_{12}}{\partial \tau} + \ldots \bigg] \  . \notag
\end{eqnarray}

The solution \pref{approx12int} for $\varrho_{12}^{\ssI}(\tau)$ is a sum of terms whose time-evolution is exponential, varying like $A \exp\left[ \left( - \tfrac{1}{\tD} + i \Phi \right) \tau \right]$, where $1/\tD = g^2 \CO$ and $\Phi = 0$ or $\Phi = 2 \omega$ (with $A$ a time-independent complex amplitude). Using this to eliminate $\partial_\tau \varrho^\ssI_{12}$ in the above terms then gives
\begin{eqnarray}
\frac{\partial \varrho^{\ssI}_{12}}{\partial \tau} & \supset & g^2 A e^{ \left( - \tfrac{1}{\tD} + i \Phi \right) \tau} \bigg[ \CO+ \left( \frac{\exd \DO}{\exd \omega} - i \frac{\exd \CO}{\exd \omega} \right) \left( - \frac{1}{\tD} + i \Phi \right) + \ldots \bigg] \\
&\ & \quad \quad  \quad \quad + g^2 e^{+ 2 i \omega \tau} A^{\ast} e^{ \left( - \tfrac{1}{\tD} - i \Phi \right) \tau} \bigg[ ( \CO - i \DO)  + \left( \frac{\exd \DO}{\exd \omega} + i \frac{\exd \CO}{\exd \omega} \right) \left( - \frac{1}{\tD} - i \Phi \right) + \ldots \bigg] \notag
\end{eqnarray}
Dropping the derivatives in the Markovian series therefore requires
\be \label{conditionONE}
|\CO|  \gg  \left| \left( \frac{\exd \DO}{\exd \omega} - i \frac{\exd \CO}{\exd \omega} \right) \left( - \frac{1}{\tD} + i \Phi \right) \right| \quad \hbox{and} \quad
|\CO - i \DO |  \gg  \left| \left( \frac{\exd \DO}{\exd \omega} + i \frac{\exd \CO}{\exd \omega} \right) \left( - \frac{1}{\tD} - i \Phi \right) \right| \,.
\ee
Since $|\CO - i \DO| \geq |\CO|$ the first of these conditions automatically ensures the second is also satisfied. Furthermore, because $|\,a+ib\,| \geq |a|$ and $|\,a+ib\,| \geq |\,b\,|$ it follows from \pref{conditionONE} that 
\be 
|\CO| \gg \left| \frac{1}{\tD} \frac{\exd\Delta_{\Omega}}{\exd\omega} -  \Phi \, \frac{\exd\mathcal{C}_{\Omega}}{\exd \omega}\right|  \quad \quad  \mathrm{and} \quad \quad |\CO| \gg\left| \frac{1}{\tD} \frac{\exd\cC_{\Omega}}{\exd\omega} + \Phi \, \frac{\exd\Delta_{\Omega}}{\exd \omega}\right| 
 \label{conditionTWO}
\ee
are necessary conditions for being able to neglect derivatives, when deriving the Markovian approximation. 

When $\Phi =0$, the above bounds become
\be 
\frac{1}{\tD} \ll \left| \frac{\CO}{\exd\Delta_{\Omega}/\exd \omega} \right|  \quad \quad \mathrm{and} \quad \quad  \frac{1}{\tD} \ll \left| \frac{\CO}{\exd\mathcal{C}_{\Omega}/\exd \omega} \right| \,,\label{tDconditions}
\ee
which when specialized to $1/\tD = g^2 \CO$ become
\be 
g^2 \left| \frac{\exd\Delta_{\Omega}}{\exd\omega} \right| \ll 1 \quad \quad \mathrm{and} \quad \quad g^2 \left| \frac{\exd\cC_{\Omega}}{\exd\omega} \right| \ll 1 \,.\label{Phi0condition}
\ee
The first of these was encountered in \pref{11condition2} as a condition for there being a Markovian limit of $\varrho^\ssI_{11}$.

For large $\omega$ the strongest condition comes from applying the bounds \pref{conditionTWO} for the rapidly oscillating case $\Phi = 2 \omega$. For this case, and using $1/\tD = g^2 \CO$, \pref{conditionTWO} becomes
\begin{eqnarray}
1 \gg \left| g^2\frac{\exd\Delta_{\Omega}}{\exd\omega} -  \frac{2\omega}{\CO} \left(\frac{\exd\cC_{\Omega}}{\exd\omega}\right) \right|  \quad \quad  \mathrm{and} \quad \quad  1 \gg \left| g^2\frac{\exd\cC_{\Omega}}{\exd\omega} + \frac{2\omega}{\CO} \left( \frac{\exd\Delta_{\Omega}}{\exd\omega} \right) \right| \label{Phi2wcondition}
\end{eqnarray}
As we show below, once evaluated as functions of the qubit/field parameters, conditions \pref{Phi2wcondition} turn out to be {\it impossible} to satisfy once $\omega$ is larger than $1/\tau_c \simeq a$. This agrees with the intuition that rapid oscillations should eventually destroy the derivative expansion that underlies the Markovian evolution. The next sections demonstrate this explicitly by evaluating the above expressions as concrete functions of the parameters $g$, $m$, $a$ and $\omega$.

\bigskip\noindent{\it Case II: $\omega \ll g^2 \sqrt{\mathcal{C}_{\Omega}^2 + \Delta_{\Omega}^2}$}

\medskip\noindent
Next consider the very degenerate regime where the qubit gap is small enough to compete with $\cO(g^2)$ effects. In this case 
\be
  \Sigma = \sqrt{\omega^2 - g^4(\mathcal{C}_{\Omega}^2 + \Delta_{\Omega}^2)} \simeq \pm ig^2\sqrt{\mathcal{C}_{\Omega}^2 + \Delta_{\Omega}^2} \Bigl[ 1 + \cdots \Bigr] \,,
\ee
where the ellipses are order $\omega^2/(g^4\mathcal{C}_{\Omega}^2 + g^4 \Delta_{\Omega}^2 )$ and so are small (but need not be suppressed by powers of $g$) in this parameter regime. Using $\varrho_{12}^{\ssI}(\tau) = e^{+ i \omega \tau} \varrho_{12}(\tau)$, the exact solution \pref{exact12Schro} in this case becomes approximately
\begin{eqnarray}
\varrho_{12}^{\ssI}(\tau) & \simeq & e^{+i \omega \tau} e^{ - g^2 \CO \tau} \left\{ \varrho_{12}(0) \left[ \cosh\left( g^2 \sqrt{ \mathcal{C}_{\Omega}^2 + \Delta_{\Omega}^2 } \; \tau \right) -  \frac{i\omega \; \sinh\left( g^2 \sqrt{ \mathcal{C}_{\Omega}^2 + \Delta_{\Omega}^2 } \; \tau \right)}{g^2 \sqrt{ \mathcal{C}_{\Omega}^2 + \Delta_{\Omega}^2 }}  \right] \right.\quad \\
& \ & \quad \quad \quad \quad \quad \quad \quad \quad \quad \quad \quad \quad \quad \quad \quad \quad \quad \quad \left.+ \varrho_{12}^{\ast}(0) \frac{ \CO - i \DO}{\sqrt{ \mathcal{C}_{\Omega}^2 + \Delta_{\Omega}^2  }} \sinh \left( g^2 \sqrt{ \mathcal{C}_{\Omega}^2 + \Delta_{\Omega}^2 } \; \tau \right) \right\} \notag
\end{eqnarray}
where effects of order $\omega^2/(g^4\mathcal{C}_{\Omega}^2 +g^4 \Delta_{\Omega}^2)$ are neglected.  This again has the form of a sum of exponential solutions,
\begin{eqnarray}
\varrho_{12}^{\ssI}(\tau) &\simeq & A_1 \; e^{ - \tfrac{\tau}{\xi_1} + i \Phi_1 \tau } + A_2 \; e^{ + \tfrac{\tau}{\xi_2} + i \Phi_2 \tau } 
\end{eqnarray}
where  
\begin{eqnarray}
\frac{1}{\xi_1} & := & g^2 \CO \bigg[  1 + \sqrt{ 1 + \sfrac{\Delta_{\Omega}^2}{\mathcal{C}_{\Omega}^2} } \;  \bigg]  \quad \quad \quad \ \ \ \, \Phi_1 \ := \ \omega  \ , \label{DegenTimescale1} \\
\frac{1}{\xi_2} & := & g^2 \CO \bigg[  - 1 + \sqrt{ 1 + \sfrac{\Delta_{\Omega}^2}{\mathcal{C}_{\Omega}^2} } \;  \bigg]  \quad \quad \quad \Phi_2 \ := \ \omega  \ .\label{DegenTimescale2}
\end{eqnarray}
Notice the potentially worrying {\it positive} exponent for the second term (more about which later). Such a growing mode would necessarily cause problems with unitarity if it were to be trusted. (We find below -- see the next section, and Appendix \ref{App:Degenerate} -- for accelerated qubits that the validity of the Markovian limit requires $|\DO/\CO| \lsim \cO(g)$ and so $1/\xi_2$ is at most order $g^4$ and so is consistent with zero at $\cO(g^2)$.)

Again expanding the $\varrho^{\ssI}(\tau - s)$ in the interaction-picture and dropping derivatives gives us the validity conditions similar to those found previously, where
\begin{eqnarray}
|\CO| \; \gg \; \left| \frac{1}{\xi_1} \frac{\exd \DO}{\exd \omega} - \Phi_1 \frac{\exd \CO}{\exd \omega} \right| \quad \quad & \mathrm{and} &  \quad \quad |\CO| \; \gg \; \left| \frac{1}{\xi_1} \frac{\exd \CO}{\exd \omega} +  \Phi_1 \frac{\exd \DO}{\exd \omega} \right| \ , \\
|\CO| \; \gg \; \left| \frac{1}{\xi_2} \frac{\exd \DO}{\exd \omega} +  \Phi_2 \frac{\exd \CO}{\exd \omega} \right| \quad \quad & \mathrm{and} &  \quad \quad |\CO| \; \gg \; \left| \frac{1}{\xi_2} \frac{\exd \CO}{\exd \omega} -  \Phi_2 \frac{\exd \DO}{\exd \omega} \right| \ , \label{degen2}
\end{eqnarray}
which imply the more the condensed forms:
\begin{eqnarray}
\left| g^2 \DOp - \frac{\omega \COp}{\CO} \right| \ll  1 \ , \ \ \left| g^2\COp + \frac{\omega \DOp}{\CO} \right| \ll 1 \ , \ \ g^2 |\DOp| \; \sqrt{ 1 + \frac{\Delta_{\Omega}^2}{\mathcal{C}_{\Omega}^2} } \ll 1 \ , \ \ g^2 |\COp| \;  \sqrt{ 1 + \frac{\Delta_{\Omega}^2}{\mathcal{C}_{\Omega}^2} } \ll 1 \,,\quad \ \ \ \label{degenCond1}
\end{eqnarray}
where primes denote $\exd/\exd \omega$. There are the forms easiest to use -- when $\omega \ll g^2 \sqrt{ \mathcal{C}_{\Omega}^2 +\Delta_{\Omega}^2 }$ -- once the integrals are explicitly evaluated below for an accelerated qubit.

\section{Accelerated qubits in the Markovian regime}
\label{sec:AcceleratedMarkovian}

To this point little is assumed about the details of the qubit trajectory or of the state in which the scalar field is initially prepared. Because of this the key assumption --- that there exists a time-scale $\tau_c$ for which the Wightman function falls when $\tau \gg \tau_c$ --- remains merely an assumption. This section aims the make the above discussion more concrete by evaluating the functions $\CO$, $\RO$, $\SO$ and $\DO$ explicitly for a uniformly accelerated qubit in flat spacetime coupled to a free field that is prepared in the Minkowski vacuuum, $|\Omega \rangle = | \Mink \rangle$. The goal is to identify all of the conditions for validity of late-time Markovian evolution explicitly as functions of the parameters $g$, $m$, $\omega$ and $a$, where $m$ is the field's mass and $a$ is the qubit's proper acceleration.  

To this end choose the qubit to move along a uniformly accelerated trajectory 
\begin{eqnarray}
y^\mu (\tau) \ = \ \Bigl[ \tfrac{1}{a} \sinh(a\tau), \tfrac{1}{a} \cosh(a\tau), y^2, y^3 \Bigr] \label{Rindler}
\end{eqnarray}
in Minkowski spacetime, where $a>0$ and $y^2, y^3 \in \bR$ do not depend on $\tau$. With this parameterization the quantity $\tau$ is the qubit's proper time as measured using the Minkowski metric. As above the joint system's initial state is assumed to be uncorrelated at $\tau = 0$, with
\begin{eqnarray}
\rho(0) \ = \ \ket{ \Mink } \bra{ \Mink } \otimes \boldsymbol{\varrho_0} \,.\label{uncorrelated} 
\end{eqnarray}
The resulting Wightman function $\WM(\tau) = \braket{ \Mink | \phi[y(\tau)] \phi[y(0)] | \Mink }$ for a real massive field (\ref{freefield}) can be explicitly evaluated along an accelerating trajectory, giving the following closed-form result \cite{Takagi:1986kn,Langlois:2005nf}:
\begin{eqnarray}
\WM(\tau) & = &  \frac{am}{8i\pi^2} \frac{1}{ \big[ \sinh({a\tau}/{2}) - i \frac{a\epsilon}{2} \big]} K_1\left( \tfrac{2mi}{a} \big[ \sinh({a\tau}/{2}) - i \tfrac{a\epsilon}{2} \big] \right) \label{massWightman}
\end{eqnarray}
where $K_1(z)$ is a Bessel function of imaginary argument and the small-distance infinitesimal $\epsilon \to 0^{+}$ is a consequence of the Wightman boundary conditions. Finally, in the massless limit $m \to 0^{+}$ we recover
\begin{eqnarray}
\WM(\tau) & \to & -\, \frac{a^2 }{16 \pi^2 \big[ \sinh({a\tau}/{2} ) - i \frac{a\epsilon}{2} \big]^2}
\qquad \hbox{(massless limit)} \,. \label{masslessWightman}
\end{eqnarray}

Notice that \pref{massWightman} and \pref{masslessWightman} exhibit both properties (\ref{conjugateW1}) and (\ref{KMS1}) explicitly, with 
\be 
   T = \frac{1}{\beta} = \frac{a}{2\pi} \,,
\ee
 being the usual Unruh temperature.\footnote{We note in passing a subtlety of the $\epsilon$-regularization. It can be tempting to write $\WM$ with $\sinh({a\tau}/{2}) - i {a\epsilon}/{2}$ replaced by $\sinh[({a( \tau - i \epsilon)}/{2}]$ (where $\epsilon$ has units of length), with the reasoning that these are equivalent because infinitesimal $\epsilon > 0$ is important only near $\tau =0$ \cite{Takagi:1986kn}. Although this reasoning is not false for real $\tau$, this replacement can be dangerous where $\tau$ is not real because it does {\it not} preserve the KMS condition \pref{KMS1}. }

\subsection{Perturbative result}

A straightforward calculation starting from \pref{massWightman}  -- whose details we present in Appendix \ref{App:A} -- reveals the integral $\SM(\omega)$ to be
\begin{eqnarray}
\SM(\omega) \ = \ \frac{m^2}{ 4 \pi^2 a} \sinh\left( \frac{\pi \omega}{a} \right) \left\{ \Big[ K_{\frac{i\omega}{a}}\big( \tfrac{m}{a} \big) \Big]^{2} - K_{\frac{i\omega}{a} - 1}\big( \tfrac{m}{a} \big)  K_{\frac{i\omega}{a} + 1}\big( \tfrac{m}{a} \big) \right\} \,.
\end{eqnarray}
which, together with (\ref{ratioSCgeneral}), then gives
\begin{eqnarray}
\CM(\omega) \ = \ \frac{m^2}{4 \pi^2 a} \cosh \left( \frac{\pi \omega}{a} \right) \left\{ K_{\frac{i\omega}{a} - 1}\big( \tfrac{m}{a} \big)  K_{\frac{i\omega}{a} + 1}\big( \tfrac{m}{a} \big) - \Big[ K_{\frac{i\omega}{a}}\big( \tfrac{m}{a} \big) \Big]^{2} \right\}
\end{eqnarray}
and so \pref{rateSC1} implies
\begin{eqnarray}
\RM(\omega) & = & \frac{m^2}{4\pi^2 a} \, e^{- {\pi \omega }/{a}} \left\{ K_{\frac{i\omega}{a}-1}\left(\tfrac{m}{a} \right)K_{\frac{i\omega}{a}+1}\left(\tfrac{m}{a} \right) - \Big[ K_{\frac{i\omega}{a}}\left(\tfrac{m}{a} \right)\Big]^2 \right\} \,. \label{massiverate}
\end{eqnarray}
This expression for $\RM$ agrees with ones given in \cite{Takagi:1986kn,Fredenhagen}. The property $K_{\alpha}(x) = K_{-\alpha}(x)$ makes it easy to see that the detailed balance relation \pref{detailedbalance} is satisfied. 

Of particular use are the asymptotic forms for these expressions in the limits $m \gg a$ and $m \ll a$, which are found using the asymptotic expansions for the Bessel function, $K_\alpha(z)$:
\begin{eqnarray}
K_{\alpha}(z) \approx \sqrt{ \frac{\pi}{2z} } e^{-z} \left[ 1 + \frac{4\alpha^2- 1}{8z} + \frac{9 - 40 \alpha^2 + 16 \alpha^4}{128z^2} + \ldots \right] \,,
\end{eqnarray}
for $|z|\gg 1$ and $|\arg \, z| < \tfrac{3\pi}{2}$ \cite{Abr}, while for $|z| \to 0$ and $\nu \in \bC \setminus \mathbb{Z}$ \cite{NIST}  $K_{\nu}(z)$ is given by
\begin{eqnarray}
K_{\nu}(z) = \frac{\Gamma(\nu)}{2} \left( \frac{z}{2} \right)^{-\nu} \left[ 1 + \frac{z^2}{4(1-\nu)} + \mathcal{O}(z^4) \right] + \frac{\Gamma(-\nu)}{2} \left( \frac{z}{2} \right)^{\nu} \left[ 1 + \frac{z^2}{4(1+\nu)} + \mathcal{O}(z^4) \right] \,,
\end{eqnarray}
where $\Gamma(z)$ is Euler's gamma function. These lead to the asymptotic large-mass $m \gg a$ result,
\begin{eqnarray}
\RM(\omega) \ \simeq \ \frac{ a}{8 \pi } \; e^{- ({\pi \omega } + {2m})/{a}} \qquad \hbox{(for $m \gg a$)} \ .
\end{eqnarray}
whose sub-leading terms are bounded when $m/a \gg 1 + 4(\omega/a)^2$. The opposing limit for $m \ll a$ gives (see Appendix \ref{App:B} for details)
\begin{eqnarray}
\RM(\omega) \ \simeq \ \frac{1}{2 \pi} \;\frac{\omega}{e^{{2\pi}\omega/a} - 1} \left\{ 1 + \frac{m^2}{2 \omega^2}  \left[ \frac{\cos\left( \frac{2\omega}{a} \log\left( \frac{m}{2a} \right) - \zeta \right)}{\sqrt{ (\omega/a)^2 + 1 }} - 1 \right] + \cdots\right\} \quad \hbox{(for $m \ll a$)}
\end{eqnarray}
where
\be \label{zetadef}
  \zeta := \mathrm{Arg}\left[  \frac{[\Gamma(\frac{i\omega}{a})]^2}{\frac{i\omega}{a} - 1} \right] \,.
\ee
and which is valid for $(m/a)^2 \ll (\omega/a)^2 \sqrt{ (\omega/a)^2 + 1 }$. Once used in the perturbative rate expression, eq.~\pref{uprate1} these formulae reproduce standard results for the strictly massless limit \cite{Sciama:1981hr}  
\begin{eqnarray}
   \frac{\partial \varrho_{11}}{\partial \tau} \ \simeq \ g^2 \RM(\omega) \simeq \left( \frac{g^2}{2 \pi} \right) \frac{\omega}{e^{{2\pi}\omega/a} - 1} \qquad \hbox{(for $m =0$)} \,,\label{uprate}
\end{eqnarray}
where it is understood that \pref{uprate1} and \pref{uprate} only apply for the proper-time interval $1\ll a \tau \ll 1/g^2$.  They also give the correct result in the limit of an inertial observer, $a \to 0^{+}$, since the transition rate then becomes
\begin{eqnarray}
   \RM(\omega) & \to & \frac{  \sqrt{ \omega^2 - m^2 } }{2 \pi} \;\Theta(- \omega - m) \qquad \hbox{(for $a \to 0$)}
\end{eqnarray}
where the Heaviside step function -- for which $\Theta(x) = 1$ if $x > 0$ and $\Theta(x) = 0$ otherwise -- ensures the result is non-zero only for $\omega < - m$ ({\it i.e.}~never, for positive $\omega$ and $m$).

\subsection{The Markovian limit}

As argued above, straight-up perturbative expressions like  \pref{uprate1} and \pref{uprate} must eventually break down at sufficiently late times, since if taken too seriously a constant transition rate would eventually predict $\varrho_{11} > 1$ (in conflict with tr $\boldsymbol{\varrho} = 1$). The feedback that prevents this is captured by the Nakajima-Zwanzig equations, which for the accelerating qubit are \pref{rho111} and \pref{rho121}, reproduced here as
\begin{eqnarray}
\frac{\partial \varrho^{\ssI}_{11}}{\partial \tau} & = & g^2 \int_{-\tau}^{\tau} \exd s \; \WM(s) \, e^{- i \omega s} - 4 g^2 \int_0^{\tau} \exd s\ \mathrm{Re}[\WM(s) ] \, \cos(\omega s) \varrho^{\ssI}_{11}(\tau - s)  \, , \\
\frac{\partial \varrho^{\ssI}_{12}}{\partial \tau} & = & + i g^2 \DM \varrho^{\ssI}_{12}(\tau) - 2 g^2 \int_0^\tau \exd s \ \mathrm{Re}[\WM(s)] e^{+ i \omega s} \varrho_{12}^{\ssI}(\tau - s)  \\
&\ & \quad \quad  \quad \quad  \quad \quad  \quad \quad \quad  \quad \quad  \quad \quad + \ 2 g^2  e^{+ 2 i \omega \tau} \int_0^\tau \exd s \ \mathrm{Re}[\WM(s)] e^{- i \omega s} \varrho_{12}^{\ssI\ast}(\tau - s)\ , \notag 
\end{eqnarray}
with counter-term $\omega_1 = - \DM$ already chosen.

Of interest for establishing if there might be a Markovian limit is whether the Wightman function falls quickly enough for large $\tau$ (measured along a uniformly accelerating worldline). As can be seen from expressions \pref{massWightman} or \pref{masslessWightman}, for $a\tau \gg 1$ the function $\mathrm{Re}[\WM(\tau)]$ behaves as
\begin{eqnarray}
\mathrm{Re}[\WM(\tau)] \simeq - \sqrt{ \frac{a^3m}{32 \pi^3} } \; e^{ - {3}| a \tau | /4} \sin\left( \frac{m}{a} \, e^{|a\tau|/2 } + \frac{\pi}{4} \right) \qquad \hbox{(late times)}\label{latetimeWMmassive}
\end{eqnarray}
provided $m$ is large enough that $\frac{m}{a} \, e^{|a \tau|/2} \gg 1$. Alternatively, in the massless limit one finds an even faster falloff, with
\begin{eqnarray}
\mathrm{Re}[\WM(\tau)] \simeq - \frac{a^2}{4\pi^2} \;e^{-|a\tau|}  \qquad \hbox{(late times, massless limit)} \,, \label{masslessRindlerXIs}
\end{eqnarray}
for $a \tau \gg 1$. This last limit also applies for massive fields if $m$ is small enough to ensure that $\frac{m}{a} \, e^{|a \tau|/2} \ll 1$. The cross-over from \pref{latetimeWMmassive} to \pref{masslessRindlerXIs} occurs for $|a \tau| \simeq 2\ln(a/m)$.

As might have been expected for a qubit interacting with a thermal state -- as the Minkowski vacuum appears (with temperature $T = a/2\pi$) from the point of view of the qubit -- this falloff suffices to make the late-time qubit behaviour Markovian over time-scales $\tau \gg \tau_c \sim 1/a$. To see that qubit relaxation towards the thermal state falls into this regime we must check that relations \pref{tDconditions}-\pref{Phi2wcondition} (or \pref{degenCond1}) are satisfied. 

To check these note that the Markovian equations of motion for the interaction-picture components in this regime are
\begin{eqnarray}
\frac{\partial \varrho^{\ssI}_{11}}{\partial \tau} & \simeq & g^2 \RM - 2 g^2  \CM  \,\varrho^{\ssI}_{11}(\tau) \label{Rindler11} \\
\frac{\partial \varrho^{\ssI}_{12}}{\partial \tau} & \simeq & - g^2 \CM\, \varrho^{\ssI}_{12}(\tau) + g^2 e^{2 i \omega \tau} ( \CM - i \DM ) \, \varrho^{\ssI}_{12}(\tau)  \ . \label{Rindler12}
\end{eqnarray}
with solutions in the non-degenerate $\omega \gg g^2 \sqrt{ \mathcal{C}_{\ssM}^2 + \Delta_{\ssM}^2 }$ limit (see Appendix \ref{App:Degenerate} for the opposing degenerate limit)
\begin{eqnarray}
  \varrho^{\ssI}_{11}(\tau) & = & \frac{1}{e^{{2\pi}{\omega}/a} + 1} + \left[ \varrho_{11}(0) -  \frac{1}{e^{{2\pi}{\omega}/a} + 1} \right] e^{ - {\tau}/{\tT} } \label{Rindler11sol} \\
    \varrho^{\ssI}_{12}(\tau) & = & e^{ - {\tau}/{\tD} }\left[  \varrho_{12}( 0 )  + \varrho_{12}^{\ast}(0) \left( \frac{g^2 \DM}{2\omega} + i \frac{g^2 \CM}{2\omega}  \right) (1 - e^{2 i \omega \tau}) \right] \,,\label{Rindler12sol}
\end{eqnarray}
for which the relaxation rates explicitly evaluate to 
\be
\frac{1}{\tD} = \frac{1}{2 \tT} = g^2 \CM = \frac{g^2 m^2}{4 \pi^2 a} \; \mathrm{cosh} \left( \frac{\pi \omega}{a} \right) \left\{ K_{\frac{i\omega}{a} - 1}\big( \tfrac{m}{a} \big)  K_{\frac{i\omega}{a} + 1}\big( \tfrac{m}{a} \big) - \Big[ K_{\frac{i\omega}{a}}\big( \tfrac{m}{a} \big) \Big]^{2} \right\}  \,.\label{timescaleMinkowski}
\ee
The static solution to which the relaxation occurs is
\begin{eqnarray}
\lim_{\tau \to \infty} \varrho(\tau) & = & \left[ \begin{matrix} \dfrac{1}{e^{{2\pi}\omega/a} + 1} & 0 \\ 0 & \dfrac{1}{e^{-{2\pi}\omega/a} + 1} \end{matrix} \right] \,,
\end{eqnarray}
which is thermal. Formally this follows from the identity \pref{KMS1} satisfied by $\WM(\tau)$. It is also as expected physically given that the Minkowski vacuum $\ket{\Mink}$ appears thermal to accelerated observers, with temperature $T= {a}/(2\pi)$ \cite{Israel:1976ur,Takagi:1986kn,Troost:1977dw,Unruh:1983ms}. 

Unlike the asymptotic (equilibrium) Unruh temperature, the two (non-equilibrium) time-scales $\tD$ and $\tT$ in (\ref{timescaleMinkowski}) depend sensitively on all of the parameters of the problem ({\it i.e.}~$m$, $\omega$ and $g$ in addition to $a$). This is most easily illustrated using the various asymptotic limits. For instance, for large masses $m \gg a$ the two time-scales are asymptotically given by 
\be
\tD \ = \ 2 \tT \ \simeq \ \frac{8\pi}{a g^2} \, \mathrm{sech} \left( \frac{\pi \omega}{a} \right)  e^{{2m}/{a}}
\quad \hbox{(if $m \gg a$)}\,. \label{largemassMinkowski}
\ee
In the opposite limit of small scalar mass, $m \ll a$, the two time-scales approach the massless limit
\be
\tD = 2\tT \simeq \frac{4\pi}{g^2 \omega} \, \tanh\left( \frac{\pi \omega }{a} \right) \qquad \hbox{(if $m \ll a$)} \,. \label{masslessRindlertimescales}
\ee
This massless rate crosses over from a thermal result ($\xi \propto 1/a$) when $\omega \ll a$ to one that scales with the qubit's intrinsic time-scale ($\xi \propto 1/\omega$) when $\omega \gg a$. 

For later purposes it is also useful to record the sub-dominant $m/a$ corrections to \pref{masslessRindlertimescales}:
\be \label{massivexiRind}
\tD = 2 \tT  \simeq \frac{4\pi}{g^2 \omega} \tanh\left( \frac{\pi \omega }{a} \right) \left\{1 + \frac{m^2}{2 \omega^2}  \left[ 1 - \frac{\cos\left( \frac{2\omega}{a} \log\left( \frac{m}{2a} \right) - \zeta \right)}{\sqrt{ (\omega/a)^2 + 1 }} \right] + \cdots \right\} \,,
\ee
where $\zeta$ is given by \pref{zetadef} and $\gamma$ is the Euler-Mascheroni constant (the sub-leading  terms are again bounded when $(m/a)^2 \ll (\omega/a)^2 \sqrt{ (\omega/a)^2 + 1 }$). 

One final remark bears on the potentially troubling dependence that \pref{Rindler12} and \pref{Rindler12sol} have on the divergent quantity $g^2\DM/\omega$. As we argue in more detail in \S\ref{sec:LateTimeLimit} below, this dependence is actually deceptive because (as shown in detail in Appendix \ref{App:D}) the divergent part of $\DM$ goes like\footnote{Depending on the regime of interest, the logarithm in \pref{divergentpart} could instead have an argument of $\omega \epsilon \ll 1$ or $m \epsilon \ll 1$. For an explicit formula for $\DM$ see \pref{explicitDM}.}
\be 
\Delta_{\mathrm{M}}^{(\mathrm{divergent})} \simeq \frac{\omega}{2\pi^2} \log( a \epsilon ) \,, \label{divergentpart}
\ee
where $\epsilon$ is the short-distance regularization scale. This shows that the combination $g^2 \DM/\omega$ appearing in \pref{Rindler12} and \pref{Rindler12sol} is explicitly $\cO(g^2)$ and so is smaller than the order to which they have been reliably computed. The same need not be true for the finite parts of $\DM$ or $\CM$, depending on the size of $\omega$.

This brings us back to the question of when eqs.~\pref{Rindler11}-\pref{Rindler12} and their solutions can be trusted.\footnote{In existing literature which applies open quantum systems methods to Unruh-DeWitt detectors in various spacetimes \cite{Benatti:2004ee,Yu:2008zza,Yu:2011eq,Hu:2012ed,Hu:2011pd,Fukuma:2013uxa,Menezes:2017rby,Menezes,Tian:2016gzg,Chatterjee:2019kxg}, Markovian master equations are sometimes taken as a starting point. In this case there may be unstated conditions on the parameters that are not explicitly stated (see however \cite{Lin:2006jw,Moustos:2016lol})} The main conditions are $a \tau \gg 1$ and that the remaining parameters are such that the relaxation is sufficiently slow; {\it i.e.} that conditions \pref{tDconditions}-\pref{Phi2wcondition} (or \pref{degenCond1}) are satisfied. We next evaluate the explicit parameter ranges that satisfy these conditions.

\subsubsection*{Domain of validity for Markovian evolution}

Mapping out the regime of validity for the Markovian evolution in parameter space involves computing the various functions $\CM$, $\DM$, $\CMp$ and $\DMp$ as functions of these parameters (where primes denote differentiation with respect to $\omega$). Since these are not simple functions of $\omega$, $a$ or $m$, we present various limiting asymptotic forms in Table \ref{Functions1}.
\begin{table}[h]
  \centering    
     \centerline{\begin{tabular}{ r|c|c|c|c|c|c| }
 \multicolumn{1}{r}{}
 & \multicolumn{1}{c}{$\underset{\ }{\frac{\omega}{a} \ll \frac{m}{a} \ll 1}$}
 & \multicolumn{1}{c}{$\frac{m}{a} \ll \frac{\omega}{a} \ll 1$}
 & \multicolumn{1}{c}{$\frac{\omega}{a} \ll 1 \ll \frac{m}{a}$}
 & \multicolumn{1}{c}{$\frac{m}{a} \ll 1 \ll \frac{\omega}{a}$}
 & \multicolumn{1}{c}{$1 \ll \frac{\omega}{a} \ll \frac{m}{a}$}
 & \multicolumn{1}{c}{$1 \ll \frac{m}{a} \ll \frac{\omega}{a}$} \\
\cline{2-7}
$\CM(\omega)  $ & $ \stackrel{\ }{ \underset{\ }{\dfrac{a}{4\pi^2} } } $ & $ \underset{\ }{\dfrac{a}{4\pi^2} } $ & $\dfrac{a}{8\pi} e^{ -{2m}/{a}}$ & $\dfrac{\omega}{4\pi}$ & $\dfrac{a}{16\pi} e^{+ {\pi\omega}/{a}} e^{ - {2m}/{a}}$ & $\dfrac{\omega}{4\pi}$ \\
\cline{2-7}
$\CMp(\omega) $ & $ \underset{\ }{\dfrac{\omega}{6 a} }  $ & $ \dfrac{\omega}{6 a} $ & $\dfrac{\pi \omega}{8 a} e^{ -{2m}/{a}}$ & $\dfrac{1}{4\pi}$ & $ \stackrel{\ }{  \dfrac{1}{16} e^{+ {\pi\omega}/{a}} e^{ - {2m}/{a}} }$ & $\dfrac{1}{4\pi}$ \\
\cline{2-7} 
$\Delta_{\ssM}(\omega)$ & $\stackrel{\ }{ \underset{\ }{ \dfrac{\omega\log(a\epsilon)}{2\pi^2} } }$ & $\dfrac{\omega\log(a\epsilon)}{2\pi^2}$ & $\dfrac{\omega\log(e^{\gamma + 1} m \epsilon)}{2\pi^2}$ & $ \dfrac{\omega\log\left( e^{\gamma} \omega \epsilon \right)}{2\pi^2}$ & $\dfrac{\omega\log\left( e^{\gamma+1} m \epsilon \right)}{2\pi^2}$ & $\dfrac{\omega\log\left( e^{\gamma} \omega \epsilon \right)}{2\pi^2}$ \\
\cline{2-7} 
$\DMp(\omega) $ & $ \stackrel{\ }{  \underset{\ }{ \dfrac{\log( a \epsilon )}{2\pi^2}  } }$ & $\dfrac{\log( a \epsilon )}{2\pi^2} $ & $\dfrac{\log( e^{\gamma+1} m \epsilon )}{2\pi^2} $ & $\dfrac{\log( e^{\gamma+1} \omega \epsilon )}{2\pi^2}$ & $\dfrac{\log( e^{\gamma+1} m \epsilon )}{2\pi^2} $ & $\dfrac{ \log( e^{\gamma+1} \omega \epsilon )}{2\pi^2}$ \\
\cline{2-7}
\end{tabular} }
        \caption{Leading-order behaviour for the various functions $\CM$, $\CMp$, $\DM$ and $\DMp$ in various regimes of relative sizes of $\omega$, $m$ and $a$, where primes denote differentiation with respect to $\omega$. Only the divergent part of $\DM$ and $\DMp$ are quoted (see Appendix \ref{App:D} for their derivations for the behaviour of $\DM$ and $\DMp$). $\gamma$ denotes the Euler-Mascheroni constant.} \label{Functions1}
\end{table}

For definiteness, consider first the case where 
\be
  \omega \gg g^2 \sqrt{ \mathcal{C}_{\ssM}^2 + \Delta_{\ssM}^2 } \,.  \label{BigOmegaDef}
\ee
In this case dropping derivatives in the non-oscillating terms in $\varrho_{ij}$ leads to conditions \pref{Phi0condition}:
\be 
g^2 | \DMp|, \ g^2 |\CMp|   \ll  1 \label{validityM1} \,;
\ee
while the same condition for the oscillating terms gives \pref{Phi2wcondition}:
\be 
\left| g^2 \Delta_{\mathrm{M}}^{\prime} -  \frac{2\omega \mathcal{C}_{\mathrm{M}}^{\prime}}{\CM} \right| , \  \left| g^2 \mathcal{C}_{\mathrm{M}}^{\prime} + \frac{2\omega\Delta_{\mathrm{M}}^{\prime}}{\CM} \right|  \ll  1 \,. \label{validityM2}
\ee
The above conditions also assume $a\tau \gg 1$, since this is required when replacing $\tau \to \infty$ in the integration limits. 

Notice also that \pref{BigOmegaDef} also implies 
\be 
\frac{g^2 \CO}{\omega} , \ \frac{g^2 |\DO|}{\omega} \lsim \cO(g^2) \ll 1 \ . \label{validityM4}
\ee

The first observation is that these conditions cannot all be satisfied if $\omega \gg a$. To see why, first notice that the rightmost three columns of Table \ref{Functions1} show that conditions \pref{validityM1} are automatically satisfied in the perturbative regime, because $g^2/4\pi \ll 1$. Because of this condition \pref{validityM2} boils down to the demand that quantities like $|\omega \CMp/\CM|$ and $|\omega \DMp/\CM|$ should be small. But these conditions cannot be satisfied, as is also visible from the rightmost three columns of Table \ref{Functions1}. We henceforth therefore require
\begin{eqnarray}
\omega \ll a
\end{eqnarray}
as a necessary condition for dropping derivatives in the Taylor series of $\boldsymbol{\varrho}^{\ssI}(\tau - s)$. The remaining conditions are then summarized for the surviving three parameter regimes by the requirement that the top four rows of Table \ref{ValidityRelationsTable1} be much smaller than one. 

\begin{table}[h]
  \centering    
     \centerline{\begin{tabular}{ r|c|c|c|c|c|c| }
 \multicolumn{1}{r}{}
 & \multicolumn{1}{c}{$\underset{\ }{ \frac{\omega}{a} \ll \frac{m}{a} \ll 1}$}
 & \multicolumn{1}{c}{$\frac{m}{a} \ll \frac{\omega}{a} \ll 1$}
 & \multicolumn{1}{c}{$\frac{\omega}{a} \ll 1 \ll \frac{m}{a}$} \\
\cline{2-4}
$g^2 |\CMp(\omega) | $ & $ \stackrel{\ }{ \underset{\ }{\dfrac{g^2 \omega}{6 a} } } $ & $ \underset{\ }{\dfrac{g^2\omega}{6 a} } $ & $\dfrac{g^2 \pi \omega}{12 a} e^{ -{2m}/{a}}$ \\
\cline{2-4} 
$  g^2 | \DMp(\omega) |  $ & $ \stackrel{\ }{ \underset{\ }{ \dfrac{g^2}{2\pi^2} | \log( a \epsilon ) | } }$ & $\dfrac{g^2}{2\pi^2} |\log( a \epsilon )|$ & $\dfrac{g^2}{2\pi^2} |\log( e^{\gamma+1} m \epsilon )|$ \\
\cline{2-4}
$  \left| g^2 \DMp - \dfrac{2 \omega \CMp}{\CM} \right|   $ & $\ \stackrel{\ }{ \underset{\ }{\left| \dfrac{g^2}{2\pi^2}\log(a\epsilon) - \dfrac{4\pi^2\omega^2}{3a^2} \right|} }$ & $\left| \dfrac{g^2}{2\pi^2}\log(a\epsilon) - \dfrac{4\pi^2\omega^2}{3a^2} \right|$ & $\left| \dfrac{g^2}{2\pi^2} \log( e^{\gamma+1} m \epsilon ) - \dfrac{2\pi^2\omega^2}{a^2} \right|$ \\
\cline{2-4} 
$  \left| g^2 \CMp + \dfrac{2\omega \DMp }{\CM} \right|  $ & $\underset{\ }{\left| \dfrac{g^2 \omega}{6 a} + \dfrac{2 \omega \log(a \epsilon)}{a} \right|}$ & $\left| \dfrac{g^2 \omega}{6 a} + \dfrac{2 \omega \log(a \epsilon)}{a} \right|$ & $ \stackrel{\ }{ \left| \dfrac{\pi g^2 \omega}{8a}e^{-\tfrac{2m}{a} } + \dfrac{8\omega}{\pi a} \log( e^{\gamma+ 1} m \epsilon ) e^{\tfrac{2m}{a}} \right| }$ \\
\cline{2-4}
$  \dfrac{g^2 \CM(\omega)}{\omega}  $ & $ \underset{\ }{\dfrac{g^2 a}{4\pi^2\omega} }  $ & $ \underset{\ }{\dfrac{g^2 a}{4\pi^2\omega} } $ & $ \stackrel{\ }{ \dfrac{g^2 a}{8\pi\omega} e^{ -{2m}/{a}} }$ \\
\cline{2-4}
$   \dfrac{g^2 |\DM(\omega)|}{\omega}  $ & $\underset{\ }{\dfrac{g^2}{2\pi^2} |\log( a \epsilon )|}$ & $\dfrac{g^2}{2\pi^2} |\log( a \epsilon )|$ & $ \stackrel{\ }{ \dfrac{g^2}{2\pi^2} |\log( e^{\gamma+1} m \epsilon )| }$ \\
\cline{2-4} 
\end{tabular} }
        \caption{Asymptotic form for the quantities that must be small if the Markovian approximation is to be good, in various regimes for the relative sizes of $\omega$, $m$ and $a$. Only $\omega \ll a$ is considered because the rightmost three columns of Table \ref{Functions1} show that all Markovian conditions cannot be satisfied unless this is true. The first four rows express conditions \pref{validityM1} and \pref{validityM2}, while the bottom two rows are assumptions about the regime for $\omega$ that are assumed when deriving the top four rows.} \label{ValidityRelationsTable1}
\end{table}

Some of the conditions given in Table \ref{ValidityRelationsTable1} are automatically satisfied in perturbation theory, where the presence of the divergence in $\DM$ means that the perturbative treatment only holds when the small-distance cutoff  $\epsilon$ is chosen so that 
\be
  \frac{g^2}{4\pi} \ll \frac{1}{ | \log(a \epsilon) |} \ll 1 \,.
\ee
The rest of the conditions are generically satisfied throughout the entire range 
\be
   \frac{g^2}{4\pi} \ll \frac{\omega}{a} \ll 1 \,,
\ee
where the first inequality expresses our starting assumption, \pref{BigOmegaDef} (as re-expressed in \pref{validityM4}).

For completeness, we also include in Table \ref{ValidityRelationsTable2} the same constraints as above, though now written as a condition on the relaxation time-scales $\xi$ (which is more convenient for the discussion of later sections). (That is, Table \ref{ValidityRelationsTable2} merely repackages information that is already presented in Table \ref{ValidityRelationsTable1}).

\begin{table}[h]
  \centering    
     \centerline{\begin{tabular}{ r|c|c|c|c|c|c| }
 \multicolumn{1}{r}{}
 & \multicolumn{1}{c}{$\underset{\ }{ \frac{\omega}{a} \ll \frac{m}{a} \ll 1 }$}
 & \multicolumn{1}{c}{$\frac{m}{a} \ll \frac{\omega}{a} \ll 1$}
 & \multicolumn{1}{c}{$\frac{\omega}{a} \ll 1 \ll \frac{m}{a}$} \\
\cline{2-4}
$\dfrac{1}{\xi}\ \ll \ \  \underset{\ }{\left| \dfrac{\CM}{\DMp} \right|}  $ & $\dfrac{a}{| \log(a\epsilon) |}$ & $\dfrac{a}{| \log(a\epsilon) |}$ & $ \stackrel{\ }{ \dfrac{\pi a e^{-2m/a}}{4|\log(e^{\gamma+1}m \epsilon)|} }$ \\
\cline{2-4}
$\dfrac{1}{\xi} \ \ll \ \   \underset{\ }{\left| \dfrac{\CM}{\CMp} \right| }  $ & $\stackrel{\ }{ \dfrac{a^2}{\pi^2\omega} }$ & $\dfrac{3a^2}{2\pi^2\omega}$ & $\dfrac{3a^2}{2\pi^2\omega}$ \\
\cline{2-4}
\end{tabular} }
        \caption{The Markovian conditions expressed as constraints on the relaxation time-scales $\xi$. We only quote these constraints in the allowed $\omega \gg a$ regime.} \label{ValidityRelationsTable2}
\end{table}

\subsection{Field self-interactions and resummation}

Up until this point the scalar field has been regarded as being non-interacting, apart from its coupling to the qubit itself. This section briefly discusses some implications for late-time physics that arise once a scalar-field self-interaction is also added, of the form
\begin{eqnarray}
  H_{\lambda} & := &  \frac{\lambda}{4!} \int_{\Sigma_t} d^{3}\bx \ \phi^4 \otimes \boldsymbol{I} \,,
\label{HintLambda}
\end{eqnarray}
where again, $\Sigma_t$ is a sheet of constant Minkowski time $t$ and $\boldsymbol{I}$ is the $2 \times 2$ unit operator acting on the qubit sector. The dimensionless coupling $\lambda$ is assumed small enough to justify a perturbative treatment. 

The reason for considering $H_\lambda$ is that ultimately our interest is in field interactions and not in qubits. In particular, field self-interactions are also known to cause the phenomenon of secular growth, in which powers of $\lambda$ can sometimes arise in perturbation theory systematically multiplied by growing functions of time, $t$. Whenever this happens perturbation theory breaks down at late times, undermining the validity of inferences based purely on a non-interacting scalar field. In particular this kind of secular breakdown of perturbation theory is known to happen for thermal systems built from massless (or very light) bosons. In the presence of interactions like $H_\lambda$ corrections to scalar field propagators, $\langle \phi(\bfx,t) \phi(\bfx, t') \rangle$, at order $\lambda$ acquire contributions of order $\lambda T^3 (t-t')$ \cite{Burgess:2018sou}. 

Since the Minkowski vacuum behaves as a thermal state from the point of view of accelerated observers, one might worry that the late-time secular growth endemic to thermal systems might also occur for late-time corrections to the Minkowski propagator evaluated along accelerated world-lines. This question is examined in \cite{Burgess:2018sou}, where it is shown that secular growth can arise for accelerated observers in some circumstances, for sufficiently light scalar fields. This study also argued that when such secular growth does occur for massless fields its effects at late times can be resummed simply by recasting the Feynman rules to perturb around a scalar Hamilton whose mass is shifted by the amount
\begin{eqnarray} \label{massshifta}
\delta m^2 =  \frac{\lambda a^2}{96\pi^2} \,.
\end{eqnarray}
In particular, the leading late-time corrections for massless fields are simply obtained in such a resummation by using correlation functions appropriate for a massive scalar with a mass given by \pref{massshifta}. 

Applying this reasoning to the qubit evolution studied here shows how secular growth can feed through to affect physical results. In particular the resummation it requires changes the late-time behaviour of the Wightman function and so alters the response to it that is felt by an accelerating qubit. We illustrate this in the present section by computing the leading $\lambda$-dependent changes to qubit evolution at late times, for a massless scalar field self-interacting through the Hamiltonian \pref{HintLambda}.

Inclusion of $H_\lambda$ does not modify the Nakajima-Zwanzig equation itself for the qubit, which turns out not to explicitly depend on the operator $H_\lambda$ at second order in $g$. The reason for this lies in the observation that the commutators between the qubit coupling and the scalar self-interaction vanish. As a result the earlier analysis done for free scalars captures well the leading late-time corrections due to scalar self-interactions. The qubit's late-time steady state remains the thermal one, but the relaxation time-scales now depend on $\lambda$ by replacing $m$ using \pref{massshifta} in eqs.~\pref{massivexiRind}:
\be \label{INTxiRind2}
\tD = 2 \tT \simeq \frac{4\pi}{g^2 \omega} \tanh\left( \frac{\pi \omega }{a} \right) \left\{ 1 + \frac{\lambda a^2}{192 \pi^2 \omega^2}  \left[  1 - \frac{\cos\Big[ \frac{\omega}{a} \log\left( \frac{\lambda}{384 \pi^2} \right) - \zeta \Big]}{\sqrt{ ( \omega / a )^2 + 1 }} \right] \right\} \,,
\ee
with $\zeta$ given by \pref{zetadef}. This provides the leading corrections to \pref{masslessRindlertimescales} in powers of the scalar self-interaction. Notice that although suppressed by $\lambda$ the correction is also enhanced by $\omega/a$ in the Markovian regime (for which $\omega \ll a$). 

 \section{Controlling the late-time limit}
 \label{sec:LateTimeLimit}
 
In this section we circle back to discuss in more detail the justification for trusting the solutions to the Markovian evolution out to times that are of order $a \tau \simeq \cO(1/g^2)$, as must be done if the exponential form of the falloff to the static solution at late times is to be believed.

\subsection{Late times and Lindblad form}
 
In the preceding sections, we began with the Nakajima-Zwanzig equation \pref{INTpictureNZ} and sought the evolution of $\boldsymbol{\varrho}(\tau)$ on time-scales long compared to the width of the scalar-field's Wightman function. In this regime \pref{INTpictureNZ} reduces to the Markovian equations \pref{rho11three} and \pref{intpic12Markov}, which re-stated in terms of the Schr\"odinger-picture components state
\begin{eqnarray}
\frac{\partial \varrho_{11}}{\partial \tau} & \simeq & g^2 \RO - 2 g^2 \CO \varrho_{11}(\tau) \ , \label{Schrodinger11} \\
\frac{\partial \varrho_{12}}{\partial \tau} & \simeq & - ( i \omega + g^2 \CO ) \varrho_{12}(\tau) + g^2 (\CO - i \DO) \varrho_{12}^{\ast}(\tau) \,. \label{Schrodinger12}
\end{eqnarray}
As described above this assumes a choice of counter-term that ensures that $\omega$ continues to denote the qubit's physical energy gap, including any shifts to this gap due to the qubit/field interaction. 

The solutions to \pref{Schrodinger11} and \pref{Schrodinger12} describe a slow exponential relaxation towards a static late-time thermal density matrix of the form
\be \label{VarRhoThermal}
   \boldsymbol{\varrho}^{\rm static} =  \left[ \begin{matrix} \dfrac{1}{e^{\beta\omega} + 1} & 0 \\ 0 & \dfrac{1}{e^{-\beta\omega} + 1} \end{matrix} \right]\,,
\ee
with temperature $T = 1/\beta = a/(2\pi)$. At least, they do so if you really believe them out to time intervals $\tau \gg \xi$ that are of order $1/g^2$ in size. Given that all inferences have been based on perturbation theory in $g$, why should solutions of the form $\varrho_{ij} \propto \exp[- \tau/\xi]$ be regarded as being more accurate than the result $\varrho_{ij} \propto  1 - (\tau/\xi)$ that explicitly emerges from perturbation theory? This section fleshes out the arguments of \cite{OpenEFT1} that the exponential can be justified along the lines of the argument that justifies \pref{ExpDecay} by starting from \pref{ExpDecayDiffer}.

To this end let us formalize the argument leading from \pref{ExpDecayDiffer} to \pref{ExpDecay}. The starting point is a perturbative calculation of $\boldsymbol{\varrho}(\tau)$ as a function of an initial condition $\boldsymbol{\varrho}(\tau_0)$, along the lines of \pref{PertVarRho}. This perturbative solution necessarily breaks down at late times (because, for instance, it predicts a constant transition rate which eventually becomes inconsistent with qubit unitarity) and so is restricted to some interval $\tau - \tau_0 \ll \tau_p$, where $\tau_p$ is the time-scale beyond which perturbation theory fails. Within this interval differentiating the perturbative prediction allows the derivation of a differential evolution equation for $\partial_\tau \boldsymbol{\varrho}$, such as the Nakajima-Zwanzig equation \pref{INTpictureNZ} or its Markovian approximation \pref{Markov1st}.

For the purposes of understanding late times there is an important distinction between the Nakajima-Zwanzig result \pref{INTpictureNZ} and its Markovian approximation \pref{Markov1st}. This is because the Nakajima-Zwanzig result also refers explicitly to the initial and final times, $\tau_0$ and $\tau$, and on the history of the evolution that happens in between them. By contrast, a Markovian equation like \pref{Markov1st} or \pref{Schrodinger11} and \pref{Schrodinger12}, however, refers only to $\boldsymbol{\varrho}$ and $\partial_\tau \boldsymbol{\varrho}$ at the time $\tau$, with calculable $\tau$-independent coefficients. This means the Markovian equation could equally well have been justified by a perturbative calculation that starts at {\it any} time, $\tau_1$ say, provided the subsequent evolution is also over a window $\tau - \tau_1 \ll \tau_p$.

Now comes the main point. Since a Markovian evolution equation makes no intrinsic reference to a specific time or specific initial conditions, it can be separately justified in a family of overlapping time domains, $\cS_i$, each one of which lies over an interval much smaller than $\tau_p$ (to justify its perturbative derivation). But since it is the same equation in each of the $\cS_i$ the domain of validity of the solutions to this equation is the {\it union} of the domains $\cS = \cup_i \cS_i$, and so can apply over times $\tau \gg \tau_p$.

In the end of the day the result is a renormalization-group like argument. Although the initial perturbative evolution might require both $g$ and $g^2 \tau$ to be small, the differential evolution equation obtained from it neglects only powers of $g$ and makes no assumptions about the size of $g^2 \tau$. Consequently its solutions can resum effects to all orders in $g^2 \tau$, while still neglecting contributions of order $g^n \tau$ for $n > 2$. 

For open systems the differential evolution to which one is led in this way is (in the Schr\"odinger picture) of the Lindblad form \cite{Gorini:1975nb,Lindblad:1975ef,Gorini:1976cm,Redfield,DaviesOQS,Alicki,Kubo,Gardiner,Weiss,Breuer:2002pc,Rivas,Schaller},
\begin{eqnarray}
    \frac{\partial \boldsymbol{\varrho} (\tau)}{\partial \tau} & = & - i \left[ \mathfrak{h} , \boldsymbol{\varrho} (\tau) \right] + \sum_{j,k = 1}^{3} c_{jk} \left( \boldsymbol{F}_{j} \boldsymbol{\varrho} (\tau)       \boldsymbol{F}_{k}^{\dagger} - \frac{1}{2} \left\{ \boldsymbol{F}_{k}^{\dagger} \boldsymbol{F}_{j}, \boldsymbol{\varrho} (\tau) \right\} \right) \label{ExplicitLindblad}
\end{eqnarray}
for some set of operators $\boldsymbol{F}_i$ and a  Kossakowski matrix $\mathfrak{c} = [c_{jk}]$ full of coefficients. Provided these operators and coefficients do not themselves depend on time the domain of validity of this equation can be promoted to the union of domains, $\cS$, and thereby to times much longer than the perturbative domain, $\cS_i$, from which it might have been initially derived.  Eq.~\pref{ExplicitLindblad} has the property that it preserves the positivity and normalization of $\boldsymbol{\varrho}$, provided only that the Kossakowski matrix is hermitian and positive semi-definite \cite{Lindblad:1975ef, Gorini:1976cm}. 

\subsection{Positivity issues}
\label{sec:PositivityIssues}

In the present example of the accelerating qubit eqs.~\pref{Schrodinger11} and \pref{Schrodinger12} indeed have the form of \pref{ExplicitLindblad}, with $\boldsymbol{F}_{j} = \tfrac{1}{2} \boldsymbol{\sigma}_j$ given by Pauli matrices and the entries of the Kossakowski matrix $\mathfrak{c} = [c_{jk}]$ given explicitly by
\begin{eqnarray}
\mathfrak{c} & = & \left[ \begin{matrix} 4 g^2 \CO  & \  2g^2 (\DO - i \SO ) & 0 \\ 2g^2 (\DO + i \SO )  & 0 & 0 \\ 0& 0 & 0 \end{matrix} \right] \,,\label{KossaMatrix}
\end{eqnarray}
which eliminates $\RO$ using the identity \pref{RCSsum}. 

As mentioned above, the Kossakowski matrix must be hermitian and positive semi-definite to ensure that the evolution of $\boldsymbol{\varrho}(\tau)$ is unitary ({\it ie.} that the eigenvalues of $\boldsymbol{\varrho}(\tau)$ remain real and bounded between $0$ and $1$ as required for their interpretation as probabilities). Inspection of \pref{KossaMatrix}, however, reveals the three eigenvalues for this matrix to be
\begin{eqnarray}
\lambda^\mathfrak{c}_{1} & = & 0 \ ,\nn \\
\lambda^\mathfrak{c}_{2} & = & 2 g^2 ( \CO + \sqrt{ \mathcal{C}_{\Omega}^2 + \mathcal{S}_{\Omega}^2 + \Delta_{\Omega}^2 } \ ) \\
\mathrm{and} \quad \lambda^\mathfrak{c}_{3} & = & 2 g^2 ( \CO - \sqrt{ \mathcal{C}_{\Omega}^2 + \mathcal{S}_{\Omega}^2 + \Delta_{\Omega}^2 }\ ) \ .\nn
\end{eqnarray}
What is, at first sight, alarming about the above is that $\lambda^\mathfrak{c}_{3}$ is negative, implying that in general our Markovian equations of motion violate positivity of the reduced density matrix. 

If true, this would be alarming because it would imply the positivity of $\boldsymbol{\varrho}(\tau)$ is eventually violated. As we argue below, however,  the negative eigenvalue of the Lindblad equation corresponding to \pref{Schrodinger11} and \pref{Schrodinger12} is not reliable, since it is of the same size as contributions that are neglected when deriving \pref{ExplicitLindblad}.

\subsubsection*{Positivity and the Markovian accelerated qubit}

Although formally, \pref{ExplicitLindblad} is not positivity-preserving for generic values of $\CO$, $\SO$ and $\DO$, in this section we demonstrate that positivity is preserved in the Markovian limit (for the accelerating qubit), provided we ruthlessly restrict to the domain of the approximations used in its derivation.

To see how this works compare the size of the negative and positive eigenvalues of the Kossakowski matrix, \pref{KossaMatrix}, for the accelerated qubit:
\begin{eqnarray}
\lambda^{\mathfrak{c}}_{2} & = & 2 g^2 ( \CM + \sqrt{ \mathcal{C}_\ssM ^2 + \mathcal{S}_\ssM ^2 + \Delta_\ssM ^2 }\ ) \\
\lambda^{\mathfrak{c}}_{3} & = & 2 g^2 ( \CM - \sqrt{ \mathcal{C}_\ssM ^2 + \mathcal{S}_\ssM ^2 + \Delta_\ssM ^2 }\ ) \ .\nn
\end{eqnarray}
Recall that the validity of these equation presuppose that $\omega \ll a$, and consider, for concreteness' sake, the case $m \ll a$ (without assuming which of $m$ or $\omega$ is larger). Keeping in mind that $\SM = - \tanh\left( \frac{\pi\omega}{a} \right) \CM$ and using the results of Table \ref{ValidityRelationsTable1} shows that for $m \ll a$ we have 
\be
  \DM \simeq \frac{\omega}{2\pi^2} \log(a\epsilon) \; \sim \; \SM \simeq -\frac{\omega}{4\pi} 
\ee
and so both are smaller than 
\be
   \CM \simeq \frac{a}{4\pi^2} \,. 
\ee
Consequently these relations imply $\lambda^{\mathfrak{c}}_2 \simeq 4 g^2 \CM$ but also give $\lambda^\mathfrak{c}_3 \simeq - g^2 ( \mathcal{S}_{\ssM}^2 + \Delta_{\ssM}^2) /\CM \sim \mathcal{O}(g^2 \omega^2 / a)$, showing that the negative eigenvalue is actually consistent with zero within the approximations being used.\footnote{These arguments follow through in both the non-degenerate $\omega \gg g^2 \sqrt{ \mathcal{C}_{\ssM}^2 + \Delta_{\ssM}^2 }$ and degenerate  $\omega \gg g^2 \sqrt{ \mathcal{C}_{\ssM}^2 + \Delta_{\ssM}^2 }$ limits, since the above arguments rely on $\omega \ll a$ and $| \DM / \CM | \ll 1$ (which is true in both cases -- see Appendix \ref{App:Degenerate}).}

In the literature the issue of non-positivity of the Lindblad equation is usually addressed using an additional approximation, called the {\it rotating-wave approximation} (RWA). Appendix \ref{App:RotWav} summarizes this approximation and its relation to the description given here in the main text.\footnote{See \cite{Whitney} for another example of a master equation which does not make use of the RWA, and yet describes a positivity-preserving solution.}

\section{Conclusions}
\label{sec:Conc}

Open EFT methods have been proposed as useful tools when exploring late-time quantum physics in gravitational backgrounds \cite{OpenEFT1, OpenEFT2, OpenEFT3,OpenEFT4,OpenEFT5,OpenEFT6,OpenEFT7,OpenEFT8,Agon:2014uxa, Boyanovsky:2015tba, Boyanovsky:2015jen, Nelson:2016kjm, Hollowood:2017bil, Shandera:2017qkg, Agon:2017oia, Martin:2018zbe, Martin:2018lin}. We here use these tools for the toy model of a quantum mechanical two-level system coupled to a real scalar field and find its late-time evolution that is inaccessible using ordinary perturbative methods. Although a wealth of physics can be gleaned by studying perturbative excitation probabilities and rates (like (\ref{perturbation1}) and (\ref{uprate1})), we argue that perturbative approaches generically miss out on the physics of late times.

The state of the two-level Unruh-DeWitt detector is known to depend on its trajectory through the spacetime of study. In this work, we first pick a generic trajectory in a static spacetime and derive the Nakajima-Zwanzig equation \pref{rho111}-\pref{rho121}, truncated at second-order in the qubit-field coupling. Although the diagonal and off-diagonal components of $\boldsymbol{\varrho}$ evolve independently, the resulting integro-differential equations are notoriously difficult to solve. By specializing to a trajectory whose correlation functions fall off exponentially fast for $\tau \gg \tau_c$, and furthermore satisfy the KMS relation \pref{KMS1} with temperature $\beta^{-1}$, the equations of motion can be greatly simplified by taking the Markovian approximation. 

In the Markovian limit, the evolution of the qubit is assumed to be extremely slow compared to the width of the correlation functions evaluated along the trajectory of the qubit. Here the convolutions of $\boldsymbol{\varrho}(\tau - s)$ in the Nakajima-Zwanzig equation are replaced with a dependence only on $\boldsymbol{\varrho}(\tau)$ so that the equations of motion contain no dependence on the history of the state (so called-memory effects). This is justified by considering a Taylor series \pref{MarkovSeries} of $\boldsymbol{\varrho}(\tau - s)$ in powers of $s$ and dropping all derivatives in the expansion. By further assuming that $\tau \gg \tau_c$ the integrals in the equation of motion can be replaced with one-sided Fourier transforms of the qubit correlation function. 

In this simplified Markovian regime, the final asymptotic state for the qubit is found to be thermal and the solutions decay with two time-scales. By constraining the derivatives in the Taylor series of $\boldsymbol{\varrho}(\tau - s)$ to be small, explicit conditions on the parameters in the problem are also derived: these conditions are generically written down in terms of the relevant one-sided Fourier transforms that appear in the Markovian regime. 

We apply the above framework to the concrete example of a uniformly accelerated qubit moving through the Minkowski vaccum. The acceleration parameter $a$ has long ago been identified as proportional to the Unruh temperature for this system, and the qubit is found to settle to the asymptotic thermal state defined by the Unruh temperature. The corresponding relaxation time-scales \pref{timescaleMinkowski} depend on $a$, the energy gap of the qubit $\omega$, and mass $m$ of the underlying field and the dimensionless qubit-field coupling $g$. We also develop asymptotic forms for the relaxation time-scales in the limit of large \pref{largemassMinkowski} and small field masses (\ref{massivexiRind}), as well as for a massless field \pref{masslessRindlertimescales}. Interacting field theories are also known to exhibit secular perturbative breakdown: for a $\lambda \phi^4$-interacting theory, the lowest-order secularly growing loop corrections can be resummed to introduce a small mass shift to (almost) massless field theories. This paper accounts for the effect of these mass shifts in how the approach to equilibrium is adjusted in our formulae for the relaxation time-scales (\ref{INTxiRind2}) for the accelerated qubit.

The above Markovian description for the accelerated qubit only applies in a narrow regime of parameter space outlined by the validity conditions \pref{validityM1}-\pref{validityM4}. These conditions can be explicitly stated in the concrete example of an accelerated qubit, and in particular imply that the qubit gap must be small compared to the Unruh temperature with $\omega \ll a$ in order for the Markovian approximation to apply.

The Markovian equation of motion for the qubit can be brought into a Lindblad form, although for generic values of parameters which appear in this equation, it is not ensured to preserve positivity of the reduced density matrix throughout its entire evolution. Interestingly, we find that the validity conditions for the Markovian regime restrict the parameters in the equation in such a way that the solution is in fact always positive (within the approximations taken in this work) and there is no need to take the commonly-used `rotating-wave' approximation.

In short, we find that an open quantum systems approach provides invaluable insights into the classic framework of the Unruh-DeWitt detector, particularly if the emphasis is on late-time behaviour. The Markovian description is sometimes valid, and is quite restrictive on the parameters in the problem. We believe that wider application of open quantum systems methods will result in a deeper understanding of late-time quantum field theory in other gravitational backgrounds.

\section*{Acknowledgements}
This work was partially supported by funds from the Natural Sciences and Engineering Research Council (NSERC) of Canada. Research at the Perimeter Institute is supported in part by the Government of Canada through NSERC and by the Province of Ontario through MRI.

\appendix

\section{The Nakajima-Zwanzig equation}
\label{App:NakZwan}

A better description of perturbative evolution at late times is given by the Nakajima-Zwanzig equation \cite{Nakajima,Zwanzig,Kubo,Gardiner,Breuer:2002pc,Weiss,Rivas}, whose derivation is briefly sketched here. The logic of this equation is to project the evolution equation, given in the interaction picture by  
\be
 \partial_t \rho = \cL_t(\rho) \quad \hbox{where} \quad  \cL_t(\rho) := -\mathi \Bigl[V(t) ,  \rho \,\Bigr] \,,
\ee
onto the uncorrelated form $\rho_{\rm vac} \otimes \boldsymbol{\varrho}(t)$ where the quantum field density matrix is the projector onto the vacuum state: $\rho_{\rm vac} := |\Omega \rangle \langle \Omega |$. 

This projection is accomplished by defining a projection operator whose action on an arbitrary hermitian operator $\cO$ is 
\be \label{projPdef}
   \cP(\cO) := \rho_{\rm vac} \otimes  \underset{\phi}{\mathrm{Tr}}(\cO)  \,.
\ee 
Because $\Tr_\phi \,\varrho_{\rm vac} = 1$ this definition defines a projection operator since $\cP^2 = \cP$. It also satisfies $\cP(\rho_{\rm vac} \otimes \mathfrak{a}) = \rho_{\rm vac} \otimes \mathfrak{a}$ for any hermitian $\mathfrak{a}$ acting purely within the qubit Hilbert space. Consequently $\cP(\rho_0) = \rho_0$ for uncorrelated initial states $\rho_0 = \rho_{\rm vac} \otimes \boldsymbol{\varrho_0}$ and, more generally, $\cP[\rho(t)] = \rho_{\rm vac} \otimes \boldsymbol{\varrho}(t)$, where $\boldsymbol{\varrho}(t) = \Tr_\phi \,\rho$ is the reduced density matrix whose time-evolution is sought. 

Because both $\cP$ and $\cL$ act linearly, the projection of the evolution equation can be found explicitly by using the pair of equations
\bea\label{NZintro}
  &&\cP ( \partial_t \rho) = \cP \cL_t (\rho) = \cP \cL_t  \cP (\rho) + \cP \cL_t \cQ (\rho) \\
  \hbox{and}
  &&\cQ (\partial_t \rho) = \cQ \cL_t (\rho) = \cQ \cL_t \cP (\rho) + \cQ \cL_t \cQ (\rho) \,,\nn
\eea
where $\cQ := 1 - \cP$ is also a projection operator. The idea is to use the second of these equations to eliminate the second term on the right-hand side of the first equation, thereby obtaining a result depending explicitly only on $\cP(\rho)$, leading to the result
\be
 \cQ[\rho(t)] = \cG(t,t_0) \cQ(\rho_0) + \int_{t_0}^t \exd s \, \cG(t,s) \, \cQ \cL_s \cP[ \rho(s)] \,,
\ee
where the quantity $\cG(t,s)$ is given explicitly by
\bea 
 \cG(t,s) &=& 1+ \sum_{n=1}^\infty \int_s^t \exd s_1 \cdots  \int_s^{s_{n-1}} \exd s_n \, \cQ \cL_{s_1}\cdots  \cQ \cL_{s_n}  \nn\\
 &=& 1+ \sum_{n=1}^\infty\frac{1}{n!} \int_s^t \exd s_1 \cdots  \int_s^{t} \exd s_n \, \mathfrak{P} \Bigl[ \cQ \cL_{s_1}\cdots  \cQ \cL_{s_n} \Bigr] \,. 
\eea
Here $\mathfrak{P}$ denotes path-ordering (or time-ordering) of the $\cQ \cL_{s_i}$. Once this solution is inserted into the first of eqs.~\pref{NZintro} one obtains the Nakajima-Zwanzig equation,
\be \label{eq:NakajimaZwanzig}
  \cP[\partial_t \rho(t)] = \cP \cL_t \cP [\rho(t)] + \cP \cL_t \cG(t,t_0) \cQ(\rho_0) +  \int_{t_0}^t \exd s \, \cK(t,s) [\rho(s)] \,,
\ee
which defines the kernel $\cK(t,s) = \cP \cL_t \cG(t,s) \cQ \cL_s \cP$. The second term on the right-hand side vanishes for uncorrelated initial conditions, $\rho_0 = \rho_{\rm vac} \otimes \boldsymbol{\varrho_0}$, since these imply $\cP(\rho_0) = \rho_0$ and so $\cQ(\rho_0) = 0$. 

Since eq.~\pref{eq:NakajimaZwanzig} is an exact consequence of the original Liouville equation for $\rho(t)$ it is typically no easier to solve. It is nonetheless convenient to expand it out order-by-order in $V$, and it is useful when doing so to expand the interaction-picture interaction hamiltonian, $V(t)$, in a basis of operators in product form, 
\be
   V(t) = \sum_n \cA_{n} (t) \otimes \mathfrak{b}_{n}(t) \,.
\ee
Keeping only terms out to second order in $V$ it suffices to approximate the kernel by its leading (second-order in $V$) part, $\cK \simeq \cK_2 = \cP \cL_t \cQ \cL_s \cP$. For an uncorrelated initial condition, $\rho(t_0) = \rho_{\rm vac} \otimes \boldsymbol{\varrho_0}$, eq.~\pref{eq:NakajimaZwanzig} reduces to the following evolution equation for the reduced density matrix:
\bea \label{NakaZwanExplicit}
  \partial_t \,\boldsymbol{\varrho}(t) &=& - \mathi \sum_n \Bigl[ \mathfrak{b}_n(t) , \boldsymbol{\varrho}(t) \Bigr] \, \llangle \, \cA_n(t) \, \rrangle + (-\mathi)^2 \sum_{mn}  \int_{t_0}^t \exd s \Biggl\{ \Bigl[ \mathfrak{b}_m(t) , \mathfrak{b}_n(s) \boldsymbol{\varrho}(s) \Bigr] \, \llangle \, \delta \cA_m(t) \delta \cA_n(s) \, \rrangle  \nn \\
  && \qquad\qquad\qquad\qquad  -  \Bigl[ \mathfrak{b}_m(t) , \boldsymbol{\varrho}(s) \mathfrak{b}_n(s)  \Bigr] \, \llangle \, \delta \cA_n(s) \delta \cA_m(t) \, \rrangle \,\Biggr\} + \cO(V^3)\,, 
\eea
where $\llangle \,(\cdots)\, \rrangle = \Tr_\phi[(\cdots) \rho_{\rm vac}]$. This is the equation used in the main text.

Notice that if the reduced $\boldsymbol{\varrho}(t)$ appearing on the right-hand-side of \pref{NakaZwanExplicit} is re-expressed in terms of its initial value, again dropping all terms beyond $V^2$, then \pref{NakaZwanExplicit} agrees with the $B$-sector trace of the differential version of eq.~\pref{2ndOrderLiouville}. It is the keeping of the full reduced density matrix on the right-hand side of \pref{NakaZwanExplicit} that extends it domain of validity and allows it to be used to control the late-time limit.   

\section{Evaluating the integral $\SM$}
\label{App:A}

Here we compute the massive integral $\SM(\omega)$ in (\ref{Srate}) assuming that $\omega >0$. We emphasize that this matches the expression for $\RM(\omega) = - \frac{2}{\exp(\frac{2\pi}{a}\omega) - 1} \SM(\omega)$ given in \cite{Takagi:1986kn}. We find it is easier to compute $\SM(\omega)$ in the form
\begin{eqnarray} 
\SM(\omega) & = & - i \int_{-\infty}^{\infty} \exd\tau \ \WM(\tau) \sin( \omega \tau) \label{SwithRe}
\end{eqnarray} 
which is equivalent to (25) since $\mathrm{Re}[\WM(\tau)]$ is even in $\tau$ and hence does not contribute to the Fourier-sine transform. We evaluate
\begin{eqnarray}
\SM(\omega) & = & - \frac{am}{8\pi^2} \lim_{\epsilon \to 0^{+}} \int_{-\infty}^{\infty} \exd\tau \  \frac{\sin( \omega \tau)}{\sinh(\tfrac{a\tau}{2}- i \epsilon)} K_1\left( \tfrac{2m}{a} i\big[ \sinh(\tfrac{a\tau}{2}- i \epsilon ) \big] \right)
\end{eqnarray}
c.f. \pref{massWightman} (here for simplicity we take $\epsilon$ as a {\it dimensionless} regulator whose limit $\epsilon \to 0^{+}$ can be safely taken after integration). First switching the integration variable as $\tau \to z = \tfrac{a\tau}{2}$ and then allowing $z$ be complex-valued, the integral (\ref{SwithRe}) is equivalent to the integral over the contour depicted in Figure \ref{fig:contour1} below.
\begin{figure}[H]
\begin{center}
\includegraphics[scale=0.30]{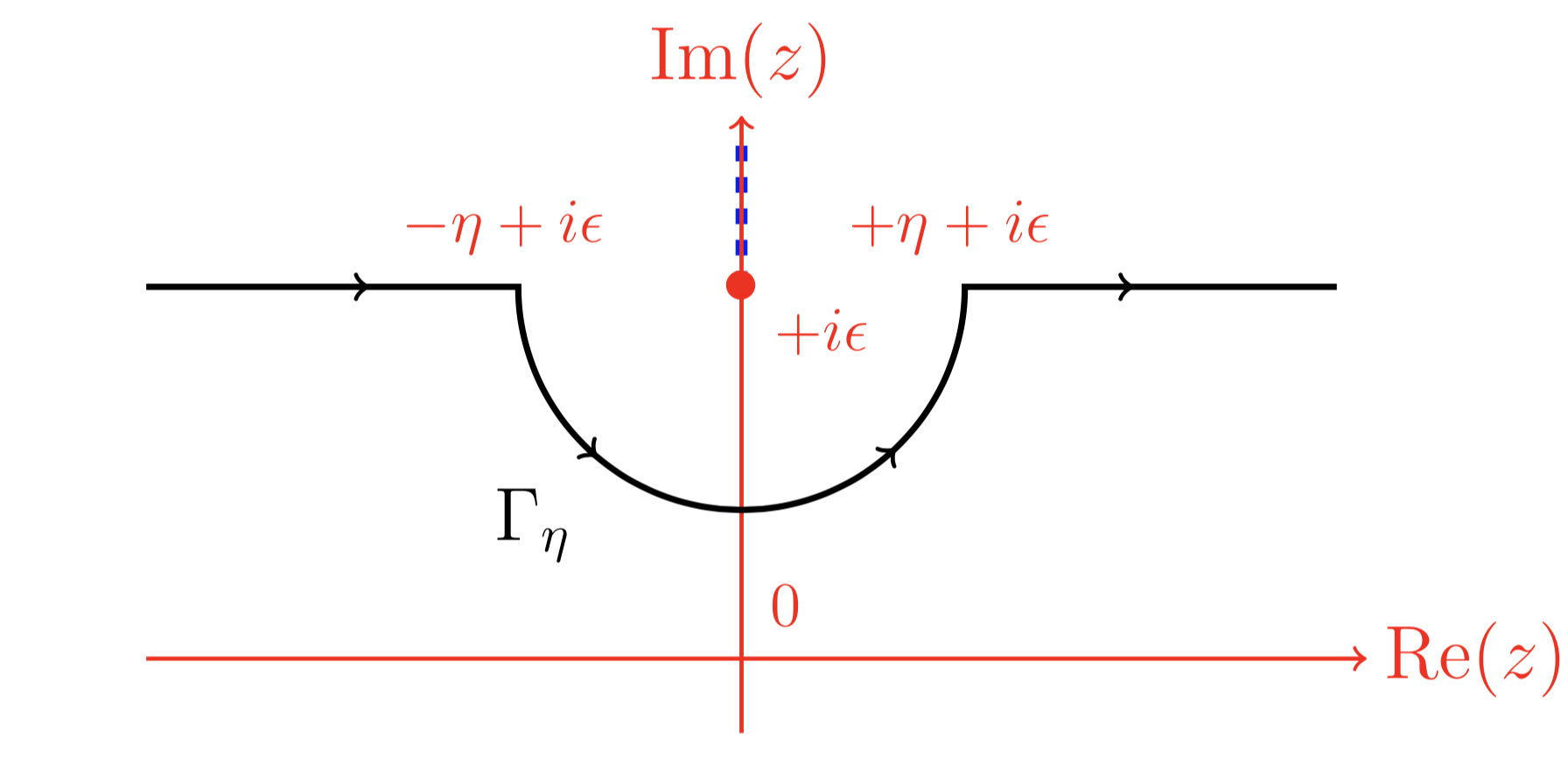}
\caption{The contour integral along $(-\infty+i\epsilon, \eta+i\epsilon] \cup \Gamma_{\eta} \cup [+\eta+i\epsilon,+\infty+i\epsilon)$ is equivalent to (\ref{SwithRe}) by the Cauchy theorem. Note the branch point at $z=+i\epsilon$ and the branch cut running upwards from there (stemming from the fact that $K_1(x)$ has a branch point at $x=0$ and a branch cut for all $\mathrm{Re}(x) < 0$).} \label{fig:contour1} 
\end{center}
\end{figure}
The contour $\Gamma_{\eta} : [0,\pi] \to \bC$ is here a semicircular contour of radius $\eta >0$ centred at $z=+i\epsilon$ which we parametrize as $\Gamma_{\eta}(\theta) = - \eta e^{i \theta} + i\epsilon$, and hence we write (\ref{SwithRe}) in the form
\begin{eqnarray}
\SM(\omega) & = & - \frac{m}{4\pi^2} \lim_{\eta, \epsilon \to 0^{+}} \bigg[ \int\limits_{-\infty+i\epsilon}^{-\eta+i\epsilon} +  \int\limits_{\Gamma_{\eta}} +  \int\limits_{\eta+i\epsilon}^{\infty+i\epsilon} \ \bigg] \exd z \ \frac{\sin( \frac{2\omega}{a} z)}{\sinh(z- i\epsilon)} K_1\left( \tfrac{2m}{a} i \sinh(z - i \epsilon) \right) \ .
\end{eqnarray}
We first examine the contour integral over $\Gamma_{\eta}$ where
\begin{eqnarray}
\lim_{\eta, \epsilon \to 0^{+}} \int_{\Gamma_{\eta}} \exd z \ \frac{\sin( \frac{2\omega}{a} z)}{\sinh(z-i\epsilon)} K_1\left( \tfrac{2m}{a} i \sinh(z-i\epsilon) \right) & = &  \lim_{\eta \to 0^{+}} \int_{\Gamma_{\eta}} \exd \theta\  \tfrac{\left( - i \eta e^{i \theta} \right) \sin( - \frac{2\omega}{a} \eta e^{i \theta})K_1\left( \frac{2m}{a} i \sinh(- \eta e^{i \theta}) \right)}{\sinh(- \eta e^{i\theta})} \ \ \ \ \ \ \ \ \\
& = & \lim_{\eta \to 0^{+}} \int_{0}^{\pi} \exd \theta\ \bigg( \frac{\omega}{m} + \mathcal{O}(\eta^2) \bigg) \\
& = & \frac{\pi\omega}{m} \ ,
\end{eqnarray}
which means that
\begin{eqnarray}
\SM(\omega) & = & - \frac{\omega}{4\pi} - \frac{m}{4\pi^2} \lim_{\eta, \epsilon \to 0^{+}} \bigg[ \int\limits_{-\infty+i\epsilon}^{-\eta+i\epsilon} +  \int\limits_{\eta+i\epsilon}^{\infty+i\epsilon} \  \bigg] \exd z \ \frac{\sin( \frac{2\omega}{a} z)}{\sinh(z-i\epsilon)} K_1\left( \tfrac{2m}{a} i \sinh(z-i\epsilon) \right) \ .
\end{eqnarray}
By shifting the integration variable by $-i\epsilon$ and then taking the limit $\epsilon \to 0^{+}$ the above becomes more simply 
\begin{eqnarray}
\SM(\omega) & = & - \frac{\omega}{4\pi} - \frac{m}{4\pi^2} \lim_{\eta \to 0^{+}} \bigg[ \int_{-\infty}^{-\eta} +  \int_{\eta}^{\infty} \  \bigg] \exd z \ \frac{\sin( \frac{2\omega}{a} z)}{\sinh(z)} K_1\left( \tfrac{2m}{a} i \sinh(z) \right) \ ,  \ \ 
\end{eqnarray}
and then switching the integration variable $z \to -z$ in the first integral we get
\begin{eqnarray}
\SM(\omega) & = & - \frac{\omega}{4\pi} - \frac{m}{4\pi^2} \lim_{\eta \to 0^{+}} \int_{\eta}^{\infty} \exd z \ \frac{\sin( \frac{2\omega}{a} z)}{\sinh(z)} \bigg[ K_1\left( \tfrac{2m}{a} i \sinh(z) \right) + K_1\left( - \tfrac{2m}{a} i \sinh(z) \right) \bigg] \ . \ \ 
\end{eqnarray}

Next using the connection formula $i \pi J_{\nu}(x) = e^{- \nu\frac{i\pi}{2}}K_{\nu}(x e^{-\frac{i\pi}{2}})  - e^{\nu\frac{i\pi}{2}}K_{\nu}(x e^{\frac{i\pi}{2}})$ valid for all $|\mathrm{arg}(x)|<\tfrac{\pi}{2}$ \cite{NIST}, the above can be expressed as
\begin{eqnarray}
\SM(\omega) & = & - \frac{\omega}{4\pi}  +  \frac{m}{4\pi} \lim_{\eta \to 0^{+}} \int_{\eta}^{\infty} \exd z \ \frac{\sin( \frac{2\omega}{a} z)}{\sinh(z)} J_1\left( \tfrac{2m}{a} \sinh(z) \right)\ .
\end{eqnarray}
The integrand is here regular at $z=0$ and so we may take the limit $\eta \to 0^{+}$ giving
\begin{eqnarray}
\SM(\omega) & = & - \frac{\omega}{4\pi} + \frac{a}{8\pi} I(\tfrac{2\omega}{a}, \tfrac{m\omega}{a}) \ ,
\end{eqnarray}
where we define the integral for $\Omega >0$ and $M>0$
\begin{eqnarray}
I(\Omega, M) & := & M \int_0^\infty \exd z\ \frac{\sin(\Omega z) }{\sinh(z)} J_{1}\big( M \sinh(z) \big) \ .\label{integralJ}
\end{eqnarray}

We will now evaluate this integral exactly by relating it to an ordinary differential equation. To begin, we compute the derivative
\begin{eqnarray}
\frac{\exd}{\exd M} \left[ M J_{1}\big(M \sinh(z)\big) \right] \ = \ J_{1}\big( M \sinh(z) \big) + \frac{M}{2} \sinh(z) \left[ J_{0} \big(M \sinh(z) \big) - J_{2} \big(M \sinh(z) \big) \right] \ .
\end{eqnarray}
Using the recurrence relation $J_{\alpha}(x) =\frac{x}{2\alpha} J_{\alpha - 1}(x) + \frac{x}{2\alpha}  J_{\alpha+1}(x)$ \cite{Abr} the above can be written as
\begin{eqnarray}
\frac{\exd}{\exd M} \left[ M J_{1}\big(M \sinh(z)\big) \right] \ = \ M \sinh(z) J_{0}\big( M \sinh(z) \big) \, 
\end{eqnarray}
and with this identity we find that
\begin{eqnarray}
\frac{\exd}{\exd M} I(\Omega, M) & = & M \int_0^\infty \exd z \ J_0\big(M\sinh(z)\big) \sin(\Omega z) \ .
\end{eqnarray}
We quote integral (6.679.4) in \cite{grad} for $a>0$ and $b>0$
\begin{eqnarray}
\int_0^\infty \exd x\ J_0\big( 2a \sinh(\tfrac{x}{2}) \big) \sin(b x) \ = \ \frac{2}{\pi} \sinh(\pi b) \big[ K_{ib}(a) \big]^{2}\ , 
\end{eqnarray}
which leads us to the ordinary differential equation
\begin{eqnarray}
\frac{\exd}{\exd M} I(\Omega, M) & = & \frac{M}{\pi} \sinh\left(\frac{\pi \Omega}{2}\right) \big[ K_{\frac{i\Omega}{2}}\big( \tfrac{M}{2} \big) \big]^{2} \ .
\end{eqnarray}
The above differential equation may be integrated up to an integration constant $c_0$ where
\begin{eqnarray}
I(\Omega, M) \ = \ \frac{M^2}{2 \pi} \sinh\left( \frac{\pi \Omega}{2} \right) \bigg( \big[ K_{\frac{i\Omega}{2}}\big( \tfrac{M}{2} \big) \big]^{2} - K_{\frac{i\Omega}{2} - 1}\big( \tfrac{M}{2} \big)  K_{\frac{i\Omega}{2} + 1}\big( \tfrac{M}{2} \big) \bigg) + c_0 \ .
\end{eqnarray}
An expansion of the integrand in (\ref{integralJ}) near $M=0$ shows a $\mathcal{O}(M^2)$ dependence, which demands that $\lim\limits_{M \to 0^{+}} I(\Omega, M) = 0$. It is this observation that allows us to determine the integration constant as
\begin{eqnarray}
c_0 & = & - \lim_{M \to 0^{+}} \left\{ \frac{M^2}{2 \pi} \sinh\left( \frac{\pi \Omega}{2} \right) \bigg( \big[ K_{\frac{i\Omega}{2}}\big( \tfrac{M}{2} \big) \big]^{2} - K_{\frac{i\Omega}{2} - 1}\big( \tfrac{M}{2} \big)  K_{\frac{i\Omega}{2} + 1}\big( \tfrac{M}{2} \big) \bigg)  \right\} = \Omega \ .
\end{eqnarray}

The integral (\ref{integralJ}) therefore evaluates to
\begin{eqnarray}
I(\Omega, M) \ = \ \frac{M^2}{2 \pi} \sinh\left( \frac{\pi \Omega}{2} \right) \bigg( \big[ K_{\frac{i\Omega}{2}}\big( \tfrac{M}{2} \big) \big]^{2} - K_{\frac{i\Omega}{2} - 1}\big( \tfrac{M}{2} \big)  K_{\frac{i\Omega}{2} + 1}\big( \tfrac{M}{2} \big) \bigg) + \Omega \ .
\end{eqnarray}

The above expression matches numerical tests (extra care must be taken when performing numerical integration of the left hand side as the integrand is heavily oscillatory). We conclude that
\begin{eqnarray}
\SM(\omega) \ = \ \frac{m^2}{ 4 \pi^2 a} \sinh\left( \frac{\pi \omega}{a} \right) \bigg( \big[ K_{\frac{i\omega}{a}}\big( \tfrac{m}{a} \big) \big]^{2} - K_{\frac{i\omega}{a} - 1}\big( \tfrac{m}{a} \big)  K_{\frac{i\omega}{a} + 1}\big( \tfrac{m}{a} \big) \bigg) \ .
\end{eqnarray}

\section{Small-mass asymptotics for the Minkowski rate integral}
\label{App:B}

Here we provide details of the small-$M$ expansion for the function 
\begin{eqnarray}
f(M,\Omega) = M^2 \bigg( K_{i\Omega - 1}\left( M \right)K_{i\Omega + 1}\left( M \right) - K_{i\Omega }\left( M \right)^2 \bigg)
\end{eqnarray}
which appears in our expression for the massive rate integral (\ref{massiverate}) in Minkowski space as $\RM(\omega) = \frac{a}{4\pi^2} e^{-\frac{\pi\omega}{a}} f(\tfrac{m}{a},\tfrac{\omega}{a})$. Using the expansion \cite{NIST} of $K_{\nu}(z)$ valid for $|z| \to 0$ and $\nu \in \bC \setminus \mathbb{Z}$
\begin{eqnarray}
K_{\nu}(z) = \frac{\Gamma(\nu)}{2} \left( \frac{z}{2} \right)^{-\nu} \left[ 1 + \frac{z^2}{4(1-\nu)} + \mathcal{O}(z^4) \right] + \frac{\Gamma(-\nu)}{2} \left( \frac{z}{2} \right)^{\nu} \left[ 1 + \frac{z^2}{4(1+\nu)} + \mathcal{O}(z^4) \right]
\end{eqnarray}
we expand $f$ for $0 < M \ll 1$ as
\begin{eqnarray}
f(M,\Omega) & \simeq & \frac{\pi \Omega}{\sinh(\pi \Omega)} - \frac{\pi M^2}{2\Omega\sinh(\pi\Omega)} + \frac{M^2}{4} \bigg( \left( \frac{M^2}{4} \right)^{i\Omega} \big[ \Gamma(-1-i\Omega) \Gamma(1-i\Omega) - \Gamma(-i \Omega)^2 \big]  \ \ \ \ \ \  \\
& \ & \ \ \ \ \ \ \ \ \ \ \ \ \ \ \ \ \ \ \ \ \  + \left( \frac{M^2}{4} \right)^{-i\Omega} \big[ \Gamma(-1+i\Omega) \Gamma(1+i\Omega) - \Gamma(i \Omega)^2 \big] \bigg) \ , \notag 
\end{eqnarray}
where we have used $|\Gamma(i\Omega)|^2 = \frac{\pi}{\Omega \sinh(\pi\Omega)}$. We use the property $\Gamma(z+1) = z \Gamma(z)$ to write the above as
\begin{eqnarray}
f(M,\Omega) & \simeq & \frac{\pi \Omega}{\sinh(\pi \Omega)} - \frac{\pi M^2}{2\Omega\sinh(\pi\Omega)}  + \frac{M^2}{4} \bigg( \left( \tfrac{M^2}{4} \right)^{i\Omega} \frac{\Gamma(-i\Omega)^2}{-i\Omega- 1}  + \left( \tfrac{M^2}{4} \right)^{-i\Omega} \frac{\Gamma(i\Omega)^2}{i\Omega- 1}  \bigg) \ .
\end{eqnarray}
In polar form the above becomes
\begin{eqnarray}
f(M,\Omega) & \simeq & \frac{\pi \Omega}{\sinh(\pi \Omega)} - \frac{\pi M^2}{2\Omega\sinh(\pi\Omega)} \\
& \ &  \ \ \ \ \ \ \ + \frac{M^2}{4} \left| \frac{\Gamma(i\Omega)^2}{i\Omega- 1}  \right| \left[ e^{i \left( 2 \Omega \log\left( \tfrac{M}{2} \right) - \mathrm{Arg}\left[ \frac{\Gamma(i\Omega)^2}{i\Omega - 1} \right] \right) }  + e^{-i \left( 2 \Omega \log\left( \tfrac{M}{2} \right) - \mathrm{Arg}\left[ \frac{\Gamma(i\Omega)^2}{i\Omega - 1} \right] \right) }  \right] \ , \notag
\end{eqnarray}
where we have used $\Gamma(z)^{\ast} = \Gamma(z^{\ast})$. After some simplification this gives 
\begin{eqnarray}
f(M,\Omega) & \simeq & \frac{\pi \Omega}{\sinh(\pi \Omega)} + \frac{\pi M^2}{2\Omega\sinh(\pi\Omega)} \left[ \frac{\cos\left( 2 \Omega \log\left( \frac{M}{2} \right) - \mathrm{Arg}\left[ \frac{\Gamma(i\Omega)^2}{i\Omega - 1} \right] \right)}{ \sqrt{\Omega^2 + 1}}  - 1 \right] \ .
\end{eqnarray}

\section{$\epsilon$-dependence of divergences in $\DM$ and $\Delta_\ssM ^{\prime}$}
\label{App:D}

Here we explore the $\epsilon$-dependence of the ultraviolet divergences in $\DM$ for the example of the accelerated qubit (from this the $\epsilon$-dependence of $\DMp$ immediately follows by differentiation). Using the Wightman function \pref{massWightman}, but with a small-{\it distance} regulator $\epsilon$, the integral \pref{DivergentShift} is explicitly\footnote{We replace $\sinh(\tfrac{as}{2}) - i a\epsilon/2 \to \sinh( {a[s-i\epsilon]}/{2} )$ relative to the form in \pref{massWightman}.}
\begin{eqnarray}
\Delta_\ssM  & = & 2 \lim_{\epsilon \to 0^{+}} \int_{0}^{\infty} \exd s\ \sin(\omega s) \; \mathrm{Re}\left[ \frac{am}{8 i \pi^2 } \frac{ K_{1}\left( \tfrac{2m i}{a} \sinh\frac{a[s - i \epsilon]}{2} \right)}{\sinh\frac{a[s - i \epsilon]}{2}} \right] \ .
\end{eqnarray}
We cannot take the limit $\epsilon \to 0^{+}$ here, so we keep $\epsilon$ small but finite (in the sense that $a\epsilon, \omega \epsilon, m \epsilon \ll 1$). For $s$ approaching the coincident limit, the Wightman function has the behaviour
\begin{eqnarray}
\WM(s) & \simeq & - \frac{1}{4\pi^2 (s- i \epsilon)^2} \ . \label{smallsWM}
\end{eqnarray}
We subtract and add \pref{smallsWM} in the expression for $\DM$ giving
\begin{eqnarray}
\Delta_\ssM  = 2 \int_{0}^{\infty} \exd s\ \sin(\omega s) \; \mathrm{Re}\left[  \frac{am}{8 i \pi^2 } \frac{ K_{1}\left( \tfrac{2m}{a} \sinh\frac{a[s - i \epsilon]}{2} \right)}{\sinh\frac{a[s - i \epsilon]}{2}} + \frac{1}{4\pi^2 (s- i \epsilon)^2 } - \frac{1}{4\pi^2 (s- i \epsilon)^2 } \right] \; . \ \ \ \  
\end{eqnarray}
We split this apart into two integrals such that
\begin{eqnarray}
\Delta_\ssM  & = & \Delta_\ssM ^{(\mathrm{divergent})} + \Delta_\ssM ^{(\mathrm{finite})} \label{DeltaSplit}
\end{eqnarray}
where 
\begin{eqnarray}
\Delta_\ssM ^{(\mathrm{divergent})} & = & 2 \int_{0}^{\infty} \exd s\ \sin(\omega s) \; \mathrm{Re}\left[ - \frac{1}{4\pi^2 (s- i \epsilon)^2 } \right] \\
\mathrm{and}\quad \quad \Delta_\ssM ^{(\mathrm{finite})} & = & 2 \int_{0}^{\infty} \exd s\ \sin(\omega s) \; \mathrm{Re}\left[  \frac{am}{8 i \pi^2 } \frac{ K_{1}\left( \tfrac{2m}{a} \sinh\frac{a[s - i \epsilon]}{2} \right)}{\sinh\frac{a[s - i \epsilon]}{2}} + \frac{1}{4\pi^2 (s- i \epsilon)^2 } \right] \; . \label{finiteDM}
\end{eqnarray}
which is justified since $\epsilon$ is finite here (and hence both integrals converge). We first compute the divergent part $\Delta_\ssM ^{(\mathrm{divergent})}$. To this end, we quote the integral (3.722.1) in \cite{grad}, 
\begin{eqnarray}
\int_{0}^{\infty} dx\ \frac{\sin( a x )}{x+\beta} \ = \ \sin(\beta a) \mathrm{ci}(\beta a) - \cos(\beta a) \mathrm{si}(\beta a )  
\end{eqnarray}
which is valid for $a>0$ and $|\arg(\beta ) |<\pi$, where $\mathrm{ci}$ and $\mathrm{si}$ are respectively the cosine integral and sine integral functions \cite{grad}, defined by\footnote{Note these definitions are valid for $z \in \bC$ in the complex plane so long as the contour connecting the limits on the integral does not intersect $(-\infty, 0]$.}
\begin{eqnarray}
\mathrm{ci}(z) = \gamma + \log(z) + \int_{0}^{z} \exd t \ \frac{\cos(t) - 1}{t} \quad \quad \quad \mathrm{and} \quad \quad \quad \mathrm{si}(z) = - \frac{\pi}{2} - \int_{0}^{z} \exd t \ \frac{\sin(t)}{t} \ . \label{ciANDsi}
\end{eqnarray}
 By differentiating the above integral with respect to $\beta$, an exact expression for $\Delta_\ssM ^{(\mathrm{divergent})}$ can be explicitly computed where
\begin{eqnarray}
\Delta_\ssM ^{(\mathrm{divergent})} & = & - \frac{1}{4\pi^2} \int_{0}^{\infty} \exd s\ \frac{\sin(\omega s)}{(s - i \epsilon )^2} \ - \frac{1}{4\pi^2} \int_{0}^{\infty} \exd s\ \frac{\sin(\omega s)}{(s + i \epsilon )^2}  \\
& = & \frac{\omega}{2\pi^2} \bigg[ \cosh(\omega \epsilon) \mathrm{chi}\left( \omega \epsilon \right) - \sinh(\omega \epsilon) \mathrm{shi}\left( \omega \epsilon \right) \bigg] \ .
\end{eqnarray}
where $\mathrm{chi}$ and $\mathrm{shi}$ are the hyperbolic cosine and sine integral functions \cite{grad}, respectively (defined analagous to \pref{ciANDsi} in the obvious way). Using $\mathrm{chi}(z) \simeq \gamma + \log(z) + \mathcal{O}(z^2)$ and $\mathrm{shi}(z) \simeq z + \mathcal{O}(z^3)$ in the $0< z \ll 1$ limit, for $\omega \epsilon \ll 1$ the above divergent piece has the form
\begin{eqnarray}
\Delta_\ssM ^{(\mathrm{divergent})} & \simeq & \frac{\omega}{2\pi^2} \bigg[ \log( e^{\gamma} \omega \epsilon ) \ + \  \mathcal{O}(\omega^2 \epsilon^2) \bigg] \ .
\end{eqnarray}

For $\Delta_\ssM^{(\mathrm{finite})}$ defined in the integral \pref{finiteDM}, the limit $\epsilon \to 0^{+}$ can be safely taken, where
\begin{eqnarray}
\Delta_\ssM ^{(\mathrm{finite})} & = & \frac{a}{4\pi^2} \int_0^\infty \exd z \ \sin\left( \tfrac{2\omega}{a} z \right) \bigg[ \frac{\pi m}{a} \frac{Y_{1}\left( \tfrac{2m}{a} \sinh(z) \right)}{\sinh(z)} + \frac{1}{z^2} \bigg] \label{deltafinite1}
\end{eqnarray}
where the connection formula $-\pi Y_{\nu}(x) = e^{- \nu \frac{i\pi}{2} } K_{\nu}\big( x e^{- \frac{i\pi}{2} } \big) + e^{ \nu \frac{i\pi}{2} } K_{\nu}\big( x e^{\frac{i\pi}{2} }  \big)$ valid for $|\arg(x)| \leq \pi /2$ \cite{NIST} has been used to relate the integrand in \pref{finiteDM} to the Bessel function of the second kind $Y_{1}$. Note that a change of variables $s \to z = {as}/{2}$ has also been made. 

We evaluate the function $\Delta_\ssM^{(\mathrm{finite})}$ by computing it as the limit
\begin{eqnarray}
\Delta_\ssM ^{(\mathrm{finite})} & = & \frac{a}{4\pi^2}  \int_0^\infty \exd z \ \sin\left( \tfrac{2\omega}{a} z \right) \lim_{p \to 1^{-}} \left\{ \frac{\pi m}{a} \frac{Y_{p}\left( \tfrac{2m}{a} \sinh(z) \right)}{\sinh(z)} + \left( \frac{m}{a} \right)^{1-p} \frac{\Gamma(p)}{z^{1+p}} \right\} \label{splitdefinside} \ ,
\end{eqnarray}
where the second $p$-dependent term is designed to continuously cancel (as a function of $p$) the leading-order behaviour of the first $p$-dependent term near $z=0$ (this follows from the leading-order behaviour $Y_{p}(x) \simeq - \frac{\Gamma(p)}{\pi}  \left( \frac{x}{2} \right)^{-p} - \frac{\cos(\pi p) \Gamma(-p)}{\pi}  \left( \frac{x}{2} \right)^{p} $ for $x \ll 1$ and $\nu \notin \mathbb{Z}$ \cite{NIST}). Such a choice ensures that the integrand is bounded by an integrable function for all $z$ being integrated (and for each $p$ in a neighbourhood of $1$), and so by the dominated convergence theorem \cite{Bartle} the limit operation can be taken outside of the integral such that
\begin{eqnarray}
\Delta_\ssM ^{(\mathrm{finite})}= \lim_{p \to 1^{-}} \left\{ \frac{m}{4\pi}  \int\limits_0^\infty \exd z \ \sin\left( \tfrac{2\omega}{a} z \right) \frac{Y_{p}\left( \tfrac{2m}{a} \sinh(z) \right)}{\sinh(z)} \ + \  \frac{a\Gamma(p) }{4\pi^2} \left( \frac{m}{a} \right)^{1 - p}  \int\limits_0^\infty \exd z \ \frac{\sin\left( \tfrac{2\omega}{a} z \right)}{z^{1+p}} \right\} \; . \quad \quad \label{splitdef}
\end{eqnarray}
Eq.~\pref{splitdef} is a useful parametrization because each of the integrals can be individually integrated for $0 < p < 1$ (and then the limit $p \to 1^{-}$ can be safely taken). The latter well-known integral in \pref{splitdef} can be evaluated (for example, with the help of formula (3.761.4) in \cite{grad}), and the first integral can be re-written with the connection formula $Y_{p}(z) = \cot(\pi p) J_{p}(z)  - \csc(\pi p) J_{-p}(z)$ \cite{NIST} so that
\begin{eqnarray}
\Delta_\ssM ^{(\mathrm{finite})} & = & \frac{\omega}{2\pi^2} \lim_{p \to 1^{-}} \bigg\{  \frac{\pi m}{2\omega} \cot(\pi p)  \int_0^\infty \exd z \ \sin\left( \tfrac{2\omega}{a} z \right) \sfrac{J_{p}\left( \tfrac{2m}{a} \sinh(z) \right)}{\sinh(z)} \label{anothersplit} \\
& \ & \quad \quad  \quad  \quad  \quad \quad \quad \quad  - \frac{\pi m}{2\omega}  \csc(\pi p)  \int_0^\infty \exd z \ \sin\left( \tfrac{2\omega}{a} z \right) \sfrac{J_{-p}\left( \tfrac{2m}{a} \sinh(z) \right)}{\sinh(z)} + \left( \sfrac{2\omega}{m} \right)^{p-1} \frac{\pi \sin\left( \frac{\pi p}{2} \right)}{ p \sin(\pi p)} \bigg\} \notag \ ,
\end{eqnarray}
where the remaining integrals can be evaluated such that
\begin{eqnarray}
\int\limits_0^\infty \exd z \  \sfrac{\sin\left( \tfrac{2\omega}{a} z \right) J_{\pm p}\left( \tfrac{2m}{a} \sinh(z) \right)}{\sinh(z)} & = & \frac{m}{\pm pa} \int\limits_0^\infty \exd z \ \sin\left( \tfrac{2\omega}{a} z \right) \bigg[ J_{\pm p-1}\left( \tfrac{2m}{a} \sinh(z) \right) + J_{\pm p+1}\left( \tfrac{2m}{a} \sinh(z) \right) \bigg] \quad \quad \quad \label{Uline2} \\
& = & \frac{m}{\pm pa} \; \mathrm{Im}\bigg[ I_{\frac{\pm p-1}{2} - i \tfrac{\omega}{a}}(\tfrac{m}{a}) K_{\frac{\pm p-1}{2} + i \tfrac{\omega}{a}}(\tfrac{m}{a}) \label{Uline3} \\
& \ & \quad \quad \quad \quad \quad \quad  \quad \quad \quad + I_{\frac{\pm p+1}{2} - i \tfrac{\omega}{a}}(\tfrac{m}{a}) K_{\frac{\pm p+1}{2}+ i \tfrac{\omega}{a}}(\tfrac{m}{a}) \bigg] \notag   \\\
& := & \pm \frac{m}{pa} \; \mathcal{U}(p, \tfrac{\omega}{a}, \tfrac{m}{a}) \label{Uline4}
\end{eqnarray}
In \pref{Uline2} we have used the recurrence relation ${2\nu J_{\nu}(z)}/{z} = J_{\nu - 1}(z) + J_{\nu + 1}(z)$ \cite{NIST}, and then in \pref{Uline3} we have used formula (6.679.1) from \cite{grad},
\begin{eqnarray}
\int_0^\infty \exd z\ \sin(2Bz) J_{2\nu}\big(  2 A \sinh z \big) & = & \mathrm{Im}\left[ I_{\nu - i B}(A) K_{\nu + i B}(A) \right] \ ,
\end{eqnarray}
which converges for $A>0$, $B>0$ and $\mathrm{Re}[\nu] > -1$ (since $0 < p < 1$ both expressions for $\pm p$ in \pref{Uline3} are valid). Writing the limit \pref{anothersplit} in terms of the function $\mathcal{U}(p, \tfrac{\omega}{a}, \tfrac{m}{a})$ defined in \pref{Uline4} then yields
\begin{eqnarray}
\Delta_\ssM ^{(\mathrm{finite})} & = & \frac{\omega}{2\pi^2} \lim_{p \to 1^{-}} \bigg\{ \dfrac{\frac{m^2}{2\omega a} \cos(\pi p) \mathcal{U}(p,\tfrac{\omega}{a}, \tfrac{m}{a}) + \frac{m^2}{2\omega a}  \mathcal{U}(-p,\tfrac{\omega}{a}, \tfrac{m}{a}) + \left( \frac{2\omega}{m} \right)^{p-1} \sin\left( \frac{\pi p}{2} \right) }{\frac{p \sin(\pi p)}{\pi}} \bigg\} \label{limitcleaner} \ .
\end{eqnarray}
In the given form, the above limit is actually in indeterminate form. To see this define the functions $f(p):=\frac{m^2}{2\omega a} \cos(\pi p) \mathcal{U}(p,\tfrac{\omega}{a}, \tfrac{m}{a}) + \frac{m^2}{2\omega a}  \mathcal{U}(-p,\tfrac{\omega}{a}, \tfrac{m}{a}) + \left( \frac{2\omega}{m} \right)^{p-1} \sin\left( \frac{\pi p}{2} \right)$ as well as $g(p):=\frac{p \sin(\pi p)}{\pi}$ (suppressing the dependence on the other variables for clarity of notation), and note that $g(1) = 0$ and that
\begin{eqnarray}
f(1) & = & \tfrac{m^2}{2\omega a} \left[ - \mathcal{U}(1,\tfrac{\omega}{a},\tfrac{m}{a}) + \mathcal{U}(-1,\tfrac{\omega}{a},\tfrac{m}{a}) \right] + 1 \\
& = & \tfrac{m^2}{2\omega a} \; \mathrm{Im} \bigg[ - I_{1 - i \tfrac{\omega}{a}}(\tfrac{m}{a}) K_{1 + i \tfrac{\omega}{a}}(\tfrac{m}{a}) + I_{-1 - i \tfrac{\omega}{a}}(\tfrac{m}{a}) K_{-1+ i \tfrac{\omega}{a}}(\tfrac{m}{a}) \bigg] + 1 \\
& = & \tfrac{m^2}{2\omega} \; \mathrm{Im} \bigg[ - \tfrac{2(-i\omega)}{m} \bigg( K_{-i \tfrac{\omega}{a}}(\tfrac{m}{a}) \sfrac{\exd I_{-i {\omega}/{a}}(\tfrac{m}{a})}{\exd m} - \sfrac{\exd K_{-i {\omega}/{a}}(\tfrac{m}{a})}{\exd m} I_{-i \tfrac{\omega}{a}}(\tfrac{m}{a}) \bigg) \bigg] + 1 \label{negative1coefficient2}  \quad \quad \quad \\
& = & 0 \label{negative1coefficient3}
\end{eqnarray}
where \pref{negative1coefficient2} follows by use of the symmetry $K_{\nu}(z) = K_{-\nu}(z)$ as well as the recurrence relations $I_{\nu \pm 1}(z) = I^{\prime}_{\nu}(z) \mp \frac{2\nu}{z} I_{\nu}(z)$ (which $K_{\nu}(z)$ also obeys) \cite{NIST}. From there \pref{negative1coefficient3} follows by use of the Wronskian relation $K_{\nu}(z) I^{\prime}_{\nu}(z) - K^{\prime}_{\nu}(z) I_{\nu}(z)  = 1/z$. Since $f(1) = g(1) = 0$ the limit \pref{limitcleaner} is in indeterminate (``${0}/{0}$'') form and can be evaluated using l'H\^{o}pital's rule \cite{Abr} where now $\Delta_\ssM ^{(\mathrm{finite})} = \frac{\omega}{2\pi^2} \lim\limits_{p \to 1^{-}} \frac{f'(p)}{g'(p)}$ giving
\begin{eqnarray}
\Delta_\ssM ^{(\mathrm{finite})} & = & \frac{\omega}{2\pi^2} \dfrac{ \left[ - \frac{m^2}{2\omega a} \left( \frac{\partial \mathcal{U}}{\partial p} |_{p=1} + \frac{\partial \mathcal{U}}{\partial p} |_{p=-1}  \right) + \log \left( \frac{2\omega}{m} \right) \right] }{\left[ -1 \right]} \\
& = & -  \frac{\omega}{2\pi^2} \log\left( \frac{2\omega}{m} \right) + \frac{m^2}{4\pi^2a} \;  \mathrm{Im}\bigg[ K_{i\tfrac{\omega}{a}}(\tfrac{m}{a}) \frac{\partial I_{\nu}(\tfrac{m}{a}) }{\partial \nu} \bigg|_{\nu = - i \tfrac{\omega}{a}}  \ + \; \frac{\partial K_{\nu}(\tfrac{m}{a}) }{\partial \nu} \bigg|_{\nu =  i \tfrac{\omega}{a}} I_{- i\tfrac{\omega}{a}}(\tfrac{m}{a}) \quad \quad \quad \quad \label{explicitDMfinite} \\
& \ & \quad \quad \quad \quad \quad  + \frac{1}{2} K_{-1 + i\tfrac{\omega}{a}}(\tfrac{m}{a}) \frac{\partial I_{\nu}(\tfrac{m}{a}) }{\partial \nu} \bigg|_{\nu = - 1 - i \tfrac{\omega}{a}} \ + \; \frac{1}{2} \frac{\partial K_{\nu}(\tfrac{m}{a}) }{\partial \nu} \bigg|_{\nu =  - 1 + i \tfrac{\omega}{a}} I_{-1 - i\tfrac{\omega}{a}}(\tfrac{m}{a}) \notag \\
& \ &  \quad \quad \quad \quad \quad  \quad \quad \quad \quad \quad + \frac{1}{2} K_{1 + i\tfrac{\omega}{a}}(\tfrac{m}{a}) \frac{\partial I_{\nu}(\tfrac{m}{a}) }{\partial \nu} \bigg|_{\nu = 1 - i \tfrac{\omega}{a}} \ + \; \frac{1}{2} \frac{\partial K_{\nu}(\tfrac{m}{a}) }{\partial \nu} \bigg|_{\nu =  1 + i \tfrac{\omega}{a}} I_{1 - i\tfrac{\omega}{a}}(\tfrac{m}{a}) \bigg] \notag
\end{eqnarray}
which means that $\DM$ is overall given by the function
\begin{eqnarray}
\DM & = &  \frac{\omega}{2\pi^2} \log\left( \frac{ e^{\gamma} m \epsilon }{2} \right) + \frac{m^2}{4\pi^2a} \;  \mathrm{Im}\bigg[ K_{i\tfrac{\omega}{a}}(\tfrac{m}{a}) \frac{\partial I_{\nu}(\tfrac{m}{a}) }{\partial \nu} \bigg|_{\nu = - i \tfrac{\omega}{a}}  \ + \; \frac{\partial K_{\nu}(\tfrac{m}{a}) }{\partial \nu} \bigg|_{\nu =  i \tfrac{\omega}{a}} I_{- i\tfrac{\omega}{a}}(\tfrac{m}{a}) \quad \quad \quad \quad\label{explicitDM} \\
& \ & \quad \quad \quad \quad \quad  + \frac{1}{2} K_{-1 + i\tfrac{\omega}{a}}(\tfrac{m}{a}) \frac{\partial I_{\nu}(\tfrac{m}{a}) }{\partial \nu} \bigg|_{\nu = - 1 - i \tfrac{\omega}{a}} \ + \; \frac{1}{2} \frac{\partial K_{\nu}(\tfrac{m}{a}) }{\partial \nu} \bigg|_{\nu =  - 1 + i \tfrac{\omega}{a}} I_{-1 - i\tfrac{\omega}{a}}(\tfrac{m}{a}) \notag \\
& \ &  \quad \quad \quad \quad \quad  \quad \quad \quad \quad \quad + \frac{1}{2} K_{1 + i\tfrac{\omega}{a}}(\tfrac{m}{a}) \frac{\partial I_{\nu}(\tfrac{m}{a}) }{\partial \nu} \bigg|_{\nu = 1 - i \tfrac{\omega}{a}} \ + \; \frac{1}{2} \frac{\partial K_{\nu}(\tfrac{m}{a}) }{\partial \nu} \bigg|_{\nu =  1 + i \tfrac{\omega}{a}} I_{1 - i\tfrac{\omega}{a}}(\tfrac{m}{a}) \bigg] \ . \notag
\end{eqnarray}
This expression can be expanded in various regimes of $\omega$, $m$ and $a$, with results quotes in Table \ref{DeltaTable} (the leading-order behaviour for $\DMp$ is achieved by differentiating the entries in this Table. 
\begin{table}[h]
  \centering    
     \centerline{\begin{tabular}{ r|c|c|c|c|c|c| }
 \multicolumn{1}{r}{}
 & \multicolumn{1}{c}{$\underset{\ }{ \frac{\omega}{a} \ll \frac{m}{a} \ll 1 }$}
 & \multicolumn{1}{c}{$\frac{m}{a} \ll \frac{\omega}{a} \ll 1$}
 & \multicolumn{1}{c}{$\frac{\omega}{a} \ll 1 \ll \frac{m}{a}$} 
 & \multicolumn{1}{c}{$\frac{m}{a} \ll 1 \ll \frac{\omega}{a}$}
 & \multicolumn{1}{c}{$1 \ll \tfrac{\omega}{a} \ll \frac{m}{a}$}
 & \multicolumn{1}{c}{$ 1 \ll \frac{m}{a} \ll \frac{\omega}{a}$} \\
\cline{2-7}
$\Delta_{\ssM}^{(\mathrm{finite})} $ & $\stackrel{\ }{ \underset{\ }{ - \frac{\omega}{2\pi^2}\log\left( \frac{e^{\gamma} \omega}{a} \right) } }$ & $- \frac{\omega}{2\pi^2}\log\left(  \frac{e^{\gamma} \omega}{a} \right)$ & $- \frac{\omega}{2\pi^2}\log\left( \frac{2\omega}{e m} \right)$ & $a \cdot \mathcal{O}(\tfrac{a}{\omega})$ & $ - \frac{\omega}{2\pi^2}\log\left( \frac{2\omega}{em} \right)$ & $a \cdot \mathcal{O}(\tfrac{a}{\omega})$ \\
\cline{2-7}
$\Delta_{\ssM} $ & $\stackrel{\ }{ \underset{\ }{ \frac{\omega}{2\pi^2}\log(a\epsilon) } }$ & $ \frac{\omega}{2\pi^2}\log(a\epsilon)$ & $\frac{\omega}{2\pi^2}\log(e^{\gamma + 1} m \epsilon)$ & $ \frac{\omega}{2\pi^2}\log\left( e^{\gamma} \omega \epsilon \right)$ & $\frac{\omega}{2\pi^2}\log\left( e^{\gamma+1} m \epsilon \right)$ & $\frac{\omega}{2\pi^2}\log\left( e^{\gamma} \omega \epsilon \right)$ \\
\cline{2-7} 
\end{tabular} }
        \caption{As given in Table \ref{Functions1}, the leading-order behaviour in various regimes for the function $\Delta_{\ssM}^{(\mathrm{finite})}$ given by \pref{explicitDMfinite} and $\DM$ given by \pref{explicitDM}. In each case the sub-leading corrections are parametrically small.} \label{DeltaTable}
\end{table}

\section{Small qubit splitting and the Markovian approximation}
\label{App:Degenerate}

We explore the degenerate $\omega \ll g^2 \sqrt{ \mathcal{C}_\ssM^2 + \Delta_\ssM^2 }$ limit in this section, applied to the example of the accelerated qubit. Naively solving the Markovian equation for $\varrho_{12}^{\ssI}(\tau)$ in this limit yields
\begin{eqnarray}
\varrho_{12}^{\ssI}(\tau) & \simeq & e^{+i \omega \tau} e^{ - g^2 \left[  \CM + \sqrt{ \mathcal{C}_\ssM ^2 + \Delta_\ssM ^2 } \right] \; \tau} \left[  \frac{\varrho_{12}(0)}{2} \bigg( 1 + i \frac{\omega}{g^2 \sqrt{ \mathcal{C}_\ssM ^2 + \Delta_\ssM ^2 }} \bigg) - \frac{\varrho_{12}^{\ast}(0)}{2} \frac{ \CM - i \DM}{\sqrt{ \mathcal{C}_\ssM ^2 + \Delta_\ssM ^2  }} \right] \label{AppDegenSolution} \\
& \ & \quad \quad \quad+ e^{+i \omega \tau} e^{ + g^2 \left[  - \CM + \sqrt{ \mathcal{C}_\ssM ^2 + \Delta_\ssM ^2 } \; \right] \tau} \left[  \frac{\varrho_{12}(0)}{2} \bigg( 1 - i \frac{\omega}{g^2 \sqrt{ \mathcal{C}_\ssM ^2 + \Delta_\ssM ^2 }} \bigg) + \frac{\varrho_{12}^{\ast}(0)}{2} \frac{ \CM - i \DM}{\sqrt{ \mathcal{C}_\ssM ^2 + \Delta_\ssM ^2  }} \right] \notag \ .
\end{eqnarray}
As outlined in \S\ref{sec:Markovian}, dropping derivatives in the Taylor series of $\varrho^{\ssI}(\tau - s)$ in the Nakajima-Zwanzig equation necessitates the bounds \pref{degenCond1}, restated here for convenience:
\begin{equation}\label{APPdegenCond}
  \begin{gathered}
\left| g^2 \DMp - \frac{\omega \CMp}{\CM} \right| \ll  1 \quad , \quad \left| g^2\CMp + \frac{\omega \DMp}{\CM} \right| \ll 1 \ ,  \\
g^2 |\DMp| \; \sqrt{ 1 + \frac{\Delta_\ssM ^2}{\mathcal{C}_\ssM ^2} } \ll 1 \quad , \quad g^2 |\CMp| \;  \sqrt{ 1 + \frac{\Delta_\ssM ^2}{\mathcal{C}_\ssM ^2} } \ll 1 \quad , \ \quad \frac{g^2 \CM}{\omega} \; \sqrt{ 1 + \frac{\Delta_\ssM ^2}{\mathcal{C}_\ssM ^2} } \ \gg \ 1\ .
  \end{gathered} 
\end{equation}
where the last bound is a re-statement of the degeneracy condition $\omega \ll g^2 \sqrt{ \mathcal{C}_\ssM^2 + \Delta_\ssM^2 }$. The first important step is to note that the last three bounds of \pref{APPdegenCond} imply the hierarchy
\begin{eqnarray}
g^2 |\DMp| \ , \ g^2 |\CMp| \quad  \ll \ \frac{1}{\sqrt{ 1 + {\Delta_\ssM ^2}/{\mathcal{C}_\ssM ^2} }} \ \ll \quad \frac{g^2 \CM}{\omega} \ . \label{statement}
\end{eqnarray}
This statement \pref{statement} implies two things. The first is that:
\begin{eqnarray}
g^2 |\DMp| & \ll & \frac{g^2 \CM}{\omega} \quad \quad \quad \implies \quad \quad \quad \frac{\omega |\DMp|}{\CM} \ \ll \ 1  \quad \quad \quad \implies \quad \quad \quad \left| \frac{\DM}{\CM} \right| \ \ll \ 1 
\end{eqnarray}
where the last implication follows because $\omega \DMp \simeq \DM$ to leading-order (this is immediately seen in Table \ref{Functions1}). The second thing that the statement (\ref{statement}) implies is that
\begin{eqnarray}
g^2 |\CMp| & \ll & \frac{g^2 \CM}{\omega} \quad \quad \quad \implies \quad \quad \quad \; \frac{\omega |\CMp|}{\CM} \ \ll \ 1 \quad \quad \quad \implies \quad \quad \quad \omega \ll a \label{simple2}
\end{eqnarray}
Where the last implication follows by examining the values of the functions $\CM$ and $\CMp$ in Table \ref{Functions1}. To make this explicit, we write down Table \ref{DegenTable1} exploring the size of $\omega \CMp / \CM$ in the various regimes (showing that only $\omega \ll a$ is allowed). For completeness we also fill out Table \ref{DegenTable2} with all the inequalities \pref{APPdegenCond} in the $\omega \ll a$ regime. We also point out that the last row of Table \ref{DegenTable2} tells us that $\omega / a$ is so small that $\omega / a \ll g^2 \ll 1$. We note finally that $ |\DM / \CM | \lesssim \mathcal{O}(g^2) \ll 1$, which is easy to see in Table \ref{DegenTable3}. 
\begin{table}[h]
  \centering    
     \centerline{\begin{tabular}{ r|c|c|c|c|c|c| }
 \multicolumn{1}{r}{}
 & \multicolumn{1}{c}{$\underset{\ }{ \frac{\omega}{a} \ll \frac{m}{a} \ll 1 }$}
 & \multicolumn{1}{c}{$\frac{m}{a} \ll \frac{\omega}{a} \ll 1$}
 & \multicolumn{1}{c}{$\frac{\omega}{a} \ll 1 \ll \frac{m}{a}$}
 & \multicolumn{1}{c}{$\frac{m}{a} \ll 1 \ll \frac{\omega}{a}$}
 & \multicolumn{1}{c}{$1 \ll \frac{\omega}{a} \ll \frac{m}{a}$}
 & \multicolumn{1}{c}{$1 \ll \frac{m}{a} \ll \frac{\omega}{a}$} \\
\cline{2-7}
$\dfrac{\omega\CMp}{\CM} $ & $\stackrel{\ }{ \underset{\ }{\dfrac{2\pi^2\omega^2}{3a^2} } } $ & $ \underset{\ }{\dfrac{2\pi^2\omega^2}{3a^2} } $ & $\dfrac{\pi^2\omega^2}{a^2}$ & $1$ & $\dfrac{\pi\omega}{a}$ & $1$ \\
\cline{2-7}
\end{tabular} }
        \caption{Leading-order behaviour for the function ${\omega\CMp}/{\CM}$ from the bound \pref{simple2} in various regimes of parameter space. Notice it is only possible to satisfy ${\omega\CMp}/{\CM} \ll 1$ in the $\omega \ll a$ regime. ({\it ie.} in the first three columns.)} \label{DegenTable1}
\end{table}

\begin{table}[h]
  \centering    
     \centerline{\begin{tabular}{ r|c|c|c|c|c|c| }
 \multicolumn{1}{r}{}
 & \multicolumn{1}{c}{$\underset{\ }{ \frac{\omega}{a} \ll \frac{m}{a} \ll 1 }$}
 & \multicolumn{1}{c}{$\frac{m}{a} \ll \frac{\omega}{a} \ll 1$}
 & \multicolumn{1}{c}{$\frac{\omega}{a} \ll 1 \ll \frac{m}{a}$} \\
\cline{2-4}
$1\ \gg \ \left| g^2 \DMp - \dfrac{2 \omega \CMp}{\CM} \right|  \simeq$ & $\stackrel{\ }{ \underset{\ }{\left| \dfrac{g^2}{2\pi^2}\log(a\epsilon) - \dfrac{4\pi^2\omega^2}{3a^2} \right|} }$ & $\left| \dfrac{g^2}{2\pi^2}\log(a\epsilon) - \dfrac{4\pi^2\omega^2}{3a^2} \right|$ & $\left| \dfrac{g^2}{2\pi^2} \log( e^{\gamma+1} m \epsilon ) - \dfrac{2\pi^2\omega^2}{a^2} \right|$ \\
\cline{2-4} 
$1\ \gg \ \left| g^2 \CMp + \dfrac{2\omega \DMp }{\CM} \right| \simeq$ & $\stackrel{\ }{ \underset{\ }{\left| \dfrac{g^2 \omega}{6 a} + \dfrac{2 \omega \log(a \epsilon)}{a} \right|} }$ & $\left| \dfrac{g^2 \omega}{6 a} + \dfrac{2 \omega \log(a \epsilon)}{a} \right|$ & $\left| \frac{\pi g^2 \omega}{8a}e^{-\tfrac{2m}{a}} + \frac{8\omega\log( e^{\gamma+ 1} m \epsilon ) }{\pi a}  e^{\tfrac{2m}{a}} \right|$ \\
\cline{2-4}
$1\ \gg \  g^2 | \DMp | \sqrt{ 1 + \sfrac{\Delta_\ssM ^2}{\mathcal{C}_\ssM ^2} } \simeq$ & $\stackrel{\ }{ \underset{\ }{ \dfrac{g^2}{2\pi^2} | \log( a \epsilon ) | } }$ & $\dfrac{g^2}{2\pi^2} |\log( a \epsilon )|$ & $\dfrac{g^2}{2\pi^2} |\log( e^{\gamma+1} m \epsilon )|$ \\
\cline{2-4}
$1\ \gg \  g^2 |\CMp | \sqrt{ 1 + \sfrac{\Delta_\ssM ^2}{\mathcal{C}_\ssM ^2} } \simeq$ & $\stackrel{\ }{ \underset{\ }{\dfrac{g^2 \omega}{6 a} } } $ & $\stackrel{\ }{ \underset{\ }{\dfrac{g^2\omega}{6 a} } } $ & $\dfrac{g^2 \pi \omega}{8 a} e^{ -{2m}/{a}}$ \\
\cline{2-4} 
$1\ \boldsymbol{\ll} \ \dfrac{g^2 \CM}{\omega} \sqrt{ 1 + \sfrac{\Delta_\ssM ^2}{\mathcal{C}_\ssM ^2} } \simeq$ & $\stackrel{\ }{ \underset{\ }{\dfrac{g^2 a}{4\pi^2\omega} } } $ & $ \underset{\ }{\dfrac{g^2 a}{4\pi^2\omega} } $ & $\dfrac{g^2 a}{8\pi\omega} e^{ -{2m}/{a}}$ \\
\cline{2-4} 
\end{tabular} }
        \caption{The leading-order behaviour for the validity relations \pref{APPdegenCond}. Note that in the last three bounds we used the fact that $\sqrt{1 + \Delta_\ssM ^2 / \mathcal{C}_\ssM ^2 } \simeq 1$ to leading-order (since $\left|\DM/\CM \right| \ll 1$). } \label{DegenTable2}
\end{table}

\begin{table}[h]
  \centering    
     \centerline{\begin{tabular}{ r|c|c|c|c|c|c| }
 \multicolumn{1}{r}{}
 & \multicolumn{1}{c}{$\underset{\ }{ \frac{\omega}{a} \ll \frac{m}{a} \ll 1 }$}
 & \multicolumn{1}{c}{$\frac{m}{a} \ll \frac{\omega}{a} \ll 1$}
 & \multicolumn{1}{c}{$\frac{\omega}{a} \ll 1 \ll \frac{m}{a}$} \\
\cline{2-4}
$\left| \dfrac{\DM}{\CM} \right|  \simeq$ & $\stackrel{\ }{ \underset{\ }{\dfrac{2\omega}{a}|\log(a\epsilon)|} }$ & $\underset{\ }{\dfrac{2\omega}{a}|\log(a\epsilon)|}$ & $\dfrac{4\omega}{\pi a} e^{2m/a} | \log(e^{\gamma+1} m \epsilon) |$ \\
\cline{2-4} 
\end{tabular} }
        \caption{The leading-order behaviour for the function $\DM / \CM$ in the $\omega \ll a$ regime. Note that $\DM /\CM \ll 1$. The third column with a $e^{2m/a}$ factor may seem alarming ({\it ie.} possibly not small), but in fact the last row in Table \ref{DegenTable2} ensures that the combination $\frac{\omega}{a} e^{2m/a} \ll g^2  / (8\pi)$ is small.} \label{DegenTable3}
\end{table}

Using the information in the above tables, the Markovian approximation demands that $\omega / a \ll 1$ as well as $1 \gg \left| \DM / \CM \right| \sim \mathcal{O}(g^2)$ as described in \S\ref{sec:Markovian}. This means that the solution \pref{AppDegenSolution} is 
\begin{eqnarray}
\varrho_{12}^{\ssI}(\tau) & \simeq & e^{+i \omega \tau} e^{ - 2 g^2 \CM \tau} \left[  \frac{\varrho_{12}(0)}{2} \bigg( 1 + i \frac{\omega}{g^2 \CM} \bigg) - \frac{\varrho_{12}^{\ast}(0)}{2} \left(1 - i \frac{\DM}{\CM} \right) \right] \\
& \ & \quad \quad \quad \quad \quad \quad \quad \quad \quad \quad \quad \quad \quad \quad + e^{+i \omega \tau} \left[  \frac{\varrho_{12}(0)}{2} \bigg( 1 - i \frac{\omega}{g^2 \mathcal{C}_\ssM } \bigg) + \frac{\varrho_{12}^{\ast}(0)}{2}\left(1 - i \frac{\DM}{\CM} \right) \right] \notag\ ,
\end{eqnarray}
where contributions $\mathcal{O}\left( \Delta_\ssM ^2 / \mathcal{C}_\ssM ^2 \right)$ have been neglected. At late times $g^2 a \tau \sim \mathcal{O}(1)$ the Schr\"odinger-picture state has the form
\begin{eqnarray}
\lim_{a \tau \sim \mathcal{O}(1/g^2)} \varrho_{12}(\tau) & \simeq &  \frac{\varrho_{12}(0)}{2} \bigg( 1 - i \frac{\omega}{g^2 \mathcal{C}_\ssM } \bigg) + \frac{\varrho_{12}^{\ast}(0)}{2}\left(1 - i \frac{\DM}{\CM} \right) \label{degenLateTimeStatic}
\end{eqnarray}
which has not yet fully decohered.

\section{Connection to the `rotating wave' approximation}
\label{App:RotWav}

In the literature, the issue of non-positivity of $\boldsymbol{\varrho}$ is usually addressed by taking an additional approximation called the {\it rotating-wave approximation} (RWA). This approximation is used when relaxation times of the qubit are very long compared to the time-scale of the system oscillations; {\it ie.} when $\omega \gg 1/\xi$. When this is so, the approximation involves coarse-graining over the fast oscillations, so that quickly oscillating factors in the interaction-picture equations of motion can be dropped by arguing that they average to zero. 

As applied to the qubit/field system considered here, the equation of motion for the diagonal component is completely unchanged by this averaging, while the off-diagonal equation in the {\it interaction picture},
\begin{eqnarray}
\frac{\partial \varrho^{\ssI}_{12}}{\partial \tau} & \simeq & - g^2 \CO \varrho^{\ssI}_{12}(\tau) + g^2 e^{+ 2 i \omega \tau} (\CO - i \DO) \varrho_{12}^{\ssI\ast}(\tau) \ , \label{RWAbefore}
\end{eqnarray}
is instead replaced by
\begin{eqnarray}
\frac{\partial \varrho^{\mathrm{RWA}\ssI}_{12}}{\partial \tau} & \simeq & - g^2 \CO \varrho^{\mathrm{RWA}\ssI}_{12}(\tau) \ , \label{RWAafter}
\end{eqnarray}
where we write $\boldsymbol{\varrho}^{\mathrm{RWA}}(\tau)$ to emphasize that this is describes evolution distinct from the Markovian equations derived in \S\ref{sec:LTL}. Replacing \pref{RWAbefore} with \pref{RWAafter} in the limit $\omega \gg 1/\xi$ is usually justified by claiming that the factor $e^{2i \omega \tau}$ in \pref{RWAbefore} oscillates extremely quickly by the time the state of the qubit changes significantly. In this sense the equations of motion in the RWA are supposed to describe a coarse-graining, since this oscillatory factor is supposed to average  to zero over any time-scales that can be resolved (as far as the evolution of the qubit is concerned).

The appeal of the RWA is that the solution $\boldsymbol{\varrho}^{\mathrm{RWA}}(\tau)$ is always positivity-preserving. In contrast to \S\ref{sec:LateTimeLimit}, the equations of motion in the RWA can be cast into the Lindblad form (in terms of the Schr\"odinger-picture state)
\begin{eqnarray}
\frac{\partial \boldsymbol{\varrho}^{\mathrm{RWA}}(\tau)}{\partial \tau} & = & - i \left[ \mathfrak{h} , \boldsymbol{\varrho}^{\mathrm{RWA}}(\tau) \right] + \sum_{j,k = 1}^{3} c^{\mathrm{RWA}}_{jk} \left( \boldsymbol{F}_{j} \boldsymbol{\varrho}^{\mathrm{RWA}}(\tau) \boldsymbol{F}_{k}^{\dagger} - \frac{1}{2} \left\{ \boldsymbol{F}_{k}^{\dagger} \boldsymbol{F}_{j}, \boldsymbol{\varrho}^{\mathrm{RWA}}(\tau) \right\} \right) \label{ExplicitLindbladRWA}
\end{eqnarray}
where $\boldsymbol{F}_{j} = \tfrac{1}{2} \boldsymbol{\sigma}_{j}$ again and the entries of the Kossakowski matrix are now instead
\begin{eqnarray}
\mathfrak{c}^{\mathrm{RWA}} & = & \left[ \begin{matrix} 2 g^2 \CO  & \  - 2 i g^2 \SO & 0 \\ 2 i g^2 \SO  & 2 g^2 \CO & 0 \\ 0& 0 & 0 \end{matrix} \right] \ .
\end{eqnarray}
This solution is positivity-preserving for any arbitrary choice of $\CO$ and $\SO$ now because the eigenvalues of $\mathfrak{c}^{\mathrm{RWA}}$ are non-negative\footnote{The eigenvalues of $\mathfrak{c}^{\mathrm{RWA}}$ are $0$ and $2g^2 ( \CO \pm \SO )$, where the latter two eigenvalues are positive because $\SO = - \tanh\left( \frac{\beta\omega}{2} \right) \CO$ on account of \pref{ratioSCgeneral} (and of course, $\CO$ is positive).}.

To the level of approximation we have taken in this work, we claim that it is {\it not} justified to take the rotating-wave approximation for this system. For convenience we re-state the solution to \pref{RWAbefore} in the interaction picture
\begin{eqnarray}
\varrho^{\ssI}_{12}(\tau) & \simeq &  \varrho_{12}(0) e^{ - g^2 \CO \tau } \ + \  \varrho_{12}^{\ast}(0) e^{ - g^2 \CO \tau } \left( \frac{g^2 \DO}{2\omega} + i \frac{g^2 \CO}{2\omega} \right)  \left( 1  - e^{ 2 i \omega \tau }  \right) \ . \label{int12solutionagain}
\end{eqnarray}
and in contrast, we state the RWA solution to \pref{RWAafter} in the interaction picture
\begin{eqnarray}
\varrho^{\mathrm{RWA}\ssI}_{12}(\tau) \simeq \varrho_{12}(0) e^{ - g^2 \CO \tau } \ , \label{RWAsolution}
\end{eqnarray}
which we see corresponds to the first term in \pref{int12solutionagain}. Recall that $g^2 \CO, g^2 \DO \ll \omega$ was assumed in the derivation of \pref{int12solutionagain} (when neglecting $\mathcal{O}(g^4)$ in the non-degenerate limit), which makes the sub-leading terms in this solution small, but {\it not} negligibly so. Although the usual assumption $\omega \gg 1/\xi = g^2 \CO$ of the RWA holds true here (at least in the non-degenerate limit), we see that the statement about rapidly oscillating factors seems here to be a red herring: clearly, even if the $e^{2i \omega \tau}$ factor is replaced with its average of zero in the solution \pref{int12solutionagain}, there are still other sub-leading corrections which remain (which do not oscillate).

This argument about dropping quickly oscillating factors in the interaction picture was most precisely outlined by Davies \cite{Spohn,Davies1,Davies2}, who showed that the Nakajima-Zwanzig equation formally reduces to the those in the RWA in the limit that $g \to 0$ while simultaneously taking $\tau \to \infty$ (such that $g^2 \tau$ is order unity)\footnote{In this limit, Davies formally showed that the Nakajima-Zwanzig equation reduces to a positivity-preserving Lindblad equation of the form $\frac{\partial \boldsymbol{\varrho}(\tau)}{\partial \tau} = - i [\mathfrak{h}, \varrho(\tau)] + g^2 K^{\natural}[\varrho(\tau)]$ where the operator $K^{\natural}$ is given by
\begin{eqnarray}
K^{\natural}[\boldsymbol{\varrho}(\tau)] & = & \lim_{T\to\infty} \frac{1}{2T} \int_{-T}^{T} \exd r\ e^{+ i \mathfrak{h} r} K[ e^{- i \mathfrak{h} r} \boldsymbol{\varrho}(\tau) e^{+ i \mathfrak{h} r} ] e^{- i \mathfrak{h} r} \notag  
\end{eqnarray}
with
\begin{eqnarray}
K[\boldsymbol{\varrho}(\tau)] & = & - \int_0^\infty \exd s\ \underset{\phi}{\mathrm{Tr}}\bigg( \big[ e^{+ i \mathcal{H} y^0(s)} \phi[y(\tau+s)] e^{- i \mathcal{H} y^0(s)} \otimes e^{+ i \mathfrak{h} s} \mathfrak{m} e^{- i \mathfrak{h} s} , \big[ \phi[y(\tau)] \otimes \mathfrak{m}, \ket{\Omega} \bra{\Omega} \otimes \boldsymbol{\varrho}(\tau) \big] \big] \bigg)  . \notag
\end{eqnarray}
which is an equivilent route to arrive at the RWA equations with \pref{RWAafter}.}. In this latter formulation of Davies, it is more clear what the rotating-wave approximation describes in this setting: in taking the limit $g \to 0$ the sub-leading terms of \pref{int12solutionagain} become neglected, while the damping factor $e^{- g^2 \CO \tau}$ must be kept since late times $g^2 a \tau \sim \mathcal{O}(1)$ are to be probed in the limit described by Davies. In our case, we are not inclined to drop the sub-leading $\mathcal{O}(g^2)$ terms in \pref{int12solutionagain} and so {\it do not} take the rotating-wave approximation.

\end{document}